\documentclass[journal,transmag]{IEEEtran}

\usepackage{amsmath,amssymb,array,graphicx,verbatim,color,cases,amsthm}
\usepackage{enumerate}
\usepackage[numbers, sort&compress]{natbib}
\usepackage{floatrow}
\newfloatcommand{capbtabbox}{table}[][\FBwidth]
\usepackage[font=small]{caption}
\usepackage{subcaption}
\usepackage{blindtext}
\usepackage{epstopdf} 
\usepackage[dvipsnames]{xcolor}
\usepackage{makecell}
\usepackage{graphics}
\usepackage{placeins}
\usepackage{afterpage}

\theoremstyle{plain}
\newtheorem{theorem}{Theorem}
\newtheorem{lemma}{Lemma}

\newtheorem{remark}{Remark}

\newcommand{\blue}[1]{\textcolor{black}{#1}}

\DeclareMathOperator*{\argmax}{arg\,max\,}

\title{Dynamic Market Mechanisms for Wind Energy \footnote{A priliminary version of this paper is presented in the 10th workshop on the economics of networks, systems and computation (NetEcon'15) \cite{NetEcon}}}

\author{Hamidreza Tavafoghi and Demosthenis Teneketzis}
\date{Working Paper\vspace*{5pt}\\This draft: 1 July 2016\\
First draft: 22 April 2015}

\newtheorem{assump}{Assumption}

\newtheorem{claim}{Claim}

\begin{document}

\maketitle
{\let\thefootnote\relax\footnote{A preliminary version of this paper was presented in the 10th workshop on the economics of networks, systems and computation (NetEcon'15) \cite{NetEcon}. This work was supported in part by the NSF under grants CNS-1238962 and CCF-1111061.\\		
		H. Tavafoghi and D. Teneketzis are with the Department of Electrical Engineering and Computer
		Science, University of Michigan, Ann Arbor, MI (e-mail:
		tavaf,teneket@umich.edu).}}

\renewcommand\thesubsection{\roman{subsection}}

\begin{abstract}
\textit{Abstract---} We investigate the problem of market mechanism design for wind energy. We consider a dynamic two-step model with \blue{one strategic seller with wind generation and one buyer, who trade energy through a mechanism determined by a designer}. The seller has private information about his technology and wind condition, which he learns dynamically over time.
We consider (static) forward and real-time mechanisms that take place at time $T\hspace*{-2pt}=\hspace*{-2pt}1$ and $T\hspace*{-2pt}=\hspace*{-2pt}2$, respectively. We also propose a dynamic mechanism that provides a coupling between the \blue{outcomes of the} forward and real-time markets. We show that the dynamic mechanism outperforms the forward and real-time  mechanisms \blue{for a general objective of the designer.} Therefore, we demonstrate the advantage of adopting dynamic market mechanisms over static market mechanisms for wind energy. \blue{The dynamic mechanism reveals information about wind generation in advance, and also provides flexibility for incorporation of new information arriving over time. We discuss how our results generalize to environments with many strategic sellers. We also study two variants of the dynamic mechanism that guarantee no penalty risk for the seller, and/or monitor the wind condition. We illustrate our results with a numerical example.}
\end{abstract}

\vspace*{-5pt}

 \section{Introduction}
 
Wind generation is intermittent and uncertain. An energy producer with wind energy (seller)  has neither complete control over his generation nor does he have an exact prediction of his generation in advance. The information about wind realization arrives dynamically over time and an accurate prediction is only available within a few (5-15) minutes of the generation time \cite{NERC2009}. The stochastic and dynamic nature of wind energy makes the integration of wind generation into grids a challenging task.

  The common practice for the integration of wind energy is to incorporate it into the existing two-settlement market architecture for conventional generators along with \textit{extra-market treatments} such as \textit{feed-in tariffs}, \textit{investment tax credit}, and \textit{production tax credit}. The two-settlement market architecture consists of forward markets (\textit{e.g.} day-ahead market) and real-time markets, where the outcome of forward markets is fed to real-time markets.
  
  For energy markets with low share of wind energy, like the U.S., it is possible to include wind energy in real-time markets, and treat it as negative load \cite{bitar2012selling}; we call this approach \textit{real-time mechanism}. One advantage of incorporating wind energy into real-time markets is that the allocation for wind generation is decided when all the information about wind generation is available. Moreover, a seller does not face any penalty risk as he commits to a certain level of generation only if he can produce it. However, in energy markets with high share of wind  energy, due to reliability concerns, inclusion of wind energy in real-time markets \blue{as negative load} is not possible.  
  
  For high shares of wind energy, the system operator needs to have information about wind generation in advance, and to incorporate wind energy as an active generation into its forward planning for power flow. Thus, wind energy is included in forward markets \cite{klessmann2008pros}; we call this approach \textit{forward mechanism}. 
  
  In forward markets, knowledge about wind generation in real-time is imperfect. Nevertheless,  a seller needs to commit to certain levels of generation in advance even without knowing the exact amount he will be able to produce in real-time. Therefore, the energy allocation decision in forward markets is determined only based on the incomplete information available at the time of forward markets, and all the new information that arrives after forward markets are closed is not incorporated into the energy allocation decision. In some countries, like the U.K., a seller is exposed to penalty risk if his real-time generation is different from his commitment in forward markets. We note that this is not an issue for a conventional generator as he has perfect knowledge of his real-time generation in advance.

The limitations of the real-time and forward market mechanisms discussed above motivate our work to study alternative market mechanisms for the integration of wind energy into grids. The U.S. department of energy encourages the development of ``rules for market evolution that enable system flexibility rationale'', \textit{i.e.} market mechanisms that give
a seller flexibility in generation, and provide opportunities for the demand side to actively
respond to changes in market over time \cite{cochran2012integrating}.  
It is desired that such mechanisms provide truthful (probabilistic) information about the seller's generation in real-time, and assign commitment to the seller in advance \cite{varaiya2011smart}. To achieve the above features, we need to study market mechanisms in a dynamic setting that accounts for the dynamic and intermittent nature of wind generation \blue{as well as the strategic behavior of the seller}. The forward and real-time mechanisms discussed above are static mechanisms in the strategic sense. That is, for each market mechanism, the sellers and buyers make simultaneous decisions only once, and their one-shot decisions determine the energy allocations and payments at that market; the outcome of the market is then assumed to be fixed and is fed as an exogenous parameter to the next market in time.

In this paper, we consider a simple two-step model that captures the dynamic and intermittent nature of wind generation. 
We propose a dynamic mechanism that provides a coupling between the forward and real-time mechanisms, and, unlike the (static) real-time and forward mechanisms, 
allows for flexible generation of wind energy, incorporates all the information that arrives over time, and provides forward commitment of the seller.  

\blue{To demonstrate the main ideas, we first consider a strategic setting with one buyer and one seller with wind generation. The buyer and the seller trade energy through a mechanism determined by a mechanism designer.}
\blue{The seller's cost depends on his private technology and the wind condition which he learns dynamically over time. Since the seller is strategic and profit maximizer,} he must be incentivized to reveal his private information about his cost function. 
We formally define such incentive payment, and utilize it in the formulation and solution of the mechanism design problems that we consider in this paper. \blue{We determine such incentive payments for different market mechanisms. We then characterize the set of feasible outcomes under each market mechanism. We show that the dynamic mechanism outperforms the real-time and forward mechanisms for a general objective of the designer. }

\blue{After we establish our results for a setting with one seller, we consider a setting with many sellers. We discuss how the problem of market design with many sellers is similar to that with one seller, and argue that our results generalize to this setting.}


\blue{Specifically,} we formulate and study \blue{three} different mechanism design problems. Two of these problems capture  the real-time and forward mechanisms. 
\blue{For the third problem}, we propose a new dynamic market mechanism 
\blue{that} dynamically couples the outcome of the real-time and forward markets. In the dynamic market mechanism, the seller is required to sequentially reveal his private information to the designer as it becomes available, and accordingly refines his commitment for energy generation over time. 

\blue{We show that the set of constraints that the designer faces due to the seller's strategic behavior and private information is less restrictive under the dynamic mechanism than under the forward and real-time mechanisms.} Consequently, we show that the proposed dynamic market mechanism outperforms the real-time and forward mechanisms. 
\blue{We further consider two variants of the dynamic mechanism; one guarantees no penalty risk to the seller, and in the other the designer monitors the wind speed. 
 By analyzing the outcome of these variants of the dynamic mechanism, we characterize the effect of penalty risk exposure and wind monitoring on the performance of the dynamic mechanism. }


\textbf{Related work:} Most of the literature on market design for wind energy assumes a static information structure and has mainly taken the forward mechanism approach. The works of \cite{bitar2012bringing,zhaowind} study the problem of optimal bidding in a forward market with an exogenous price and penalty rate. The works of \cite{lin2014forward,nayyar2013statistically} investigate the problem of mechanism design for wind aggregation among many wind producers that jointly participate in a forward market. The work in \cite{tang14VCG} studies the problem of auction design for a forward market that determines the penalty rate endogenously.

The concept of (flexible) contracts with risk in electricity market has been proposed in \cite{tan1993interruptible,bitar2013incentive,zhaowind,multicontract}. The authors of \cite{zhaowind} propose and investigate risky contracts for wind aggregation \blue {where there is no private information}. The work in \cite{tan1993interruptible} studies  the problem of efficient pricing of interruptible energy services. The authors of \cite{bitar2013incentive} look at the problem of optimal pricing for deadline-differentiated deferrable loads. The authors of \cite{multicontract} study the problem of forward risky contracts when the seller's private information is multi-dimensional and the wind is monitored. 
 
The various mechanism design problems formulated in this paper belong to the \textit{screening} literature in economics. We  formulate static and dynamic mechanism design problems. Our approach to the static mechanism design problems is similar to the one in \cite{borgers}. Our approach to the dynamic mechanism design problems is inspired by the ones in \cite{Courty2000,esHo2007optimal,krahmer14ex}. 
We provide a unified approach for all the mechanism design problems \blue{that enables us to demonstrate the advantage of dynamic mechanisms over static mechanisms for the wind energy market.} 

\textbf{Contribution:} We propose a dynamic model that enables us to provide a comparison among various market mechanisms for wind energy. We propose and analyze a dynamic market mechanism that couples the outcomes of the real-time and forward mechanisms. We show that the proposed dynamic mechanism outperforms the real-time and forward mechanisms \blue {for a general objective of the market designer}. The proposed dynamic mechanism reveals to the system designer the information required for planning in advance, incorporates the new information that arrives over time, and provides flexibility for the intermittent wind energy generation. \blue{We further study the effect of providing penalty insurance to the seller and monitoring the wind condition. We show that} the performance of the \blue{dynamic} mechanism with no penalty risk is in general inferior to the dynamic mechanism with penalty risk. \blue{Moreover, with wind monitoring, the outcome of the dynamic mechanism improves, as the seller cannot manipulate the outcome of the mechanism by misrepresenting his wind condition.}

\textbf{Organization:} 
We present our model in Section \ref{sec-model}. In Section \ref{sec-rent}, we discuss the market mechanism design problems as well as the seller's strategic behavior and private information. In Section \ref{sec-formulation}, we propose the dynamic market mechanism, and formulate \blue{three} mechanism design problems accordingly. We analyze the formulated mechanism design problems and compare them in Section \ref{sec-comp}. In Section \ref{sec-disc}, \blue{we provide some additional remarks and consider  variants of the proposed dynamic mechanism.}
We illustrate our results with an example in Section \ref{sec-example}. \blue{We discuss how our results generalize to settings with many sellers in Section \ref{sec-extension}, and} conclude in Section \ref{sec-con}. All the technical proofs can be found in Appendices A-C \cite{companion}.

\vspace*{-15pt}

\section{Model}
\label{sec-model}
Consider a buyer and a diversified seller with wind energy generation who \blue{trade energy through a mechanism determined by a designer}. \blue{We refer to the seller as ``he'' and the designer and the buyer as ``she''.}  The buyer gets utility $\mathcal{V}(\hat{q})$ from receiving energy $\hat{q}$; $\mathcal{V}(\hat{q})$ is increasing and strictly concave in $\hat{q}$. The seller has production cost $C(\hat{q};\theta)$ parametrized by his type $\theta\hspace{-2pt}\in\hspace{-2pt}\{\underline{\theta},\overline{\theta}\}$.\footnote{We assume that $C(\hat{q};\theta)$ captures the seller's operational cost, capital cost, and an exogenous opportunity cost associated with wind generation participation in his outside option. Moreover, we assume that the seller has a diversified energy portfolio, so he does not face a strict capacity constraint for generation as he can produce energy from other resources that are more expensive. Therefore, $C(\hat{q};\theta)$ is non-zero but decreasing in  the realization of wind \blue{speed}.} The seller's type depends on his technology $\tau$ \blue{(\textit{i.e.} technology of his wind turbines and the operational status of them,  the size of his wind farm and its location)} and the wind \blue{speed} $\omega$. We assume that at $T\hspace{-2pt}=\hspace{-2pt}1$ \blue{(\textit{e.g.} day ahead)}, \textit{ex-ante} the seller knows privately his technology $\tau$ for wind generation, which takes values from \blue{one of $M$ possible technologies} $\mathcal{T}\hspace{-3pt}:=\hspace{-3pt}\{\hspace{-1pt}\tau_1\hspace{-1pt},\hspace{-1pt}\tau_2,\cdots\hspace{-1pt},\hspace{-1pt}\tau_M\hspace{-1pt}\},$ \blue{$\tau_1\hspace{-1pt}<\hspace{-1pt}\tau_2\hspace{-1pt}<\hspace{-1pt}...\hspace{-1pt}<\hspace{-1pt}\tau_M$,} with probability $p_1,p_2,\cdots,p_M$, $\sum_{i=1}^Mp_i=1$, respectively.
At $T\hspace{-2pt}=\hspace{-2pt}2$ \blue{(\textit{e.g.} real-time)}, the seller receives additional information $\omega\in[\underline{\omega},\overline{\omega}]$ about the wind condition and can refine his private information about his cost function at $T\hspace{-2pt}=\hspace{-2pt}2$ as $\theta=\Theta(\tau,\omega)$. The probability distribution of wind $\omega$ is independent of the technology $\tau$ and is denoted by $G(\omega)$. We assume 
that  $C(\hat{q};\theta)$ is increasing in $\theta$, and $\Theta(\tau,\omega)$ is decreasing in $\omega$ and $\tau$. %
Define $C_\theta(\hat{q};\theta)\hspace{-2pt}:=\hspace{-2pt}\frac{\theta C(\hat{q};\theta)}{\partial \theta}$, $c(\hat{q};\theta)\hspace{-2pt}:=\hspace{-2pt}\frac{\partial C(\hat{q};\theta)}{\partial \hat{q}}$, $c_\theta(\hat{q};\theta)\hspace{-2pt}:=\hspace{-2pt}\frac{\partial c(\hat{q};\theta)}{\partial \theta}$, and $\Theta_\omega(\tau,\omega)\hspace{-2pt}:=\hspace{-2pt}\frac{\partial \Theta(\tau,\omega)}{\partial \omega}$.

\begin{assump}
The production cost $C(\hat{q};\hspace{-1pt}\theta)$ is increasing and convex in $\hat{q}$. The marginal cost $c(\hat{q};\theta)\hspace{-2pt}$ is increasing in $\theta$, \textit{i.e.}  $C_\theta(\blue{\hat{q}},\theta)\hspace{-2pt}\geq \hspace{-2pt}0$ and $c_\theta(\hat{q},\theta)\hspace{-2pt}\geq 0$. The seller's type $\Theta(\tau,\omega)$ is decreasing in $\tau$, \textit{i.e. }  $\Theta(\tau_i,\omega)\hspace{-2pt}\leq\hspace{-2pt} \Theta(\tau_j,\omega)$ for $i\hspace{-2pt}>\hspace{-2pt}j$, $\forall \omega$, and strictly decreasing in $\omega$, \textit{i.e.} $\Theta_\omega(\tau,\omega)\hspace{-2pt}<\hspace{-2pt}0$.
\label{assump-FSD}
\end{assump}

Let $F_i(\theta)$ denote the resulting conditional probability distribution of $\theta$ given $\tau_i$. 
Then, Assumption \ref{assump-FSD} states that a seller with technology $\tau_i$ expects to have a lower production cost than the one with technology $\tau_j$ for $i\hspace{-2pt}>\hspace{-2pt}j$, \textit{i.e.} $F_{\tau_j}(\theta)$ \textit{first order stochastically dominates} $F_{\tau_i}(\theta)$.

We also make the following technical assumption. 
\begin{assump}
The distribution $F_\tau(\theta)$ has \textit{non-shifting support}, \text{i.e.} $f_\tau$ is \blue{strictly positive} on the interval $\left[\underline{\theta},\overline{\theta}\right]$ for $\tau\in\blue{\mathcal{T}}$.
\label{assum-nonshift}
\end{assump}

Assumption \ref{assum-nonshift} implies that the range of achievable values of $\theta$ is the same for all technologies $\tau_i$,\hspace{-1pt} $i\hspace{-2pt}=\hspace{-3pt}1\hspace{-1pt},\hspace{-1pt}...,\hspace{-2pt}M$. However, the same realization of the wind $\omega$ results in different values of $\theta$ for different technologies. Thus, the probability distribution of $\theta$, given by $F_{\tau_i}(\theta)$, is different for different technologies (\textit{cf.} Assumption \ref{assump-FSD}).


\section{Mechanism Design and \blue{Strategic Behavior}}
\label{sec-rent}
Consider an arbitrary mechanism that \blue {determines the agreement between the buyer and the seller}. 
Let $t(\hspace{-1pt}\tau\hspace{-1pt},\hspace{-1pt}\omega\hspace{-1pt})$ denote the payment the buyer ends up paying to the seller with technology $\tau$ and wind \blue{speed} $\omega$, and $q(\hspace{-1pt}\tau\hspace{-1pt},\hspace{-1pt}\omega\hspace{-1pt})$ denote the amount of energy the seller delivers according to the mechanism. The social welfare $\mathcal{S}$, the seller's revenue $\mathcal{R}$, and the buyer's utility $\mathcal{U}$ can be written as,
\begin{align}
&\mathcal{S}:= \mathbb{E}_{\tau,\omega}\hspace*{-2pt}\left\{\mathcal{V}(q(\tau,\omega))\hspace*{-2pt}-\hspace*{-1pt}C(q(\tau,\omega);\Theta(\tau,\omega))\right\}\hspace{-1pt},\\
&\mathcal{R}:= \mathbb{E}_{\tau,\omega}\hspace*{-2pt}\left\{t(\tau,\omega)\hspace*{-2pt}-\hspace*{-1pt}C(q(\tau,\omega);\Theta(\tau,\omega))\right\}\hspace{-1pt},\label{inforent}\\
&\mathcal{U}:=\mathbb{E}_{\tau,\omega}\hspace*{-2pt}\left\{\mathcal{V}(q(\tau,\omega))\hspace*{-2pt}-\hspace*{-1pt}t(\tau,\omega)\right\}\hspace*{-2pt}=\mathcal{S}\hspace*{-2pt}-\hspace*{-2pt}\mathcal{R}.
\label{buyer-obj}
\end{align}
The social welfare $\mathcal{S}$ only depends on $q$ and is independent of $t$. Thus, by the first order optimality condition, an allocation $q^{e*}$ is socially efficient (maximizing $\mathcal{S}$) if and only if at $q^{e*}$ the marginal utility $v(\hat{q})\hspace{-2pt}:=\hspace{-2pt}\frac{\partial \mathcal{V}(\hat{q})}{\partial \hat{q}}$ is equal to the marginal cost $c(\hat{q};\theta)$, \textit{i.e.}, for all $\tau,\omega$,
\begin{align}
v(q^{e*}(\tau,\omega))=c(q^{e*}(\tau,\omega);\Theta(\tau,\omega)).
\label{eff-all}
\end{align}
If the seller is not strategic or does not have any private information, the mechanism designer can set $q(\tau,\hspace{-1pt}\omega)$ equal to $q^{e*}(\tau,\hspace{-1pt}\omega)$, 
\blue {to maximize the social welfare $\mathcal{S}$. Then, the designer can set payment $t(\tau,\hspace{-1pt}\omega)$ so as to achieve any arbitrary surplus distribution between the buyer and the seller.}

However, the strategic seller does not simply reveal his production cost function, which is his private information. Therefore, the socially efficient allocation (\ref{eff-all}) is not sinply implementable by the mechanism designer. The seller has incentives to manipulate the outcome (misrepresent his cost function, by misreporting $\tau$ and $\omega$) in order to gain a higher revenue. Thus, the mechanism's output would differ from the efficient allocation $q^{e*}$.

Due to the seller's strategic behavior, any mechanism must (i) incentivize the seller to truthfully reveal his private information, and (ii) leave a non-negative revenue to the seller so that he voluntarily participates in the mechanism. 

\blue{Following the literature on regulation and market design \cite{laffont1993theory}, we assume that the mechanism designer has the following general objective,
\begin{align}
\mathcal{W}&:=\mathcal{U}+\alpha \mathcal{R}, \label{eq:designerobj}
\end{align}
where $\alpha\hspace{-1pt}\in\hspace{-1pt}[0,1]$. When $\alpha\hspace{-1pt}=\hspace{-1pt}1$, the designer wants to maximize the social welfare $\mathcal{S}$. When $\alpha\hspace{-1pt}=\hspace{-1pt}0$, the designer seeks to maximize the utility of the buyer (demand side). For $\alpha\hspace{-1pt}\in\hspace{-1pt}(\hspace{-1pt}0\hspace{-1pt},\hspace{-1pt} 1\hspace{-1pt})$, the designer's objective is to maximize a weighted sum of the buyer's utility $\mathcal{U}$ and the seller's revenue $\mathcal{R}$. We assume that the designer knows the buyer's utility function $\mathcal{V}(\hat{q})$.\footnote{\blue{In practice, even when the designer's objective is to maximize the social welfare, we have $\alpha<1$ due to cost/loss associated with financial transaction between the buyer and the seller (see \cite{laffont1993theory} for more discussion). We note that the case where $\alpha>1$ is of no interest, as it implies that simply the money flow from the buyer to the seller increases the designer's objective.}}} 

 We invoke the revelation principle for multistage games \cite{myerson86}, and, without loss of generality, restrict attention to  direct revelation  mechanisms that are incentive compatible. In these mechanisms, \blue{the designer determines} a mechanism for the payment and the allocation \blue{$\{\hspace{-1pt}t(\hspace{-1pt}\tau\hspace{-2pt},\hspace{-1pt}\omega\hspace{-1pt})\hspace{-1pt},\hspace{-1pt}q(\hspace{-1pt}\tau\hspace{-2pt},\hspace{-1pt}\omega\hspace{-1pt})\hspace{-1pt},\hspace{-1pt}\tau\hspace{-3pt}\in\hspace{-3pt}\mathcal{T}\hspace{-1pt},\hspace{-1pt}\omega\hspace{-3pt}\in\hspace{-3pt}[\underline{\omega}\hspace{-1pt},\hspace{-1pt}\overline{\omega}]\hspace{-1pt}\}$} based on the seller's technology $\tau$ and the wind \blue{speed} $\omega$, and asks the seller to report his private information \blue{about} $\tau$ and $\omega$ over time. 
\blue{The designer determines} \blue{$\{\hspace{-1pt}t(\hspace{-1pt}\tau\hspace{-2pt},\hspace{-1pt}\omega\hspace{-1pt})\hspace{-1pt},\hspace{-1pt}q(\hspace{-1pt}\tau\hspace{-2pt},\hspace{-1pt}\omega\hspace{-1pt})\hspace{-1pt},\hspace{-1pt}\tau\hspace{-3pt}\in\hspace{-3pt}\mathcal{T}\hspace{-2pt},\hspace{-1pt}\omega\hspace{-2pt}\in\hspace{-2pt}[\underline{\omega}\hspace{-1pt},\hspace{-1pt}\overline{\omega}]\}$} so as to ensure the truthful report of the seller; this is called \textit{incentive compatibility}  (IC). A mechanism \blue{$\{\hspace{-1pt}t(\hspace{-1pt}\tau\hspace{-2pt},\hspace{-1pt}\omega\hspace{-1pt})\hspace{-1pt},\hspace{-1pt}q(\hspace{-1pt}\tau\hspace{-2pt},\hspace{-1pt}\omega\hspace{-1pt})\hspace{-1pt},\hspace{-1pt}\tau\hspace{-3pt}\in\hspace{-3pt}\mathcal{T}\hspace{-2pt},\hspace{-1pt}\omega\hspace{-2pt}\in\hspace{-2pt}[\underline{\omega}\hspace{-1pt},\hspace{-1pt}\overline{\omega}]\}$} is incentive compatible if it is always optimal for the seller to report truthfully his private information. 
The seller voluntarily participates in the mechanism if he earns a positive (expected) revenue from the agreement; this is known as \textit{individual rationality} (IR). 

Define $\mathcal{R}_{\tau\hspace*{-1pt},\omega}\hspace*{-3pt}:=\hspace*{-2pt}t(\tau\hspace*{-2pt},\hspace*{-1pt}\omega\hspace*{-1pt})\hspace*{-2pt}-\hspace*{-2pt}C(\hspace*{-1pt}q(\hspace*{-1pt}\tau\hspace*{-2pt},\hspace*{-1pt}\omega);\hspace*{-1pt}\Theta(\hspace*{-1pt}\tau\hspace*{-2pt},\hspace*{-1pt}\omega\hspace*{-1pt})\hspace*{-1pt})$ as the seller's revenue with technology $\tau$ and wind $\omega$, and $\mathcal{R}_\tau\hspace{-2pt}:=\hspace{-2pt}\mathbb{E}_\tau\{\mathcal{R}_{\tau\hspace*{-1pt},\omega}\}$ as the seller's expected revenue with technology $\tau$. Then,  $\mathcal{R}\hspace{-2pt}=\hspace{-2pt}\mathbb{E}_{\tau}\{\mathcal{R}_{\tau}\}\hspace{-2pt}=\hspace{-2pt}\mathbb{E}_{\tau\hspace*{-1pt},\omega}\{\mathcal{R}_{\tau\hspace*{-1pt},\omega}\}$. 
\blue{In the next section, we show that the IC and IR constraints can be written in terms of $\mathcal{R}_{\tau\hspace*{-1pt},\omega}$, and $\mathcal{R}_\tau$.}
\blue{We show that} any mechanism design problem can be formulated as a constrained functional optimization problem, where we determine the optimal allocation and the seller's revenue that maximize $\mathcal{W}$ subject to the IC and IR constraints. 


\section{Market Mechanisms}
\label{sec-formulation}
\blue{We consider different structures and timings of mechanisms, and formulate the resulting mechanism design problems accordingly. We consider a forward mechanism that takes place at $T=1$. We also consider a real-time mechanism that takes place at $T=2$. }
Moreover, we propose \blue{a dynamic} mechanism that takes place at $T=1$ and $T=2$. Therefore, we formulate \blue{three} mechanism design problems for the model of Section \ref{sec-model}: (A) real-time mechanism,  (B) forward mechanism, (C) dynamic mechanism. \blue{By comparing the solutions to these mechanism design problems, we demonstrate the advantage of the dynamic mechanism over the static forward and real-time mechanisms.}

\blue{We show that when the objective of the designer is to maximize $\mathcal{W}$, given by (\ref{eq:designerobj}), each of the mechanism design problems mentioned above can be formulated as a functional optimization problem with different sets of constraints. In Section \ref{sec-comp}, we determine the restrictions that each of these sets of constraints implies on the outcome of the mechanism design problems. Thus, we demonstrate how each of the three market structures  impact the market outcome.}
\subsection{Forward Mechanism}
\label{F} 
\blue{In the forward mechanism, the designer specifies the mechanism that determines the agreement between the buyer and the seller at  $T\hspace{-2pt}=\hspace{-2pt}1$; the information about wind speed $\omega$ that becomes available at $T\hspace{-2pt}=\hspace{-2pt}2$ is ignored.} This mechanism is similar to the current day-ahead integration in Europe, and the proposed firm contracts in the literature. 

\blue{Since information $\omega$ is not available at $T\hspace{-2pt}=\hspace{-2pt}1$, the allocation function $q(\hspace{-1pt}\tau\hspace{-1pt},\hspace{-1pt}\omega\hspace{-1pt})$ and the payment function $t(\hspace{-1pt}\tau\hspace{-1pt},\hspace{-1pt}\omega\hspace{-1pt})$ are independent of $\omega$. Therefore, we drop the dependence on $\omega$ and denote the allocation and payment functions for the forward mechanism by $q(\hspace{-1pt}\tau\hspace{-1pt})$ and $t(\hspace{-1pt}\tau\hspace{-1pt})$, respectively. The optimal forward mechanism is then given by the solution to the following optimization problem:}
\begin{align}
&\blue{\hspace*{60pt}\max_{q(\cdot),t(\cdot)}{\mathcal{W}}}\nonumber\\
\hspace{30pt}&\hspace{-30pt}\text{subject to}\nonumber\\
&IC\hspace{-3pt}:\mathcal{R}_{\tau}\geq\hspace{-2pt} t(\hat{\tau})\hspace{-2pt}-\hspace{-2pt}\mathbb{E}_\omega\{C(\hspace{-1pt}q(\hat{\tau});\hspace{-1pt}\Theta(\tau,\hspace{-1pt}\omega))\}\hspace{-1pt}\quad\forall \tau,\hat{\tau},\hspace{-1pt}
\label{rent-forward} \\
&IR\hspace{-3pt}: \mathcal{R}_\tau\geq 0\quad \forall \tau.\label{IR-F}
\end{align}
We note that (\ref{IR-F}) only ensures a positive expected revenue for the seller with technology $\tau$, and, therefore, exposes him to the risk of penalty (\textit{i.e.} $\mathcal{R}_{\tau,\omega}\hspace*{-2pt}<\hspace*{-2pt}0$) for low realizations of $\omega$.\blue{\footnote{\blue{For simplicity, we do not model forward mechanisms with explicit penalty rate for shortfalls here. In Section \ref{sec-disc}, we discuss how our result continue to hold when we consider forward market mechanisms with explicit penalty for shortfalls.}}   }

\subsection{Real-Time Mechanism}
\label{R-NM}
\blue{In the real-time mechanism, the designer specifies the mechanism that determines the agreement between the buyer and the seller at $T\hspace{-2pt}=\hspace{-2pt}2$, after the information about wind speed $\omega$ is available to the seller.} We assume that the wind \blue{speed $\omega$} is not monitored by the \blue{designer}. This mechanism is similar to the current real-time market integration in the U.S.
The allocation and payment functions $\{\hspace{-1pt}q(\tau\hspace{-1pt},\hspace{-1pt}\omega),\hspace{-1pt}t(\tau\hspace{-1pt},\hspace{-1pt}\omega)\hspace{-1pt}\}$ depend on $(\tau,\omega)$, and the seller reports/reveals his technology $\tau$ and his private knowledge about $\omega$ simultaneously at $T\hspace{-2pt}=\hspace{-2pt}2$. \blue{The optimal real-time mechanism is then given by the solution to the following optimization problem:} 
\begin{align}
&\blue{\hspace{70pt}\max_{q(\cdot,\cdot),t(\cdot,\cdot)}{\mathcal{W}}}\nonumber\\
&\hspace{-30pt}\text{subject to}\nonumber\\
&IC\hspace{-3pt}:\hspace{-1pt}\mathcal{R}_{\tau,\omega}\hspace{-3pt}\geq\hspace{-3pt} t(\hspace{-1pt}\hat{\tau}\hspace{-1pt},\hspace{-1pt}\hat{\omega}\hspace{-1pt})\hspace{-2pt}-\hspace{-2pt}C(\hspace{-1pt}q(\hspace{-1pt}\hat{\tau}\hspace{-1pt},\hspace{-1pt}\hat{\omega}\hspace{-1pt});\hspace{-1pt}\Theta(\hspace{-1pt}\tau\hspace{-1pt},\hspace{-1pt}\omega\hspace{-1pt})\hspace{-1pt})\quad\forall \tau\hspace{-1pt},\hspace{-1pt}\omega\hspace{-1pt},\hspace{-1pt}\hat{\tau}\hspace{-1pt},\hspace{-1pt}\hat{\omega}\hspace{-1pt},
\label{rent-R-NM}\\
&IR\hspace{-3pt}:\mathcal{R}_{\tau,\omega}\geq 0\quad \forall \tau,\omega. \label{IR-R-NM}
\end{align} 
Equation (\ref{rent-R-NM}) ensures that the seller's revenue by reporting the true value of $(\hspace{-1pt}\tau\hspace{-1pt},\hspace{-1pt}\omega\hspace{-1pt})$ is higher than the one with any misreport \blue{$(\hspace{-1pt}\hat{\tau}\hspace{-1pt},\hspace{-1pt}\hat{\omega}\hspace{-1pt})$}. Equation (\ref{IR-R-NM}) guarantees a positive revenue for the seller for all wind realizations, thus, ensuring no penalty risk for \blue{him}. 

\subsection{Dynamic Mechanism}
\label{SNMP}
\blue{In the dynamic mechanism, the designer specifies the mechanism that determines the agreement between the buyer and the seller  at $T\hspace{-2pt}=\hspace{-2pt}1$. However, unlike the forward mechanism, the dynamic mechanism determines an agreement that is contingent on the information about the wind speed $\omega$ that becomes available at $T\hspace{-2pt}=\hspace{-2pt}2$.} 
Moreover, unlike the real-time mechanism, the seller reports $\tau$ and $\omega$ sequentially; at $T\hspace{-1pt}=\hspace{-2pt}1$, he reports $\tau$, then at $T\hspace{-2pt}=\hspace{-2pt}2$, he reports $\omega$.  \blue{We assume that the wind \blue{speed $\omega$} is not monitored by the designer.} \blue{Therefore, the optimal dynamic mechanism is given by the solution to the following optimization problem:} 
\begin{align}
&\blue{\hspace{50pt}\max_{q(\cdot,\cdot),t(\cdot,\cdot)}{\mathcal{W}}}\nonumber\\
\hspace{40pt}&\hspace{-35pt}\text{subject to}\nonumber\\
\hspace{40pt}&\hspace{-45pt}IC_1\hspace{-3pt}:\hspace{-2pt}\mathcal{R}_{\tau}\hspace{-4pt}\geq\hspace{-2pt}\hspace{-2pt} \mathbb{E}_\omega\hspace{-1pt}\{t(\hspace{-1pt}\hat{\tau}\hspace{-1pt},\hspace{-1pt}\sigma(\omega\hspace{-1pt})\hspace{-1pt})\hspace{-2pt}-\hspace{-2pt}C(\hspace{-1pt}q(\hspace{-1pt}\hat{\tau}\hspace{-1pt},\hspace{-1pt}\sigma(\omega\hspace{-1pt})\hspace{-1pt})\hspace{-1pt};\hspace{-1pt}\Theta(\hspace{-1pt}\tau\hspace{-1pt},\hspace{-1pt}\omega\hspace{-1pt})\hspace{-1pt})\hspace{-1pt}\}\hspace{-8pt}\quad\forall\hspace{-1pt} \tau\hspace{-1pt},\hspace{-1pt}\hat{\tau}\hspace{-1pt},\hspace{-1pt}\sigma(\hspace{-1pt}\cdot\hspace{-1pt})\hspace{-1pt},\hspace{-3pt}
\label{rent-S-NM-P-1} \\
\hspace{40pt}&\hspace{-45pt}IC_2\hspace{-3pt}:\hspace{-2pt}\mathcal{R}_{\tau,\omega}\hspace{-2pt}\geq\hspace{-2pt} t(\tau,\hat{\omega})\hspace{-2pt}-\hspace{-2pt}C(\hspace{-1pt}q(\tau,\hspace{-1pt}\hat{\omega});\hspace{-1pt}\Theta(\tau,\hspace{-1pt}\omega))\hspace{-1pt}\quad\forall \tau,\omega,\hat{\omega}\hspace{-1pt},\label{rent-S-NM-P-2} 
\\
\hspace{40pt}&\hspace{-45pt}IR\hspace{-3pt}:\hspace{-2pt}\mathcal{R}_{\tau}\hspace{-2pt}\geq\hspace{-2pt}0\quad \forall \tau. \label{IR-S-NM-P}
\end{align}
\blue{The above dynamic mechanism design problem involves two IC constraints (\ref{rent-S-NM-P-1}) and (\ref{rent-S-NM-P-2}).} By (\ref{rent-S-NM-P-1}), the designer ensures the seller's true report of $\tau$, even when he can potentially coordinate his misreport $\hat{\tau}$ at $T\hspace*{-2pt}=\hspace*{-2pt}1$ with an arbitrary (mis)report strategy $\sigma(\omega)$ of $\omega$ at  $T\hspace*{-2pt}=\hspace*{-2pt}2$. The designer ensures the seller's truthful report of $\omega$ at $T\hspace{-2pt}=\hspace{-2pt}2$ by (\ref{rent-S-NM-P-2}), assuming that the seller already reported the true $\tau$ at time $T\hspace*{-2pt}=\hspace*{-2pt}1$. 

\section{Comparison of Market Mechanisms}
\label{sec-comp}
 \blue{In all three mechanism design problems formulated in Section \ref{sec-formulation}, the designer wants to maximize $\mathcal{W}=\mathcal{U}+\alpha \mathcal{R}=\mathcal{S}+(\alpha-1)\mathcal{R}$.} \blue{However, in each problem the designer faces a different set of constraints due to the seller's strategic behavior and private information about his cost function, as well as the specific mechanism structure and rules. In this section, we analyze these sets of constraints so as to compare the performance of the three  mechanism design problems presented in Section \ref{sec-formulation}.}  
 
 \blue{We proceed as follows. We consider the set of constraints that the designer needs to satisfy for each market mechanism:  constraints (\ref{rent-forward},\ref{IR-F}) for the forward mechanism, constraints (\ref{rent-R-NM},\ref{IR-R-NM}) for the real-time mechanism, and constraints (\ref{rent-S-NM-P-1}-\ref{IR-S-NM-P}) for the dynamic mechanism. We investigate the implications of each of these sets of constraints under the corresponding market structure. We provide a reduced form of these  constraints in Theorems \ref{thm-lemma} and \ref{thm-rent}. Using the results of Theorems \ref{thm-lemma} and \ref{thm-rent}, we show that the set of constraints in the dynamic mechanism is less restrictive than the set of constraints in the forward and real-time mechanisms. Therefore, we show that the outcome of the dynamic market mechanism outperforms those of the forward and real-time mechanisms (Theorem \ref{thm-buyer}). }

We start \blue{our analysis} by determining the condition that $\mathcal{R}_{\tau,\omega}$ must satisfy so as to ensure the seller's truthful report about $\omega$ \blue{in the real-time and dynamic mechanisms}.

\begin{theorem}
	\label{thm-lemma}
	\blue{The real-time and dynamic mechanisms are incentive compatible for $\omega$ (constraints (\ref{rent-R-NM}) and (\ref{rent-S-NM-P-2}), respectively), if the allocation function $q(\tau,\omega)$ is increasing in $\omega$, and the seller's revenue $\mathcal{R}_{\tau,\omega}$ satisfies }
	\begin{align}
	\frac{\partial\mathcal{R}_{\tau,\omega}}{\partial \omega}=C_\theta(q(\tau,\omega);\Theta(\tau,\omega))\Theta_\omega(\tau,\omega)\geq0,
	\label{rent-IC2}
	\end{align}
	where $C_\theta(q;\theta):=\frac{\partial C(q;\theta)}{\partial \theta}$ and $\Theta_\omega(\tau,\omega):=\frac{\partial \Theta(\tau,\omega)}{\partial \omega}$. The inequality (\ref{rent-IC2}) is strict if $q(\tau,\omega)>0$.\\

\end{theorem}
Theorem \ref{thm-lemma} states that, \blue{under the assumption that the seller reports truthfully his technology $\tau$, his }revenue $\mathcal{R}_{\tau,\omega}$ must be increasing in $\omega$, as in (\ref{rent-IC2}), so that \blue{he is incentivized }to report $\omega$ truthfully. 
      
 We next provide conditions on $\mathcal{R}_\tau=\mathbb{E}_\omega\{\mathcal{R}_{\tau,\omega}\}$ so as to ensure that the seller with technology $\tau_i$ reports truthfully \blue{in the forward, real-time, and dynamic mechanisms (constraints (\ref{rent-forward}), (\ref{rent-R-NM}), and (\ref{rent-S-NM-P-1}) resp.). Moreover, we determine the conditions that $\mathcal{R}_\tau$ must satisfy so as to ensure the seller's voluntary participation in the forward, real-time, and dynamic mechanisms (constraints (\ref{IR-F}), (\ref{IR-R-NM}), and (\ref{IR-S-NM-P}) resp.).}

For that matter, we define, 
\begin{align}
&\mathcal{R}^{\hspace{-1pt}T}\hspace{-2pt}(\hspace{-1pt}\tau_j\hspace{-1pt},\hspace{-2pt}\tau_i;\hspace{-1pt}q)\hspace{-3pt}:=\hspace{-5pt}\int{\hspace{-5pt}\left(\hspace{-1pt}C(\hspace{-1pt}q(\hspace{-1pt}\tau_j\hspace{-1pt},\hspace{-1pt}\omega\hspace{-1pt});\hspace{-2pt}\Theta(\hspace{-1pt}\tau_j\hspace{-1pt},\hspace{-1pt}\omega\hspace{-1pt})\hspace{-1pt})\hspace{-2pt}-\hspace{-2pt}C(\hspace{-1pt}q(\hspace{-1pt}\tau_j\hspace{-1pt},\hspace{-1pt}\omega\hspace{-1pt});\hspace{-2pt}\Theta(\hspace{-1pt}\tau_i\hspace{-1pt},\hspace{-1pt}\omega\hspace{-1pt})\hspace{-1pt})\hspace{-1pt}\right)\hspace{-2pt}dG(\omega)}\hspace{-1pt},\nonumber\\
&\mathcal{R}^{\hspace{-1pt}W}\hspace{-2pt}(\hspace{-1pt}\tau_j\hspace{-1pt},\hspace{-2pt}\tau_i;\hspace{-1pt}q)\hspace{-3pt}:=\hspace{-5pt}\int\hspace{-8pt}\int_{\omega}^{\sigma^*\hspace{-1pt}(\tau_j;\tau_i,\omega)}\hspace{-35pt}\left(C_\theta(q(\hspace{-1pt}\tau_j\hspace{-1pt},\hspace{-1pt}\omega\hspace{-1pt});\hspace{-1pt}\Theta(\hspace{-1pt}\tau_j\hspace{-1pt},\hspace{-1pt}\hat{\omega}\hspace{-1pt})\hspace{-1pt})\hspace{-2pt}-\hspace{-2pt}C_\theta(q(\hspace{-1pt}\tau_j\hspace{-1pt},\hspace{-1pt}\hat{\omega}\hspace{-1pt});\hspace{-1pt}\Theta(\hspace{-1pt}\tau_j\hspace{-1pt},\hspace{-1pt}\hat{\omega}\hspace{-1pt})\hspace{-1pt})\hspace{-1pt}\right)\nonumber\\
&\hspace{173pt}\Theta_\omega\hspace{-1pt}(\hspace{-1pt}\tau_j\hspace{-1pt},\hspace{-1pt}\hat{\omega}\hspace{-1pt})\hspace{-1pt}d\hat{\omega}dG(\omega),\nonumber
\end{align}
where 
$\sigma^{\hspace{-1pt}*}\hspace{-1pt}(\hspace{-1pt}\tau_j;\hspace{-1pt}\tau_i\hspace{-1pt},\hspace{-1pt}\omega\hspace{-1pt})$ is uniquely defined by $\Theta(\hspace{-1pt}\tau_j\hspace{-1pt},\hspace{-1pt}\omega\hspace{-1pt})\hspace{-2pt}=\hspace{-2pt}\Theta(\hspace{-1pt}\tau_i\hspace{-1pt},\sigma^{\hspace{-1pt}*}\hspace{-1pt}(\hspace{-1pt}\tau_j\hspace{-1pt};\hspace{-1pt}\tau_i\hspace{-1pt},\hspace{-1pt}\omega\hspace{-1pt})\hspace{-1pt})$\footnote{This is true because of Assumption \ref{assum-nonshift}.}. \blue{That is, $\sigma^{\hspace{-1pt}*}\hspace{-1pt}(\hspace{-1pt}\tau_j;\hspace{-1pt}\tau_i\hspace{-1pt},\hspace{-1pt}\omega\hspace{-1pt})$ denotes the wind speed that the seller with technology $\tau_j$ requires so as to have a generation cost identical to that of the seller with technology $\tau_i$ and wind speed $\omega$. Below, we elaborate on the role of $\mathcal{R}^T$ and $\mathcal{R}^W$.}

\blue{Consider a seller with technology $\tau_i$ and wind speed $\omega$. The seller can misreport his technology (say, by declaring $\tau_j$), or his wind speed (say, by declaring $\hat{\omega}$), or both (by declaring $\tau_j\hspace{-1pt},\hspace{-1pt}\hat{\omega}$). In Theorem \ref{thm-rent}, below, we prove the following result. Under the assumption that the seller reports truthfully his wind speed $\omega$, the payment that incentivizes him to report truthfully $\tau_i$ instead of misreporting $\tau_j$ is given by $\mathcal{R}^T\hspace{-1pt}(\hspace{-1pt}\tau_j\hspace{-1pt},\hspace{-2pt}\tau_i;\hspace{-1pt}q\hspace{-1pt})$. 
	The seller may also misreport his wind speed $\omega$ (say, by declaring $\hat{\omega}$) after misreporting his technology $\tau_i$ as $\tau_j$. In this case, $\mathcal{R}^W\hspace{-1pt}(\hspace{-1pt}\tau_j\hspace{-1pt},\hspace{-2pt}\tau_i;\hspace{-1pt}q\hspace{-1pt})$ represents the additional expected payment that incentivizes the seller not to misreport his technology $\tau_i$ as $\tau_j$ even when he can misreport his true wind speed. Consequently, $\mathcal{R}^T\hspace{-1pt}(\hspace{-1pt}\tau_j\hspace{-1pt},\hspace{-1pt}\tau_i;\hspace{-1pt}q\hspace{-1pt})\hspace{-1pt}+\hspace{-1pt}\mathcal{R}^W\hspace{-1pt}(\hspace{-1pt}\tau_j\hspace{-1pt},\hspace{-1pt}\tau_i;\hspace{-1pt}q\hspace{-1pt})$ represents the incentive payment that the designer must provide to the seller with technology $\tau_i$ so as to incentivize him not to misreport his technology as $\tau_j$.  }

%

\vspace{5pt}

\blue{We also define,
	\begin{align}
	&\mathcal{R}^P(\hspace{-1pt}\tau_i\hspace{-1pt};\hspace{-1pt}q\hspace{-1pt})\hspace{-3pt}:=\hspace{-3pt}\int\hspace{-4pt}\left[\hspace{-1pt}1\hspace{-2pt}-\hspace{-2pt}G(\hspace{-1pt}\omega\hspace{-1pt})\hspace{-1pt}\right] C_\theta(q(\tau_i,\omega);\Theta(\tau_i,\omega)\hspace{-1pt})\Theta_\omega(\tau_i,\omega)d\omega,\nonumber
	\end{align}
	which results from integrating (\ref{rent-IC2}), followed by an expectation with respect to $\omega$. We show below that $\mathcal{R}^P(\tau_1;q)$ determines the minimum incentive payment that the seller with technology $\tau_1$ (the worse technology) must receive in order to voluntarily participate in the real-time mechanism. The precise statement of these results is as follows.}

\begin{theorem}
	\label{thm-rent}
In the mechanism design problems formulated in Section \ref{sec-formulation}, the incentive compatibility and individual rationality constraints can be reduced to the following conditions.
\begin{enumerate}[a)]
	\item \label{class-b}\blue{For the forward mechanism, 
	\begin{align}
	q(\hspace{-1pt}\tau\hspace{-1pt},\hspace{-1pt}\omega\hspace{-1pt}) \textnormal{ is independent of } \omega \label{allocation-forward},
	\end{align}
	\vspace*{-20pt}}
	\begin{align}
	&\hspace*{-10pt}\mathcal{R}_{\tau_{i}}\hspace{-2pt}-\hspace{-2pt}\mathcal{R}_{\tau_{i-1}}\hspace{-2pt}=\hspace{-2pt} \mathcal{R}^T\hspace*{-1pt}(\tau_{i},\tau_{i-1};q)\hspace{-2pt}\geq\hspace{-2pt}0\hspace{-2pt}\quad \forall i\hspace*{-2pt}\in\hspace*{-2pt}\{2,..,M\},\label{ICbound-monitoring}\\
	&\hspace*{-10pt}\mathcal{R}_{\tau_1}= 0.\label{IRbound-monitoring}
	\end{align}
	\item \label{class-e}\blue{For the real-time mechanism,
	\begin{align}
	\hspace{-8pt}q(\hspace{-1pt}\tau\hspace{-2pt},\hspace{-1pt}\omega\hspace{-1pt}) \textnormal{\hspace{-2pt} is\hspace{-2pt} only \hspace{-2pt}dependent\hspace{-2pt} on \hspace{-2pt}} \Theta(\hspace{-1pt}\tau\hspace{-2pt},\hspace{-1pt}\omega\hspace{-1pt}) \textnormal{\hspace{-2pt} and\hspace{-2pt} \hspace{-2pt} increasing\hspace{-2pt} in \hspace{-2pt}} \omega\hspace{-1pt},\hspace{-8pt} \label{allocation-real}
	\end{align} 
	\vspace*{-18pt}}
	\begin{align}
	&\blue{\hspace{-14pt}\frac{\partial\mathcal{R}_{\tau,\omega}}{\partial \omega}=C_\theta(q(\tau,\omega);\Theta(\tau,\omega))\Theta_\omega(\tau,\omega)\geq0,}\label{IC2-real}\\
	&\hspace*{-14pt}\mathcal{R}_{\tau_{i}}\hspace{-2pt}-\hspace{-2pt}\mathcal{R}_{\tau_{i-1}}\hspace{-2pt}=\hspace{-2pt} \mathcal{R}^T\hspace*{-1pt}(\tau_{i},\tau_{i-1};q)\hspace{-2pt}+\hspace*{-2pt}\mathcal{R}^W\hspace*{-1pt}(\tau_{i},\tau_{i-1};q)\hspace{-2pt}\geq\hspace{-2pt}0\hspace{-7pt}\quad\nonumber\\&\hspace{136pt} \forall i\hspace{-1pt}\hspace{-2pt}\in\hspace{-2pt}\{\hspace{-1pt}2\hspace{-1pt},\hspace{-2pt}...,\hspace{-2pt}M\hspace{-1pt}\}\hspace*{-1pt},\hspace{-2pt}\label{ICbound-R-NM}\\
	&\hspace*{-14pt}\mathcal{R}_{\tau_1}= \mathcal{R}^P\hspace{-1pt}(\tau_1;q).\label{IRbound-R-NM}
	\end{align}
	\item \label{class-c}\blue{For the dynamic mechanism, 
	\begin{align}
		q(\hspace{-1pt}\tau\hspace{-1pt},\hspace{-1pt}\omega\hspace{-1pt}) \textnormal{ is increasing in } \omega,\label{allocation-dynamic}
	\end{align}
	\vspace*{-18pt}}
	\begin{align}
	&\blue{\hspace{-14pt}\frac{\partial\mathcal{R}_{\tau,\omega}}{\partial \omega}=C_\theta(q(\tau,\omega);\Theta(\tau,\omega))\Theta_\omega(\tau,\omega)\geq0,}\label{IC2-dynamic}\\
	&\hspace*{-14pt}\mathcal{R}_{\tau_i}\hspace{-2pt}-\hspace{-2pt}\mathcal{R}_{\tau_j}\hspace{-2pt}\geq\hspace{-2pt} \mathcal{R}^T\hspace*{-2pt}(\tau_j,\tau_i;q)\hspace{-2pt}+\hspace*{-2pt}\mathcal{R}^W\hspace*{-2pt}(\tau_j,\tau_i;q)\hspace{-2pt}\geq\hspace{-2pt}0,\hspace{-7pt}\quad\nonumber\\&\hspace{86pt} \forall i\hspace{-1pt},\hspace{-1pt}j\hspace{-1pt}\hspace{-2pt}\in\hspace{-2pt}\{\hspace{-1pt}1\hspace{-1pt},\hspace{-2pt}...,\hspace{-2pt}M\hspace{-1pt}\}\hspace*{-1pt},\hspace*{-1pt}i\hspace*{-3pt}>\hspace*{-3pt}j\hspace{-1pt},\hspace{-2pt}\label{ICbound-S-NM-P}\\
	&\hspace*{-14pt}\mathcal{R}_{\tau_1}= 0.\label{IRbound-S-NM-P}
	\end{align}	
\end{enumerate}
Moreover, $R^T\hspace*{-2pt}(\hspace{-1pt}\tau_j\hspace{-1pt},\hspace*{-3pt}\tau_i;\hspace*{-1pt}q\hspace{-1pt})\hspace*{-3pt}\geq\hspace*{-2pt}0$ (strict if $q(\hspace{-1pt}\tau_j\hspace{-1pt},\hspace{-1pt}\omega\hspace{-1pt})\hspace*{-2pt}\neq\hspace*{-2pt} 0$ for some $\omega$) and $R^W\hspace*{-1pt}(\hspace{-1pt}\tau_j\hspace{-1pt},\hspace{-2pt}\tau_i\hspace{-1pt};\hspace{-1pt}q\hspace{-1pt})\hspace*{-2pt}\geq \hspace*{-2pt}0$ (strict if $q(\hspace{-1pt}\tau\hspace{-2pt},\hspace{-1pt}\omega\hspace{-1pt})$ is dependent on $\omega$). 
\end{theorem}

\blue{We now  comment on the results of Theorem \ref{thm-rent}. }
\blue{The results of parts (a-c) reduce the set of constraints (\ref{rent-forward},\ref{IR-F}) for the forward mechanism, constraints (\ref{rent-R-NM},\ref{IR-R-NM}) for the real-time mechanism, and constraints (\ref{rent-S-NM-P-1}-\ref{IR-S-NM-P}) for the dynamic mechanism to those given by (\ref{allocation-forward}-\ref{IRbound-monitoring}), (\ref{allocation-real}-\ref{IRbound-R-NM}), and (\ref{allocation-dynamic}-\ref{IRbound-S-NM-P}), respectively. 
	} 
	
	\blue{Part (a) of Theorem \ref{thm-rent} follows from the fact that the forward mechanism takes place at $T\hspace{-2pt}=\hspace{-2pt}1$ when information about $\omega$ in not available. Therefore, the allocation function is independent of $\omega$, and, thus, $\mathcal{R}^{\hspace{-1pt}W}\hspace{-2pt}(\hspace{-1pt}\tau_i\hspace{-1pt},\hspace{-2pt}\tau_j;\hspace{-1pt}q)\hspace{-2pt}=\hspace{-2pt}0$ by its definition. Therefore, the designer only needs to provide the payment $\mathcal{R}^{\hspace{-1pt}T}\hspace{-2pt}(\hspace{-1pt}\tau_i\hspace{-1pt},\hspace{-2pt}\tau_j;\hspace{-1pt}q)$ so as to incentivize the seller to report truthfully his technology $\tau_i$ instead of misreporting $\tau_j$. Part (a) of Theorem \ref{thm-rent} states further that when the seller's technology is $\tau_i$, the payment $\mathcal{R}^{\hspace{-1pt}T}\hspace{-2pt}(\hspace{-1pt}\tau_{i-1}\hspace{-1pt},\hspace{-2pt}\tau_i;\hspace{-1pt}q)$ is enough to incentivize the seller not to misreport $\tau_i$ as $\tau_{i-1}$ or as any other technology $\tau_j$.}
		

\blue{In part (b) of Theorem \ref{thm-rent}, constraint (\ref{allocation-real}) states that the allocation function $q(\tau\hspace{-1pt},\hspace{-1pt}\omega)$ must be only a function of $\Theta(\tau\hspace{-1pt},\hspace{-1pt}\omega)$, as the seller reports both of $\tau$ and $\omega$ simultaneously at $T\hspace{-2pt}=\hspace{-2pt}2$. Therefore, the designer cannot differentiate between different pairs of $(\tau\hspace{-1pt},\hspace{-1pt}\omega)$ that correspond to the same cost function $C(q;\hspace{-1pt}\Theta(\tau\hspace{-1pt},\hspace{-1pt}\omega)\hspace{-1pt})$. Moreover, (\ref{allocation-real}) requires the allocation function to be increasing in $\omega$, which, along with constraint (\ref{IC2-real}), ensures the seller's truth telling about $\omega$ (Theorem \ref{thm-lemma}). Constraint (\ref{ICbound-R-NM}) states that when the seller's technology is $\tau_i$, the payment $\mathcal{R}^{\hspace{-1pt}T}\hspace{-1pt}(\tau_{i-1}\hspace{-1pt},\hspace{-2pt}\tau_i;\hspace{-1pt}q)\hspace{-1pt}+\hspace{-1pt}\mathcal{R}^{\hspace{-1pt}W}\hspace{-1pt}(\tau_{i-1}\hspace{-1pt},\hspace{-2pt}\tau_i;\hspace{-1pt}q)$ is enough to incentivize the seller not to misreport $\tau_i$ as $\tau_{i-1}$ or as any other technology $\tau_j$. We note that, unlike the forward mechanism where the seller with technology $\tau_1$ receives no incentive payment, in the real-time mechanism the seller with technology $\tau_1$ receives a positive expected incentive payment $\mathcal{R}^{\hspace{-1pt}P}\hspace{-1pt}(\tau_1;\hspace{-1pt}q)$, given by (\ref{IRbound-R-NM}), that ensures truth telling about $\omega$.}   

\blue{In part (c) of Theorem \ref{thm-rent}, constraints (\ref{allocation-dynamic},\ref{IC2-dynamic}) ensure the seller's truth telling about $\omega$ by Theorem \ref{thm-lemma}. Equation (\ref{ICbound-S-NM-P}) determines the incentive payment that the designer needs to provide to the seller with technology $\tau_i$, so that he does not misreport his technology as $\tau_j$, and does not misreport his wind speed $\omega$ along with $\tau_j$. We note that, unlike the real-time market, the seller with technology $\tau_1$ does not receive a positive expected incentive payment (see (\ref{IRbound-S-NM-P})).}

\blue{From the designer's point of view, the dynamic mechanism has the following advantages over the forward and real-time mechanisms: (i) In contrast to the forward mechanism, the dynamic mechanism incorporates the information about the wind speed $\omega$ that becomes available at $T\hspace{-1pt}=\hspace{-1pt}2$ into the allocation and payment functions. As a result, the set of allocation and payment functions available to the designer in the dynamic mechanism is richer than the ones available in the forward mechanism. (ii) In the real-time mechanism, the seller reports $\tau$ and $\omega$ simultaneously. Therefore, he can perfectly coordinate his reports about $\tau$ and $\omega$. In the dynamic mechanism, the seller reports $\tau$ and $\omega$ sequentially over time, thus, he cannot perfectly coordinate his reports about $\tau$ and $\omega$. As a result, in the dynamic mechanism, the designer can distinguish among different pairs $(\hspace{-1pt}\tau\hspace{-1pt},\hspace{-1pt}\omega\hspace{-1pt})$ that result in the same cost function  $C(q;\hspace{-1pt}\Theta(\tau\hspace{-1pt},\hspace{-1pt}\omega)\hspace{-1pt})$; this is not the case in the real-time mechanism (see (\ref{allocation-real}) and (\ref{allocation-dynamic})). Furthermore, in the dynamic mechanism the designer faces a less restrictive set of constraints on the seller's revenue $\mathcal{R}$ than the ones in the real-time mechanism (see (\ref{IRbound-R-NM}) and (\ref{IRbound-S-NM-P})).}
	

\blue{Using the result of Theorem \ref{thm-rent}, we can determine the adequate incentive payment to the seller that is associated with an allocation function $q(\hspace{-1pt}\tau\hspace{-2pt},\hspace{-1pt}\omega\hspace{-1pt})$ for the forward, real-time, and dynamic mechanisms. Consequently, we can omit the payment function $t(\hspace{-1pt}\tau\hspace{-2pt},\hspace{-1pt}\omega\hspace{-1pt})$ and optimize over the allocation function $q(\hspace{-1pt}\tau\hspace{-2pt},\hspace{-1pt}\omega\hspace{-1pt})$  to determine the optimal forward, real-time, and dynamic mechanisms.\footnote{For dynamic mechanisms, the set of constraints (\ref{ICbound-S-NM-P}) are in the form of inequality constraints, and it is not possible to determine a priori which of these inequality constraints are binding at the optimal solution (see \cite{battaglini2015optimal} for more discussion). Therefore, we need to make assumptions about which subset of these inequality constraints are binding, and omit the payment function using the assumed binding conditions. We then need to verify that the set of assumed binding constraints are in fact binding at the computed optimal dynamic mechanism.}  Under a set of regularity conditions,  the resulting functional optimization problems, which are in terms of the allocation function $q(\hspace{-1pt}\tau\hspace{-1pt},\hspace{-1pt}\omega\hspace{-1pt})$, can be solved point-wise separately for every pair $(\hspace{-1pt}\tau\hspace{-1pt},\hspace{-1pt}\omega\hspace{-1pt})$ as a value optimization problem. The closed form solutions of the optimal forward, real-time, and dynamic mechanisms can be found in Appendix C \cite{companion}.}
	
\blue{Using the result of Theorem \ref{thm-rent}, we can compare the sets of constraints that the designer faces under the forward, real-time, and forward mechanisms. We prove below that for any arbitrary designer's objective of the form (\ref{eq:designerobj}), the outcome is superior under the optimal dynamic mechanism than under the optimal forward or optimal real-time mechanisms.}

\begin{theorem}
	\label{thm-buyer}
\blue{The designer's objective under the optimal dynamic mechanism is higher than her objectives under the optimal forward and the real-time mechanisms, \textit{i.e.}
$\mathcal{W}^{dynamic}>\mathcal{W}^{forward}$ and $\mathcal{W}^{dynamic}>\mathcal{W}^{real-time}$.} \end{theorem}

The result of Theorem \ref{thm-buyer} demonstrates the advantage of the dynamic mechanism over the forward and real-time mechanisms. 
\blue{We note that Theorem \ref{thm-buyer} does not provide a comparison between the forward  and real-time mechanisms. That is because the designer faces different sets of constraints in the forward and the real-time mechanisms. On one hand, the forward mechanism ignores $\omega$ in its allocation function, while the real-time mechanism incorporates $\omega$. On the other hand, the incentive payments are higher in the real-time mechanism than in the forward mechanism (see (\ref{IRbound-R-NM}) and (\ref{IRbound-monitoring})) since the seller can perfectly  coordinate his simultaneous reports about $\tau$ and $\omega$ in the real-time mechanism. The impact of these constraints on the performance of the	forward and the real-time mechanisms depend on the specific parameters of the model, and thus, there is no generic ordering of the designer's objective under the forward and real-time mechanisms.}


\section{Additional Remarks}
\label{sec-disc}

\blue{In this section, we examine the following variations of the problems formulated in Section \ref{sec-formulation}, and analyzed in Section \ref{sec-comp}. First, we consider a forward mechanism with explicit penalty rate for shortfalls, and compare it with the dynamic mechanism proposed in Section \ref{sec-formulation}. Second, we consider a dynamic mechanism that guarantees no loss for the seller for every realization of wind speed $\omega$. We show that the performance of this mechanism is superior to that of the real-time mechanism and inferior to that of the dynamic mechanism proposed in Section \ref{sec-formulation}. Third, we consider a dynamic mechanism with monitoring, where the designer monitors the wind speed $\omega$. We show that the dynamic mechanism with monitoring outperforms the dynamic mechanism proposed in Section \ref{sec-formulation}, where the designer does not monitor the wind speed $\omega$. Moreover, in the dynamic mechanism with monitoring, the designer can guarantee no loss for the seller without compromising the performance of the mechanism. }

\blue{\subsection{Forward Mechanism with Penalty Rate}}

\blue{The forward mechanism formulated in Section \ref{sec-formulation} ignores the wind speed $\omega$, and determines the allocation $q(\hspace{-1pt}\tau\hspace{-1pt})$ and  payment function $t(\hspace{-1pt}\tau\hspace{-1pt})$ only as a function of $\tau$. Here, we consider a variation of the forward mechanism where the seller is not bound to produce exactly $q(\hspace{-1pt}\tau\hspace{-1pt})$ at $T\hspace{-1pt}=\hspace{-1pt}2$. However, if the seller's generation at $T\hspace{-1pt}=\hspace{-1pt}2$ falls short of his commitment $q(\tau)$, he faces a penalty charge rate $\lambda(\hspace{-1pt}\tau\hspace{-1pt})$ by the designer for each unit of generation that he falls short of producing. The penalty rate $\lambda(\hspace{-1pt}\tau\hspace{-1pt})$ is agreed at $T=1$ based on the seller's report about his technology. We refer to this variation of the forward mechanism as \textit{forward mechanism with penalty rate}. The work of \cite{tang14VCG}  studies the design of such a mechanism for wind procurement, and the works of \cite{bitar2012bringing,zhaowind} consider settings where $\lambda$ is exogenously fixed and does not depend on $\tau$. }

\blue{We note that the forward mechanism with penalty rate does not fully ignore $\omega$ at $T\hspace{-1pt}=\hspace{-1pt}2$, as it allows the seller to change his generation at $T\hspace{-1pt}=\hspace{-1pt}2$ based on the realized wind condition $\omega$. However, we argue below that such incorporation of $\omega$ into the forward mechanism with penalty rate is not efficient. Therefore, the dynamic mechanism  outperforms the forward mechanism with penalty rate. }

\blue{A forward mechanism with penalty rate that is incentive compatible and individually rational can be characterized by allocation and  payment functions $\{\hspace*{-1pt}q(\tau)\hspace*{-1pt},\hspace*{-1pt}t(\tau)\hspace*{-1pt},\hspace*{-1pt}\tau\hspace*{-1pt}\in\hspace*{-1pt}\mathcal{T}\hspace*{-1pt}\}$, along with associated penalty rates $\{\hspace*{-1pt}\lambda(\tau)\hspace*{-1pt},\hspace*{-1pt}\tau\hspace*{-1pt}\in\hspace*{-1pt}\mathcal{T}\hspace*{-1pt}\}$. Let $e(\hspace*{-1pt}\tau\hspace*{-1pt},\hspace*{-1pt}\omega\hspace*{-1pt})$, $e(\hspace*{-1pt}\tau\hspace*{-1pt},\hspace*{-1pt}\omega\hspace*{-1pt})\hspace*{-1pt}\leq\hspace*{-1pt} q(\hspace*{-1pt}\tau\hspace*{-1pt})$, denote the amount of energy that the seller with technology $\tau$ and wind speed $\omega$ actually produces at $T=2$ (given the described forward mechanism with penalty rate). The seller with technology $\tau$ and wind $\omega$ chooses $e(\hspace*{-1pt}\tau\hspace*{-1pt},\hspace*{-1pt}\omega\hspace*{-1pt})$ so as to maximize his revenue as,
\begin{align}
e(\tau,\omega)\hspace{-2pt}:=\hspace{-2pt}\argmax_{0\leq\hat{e}\leq q(\tau)}\left\{t(\tau)\hspace{-2pt}-\hspace{-2pt}C(\hat{e};\Theta(\tau,\omega))-\lambda(\tau)(q(\tau)\hspace{-2pt}-\hspace{-2pt}\hat{e})\right\}.\nonumber
\end{align} }
\blue{Using $e(\hspace*{-1pt}\tau\hspace*{-1pt},\hspace*{-1pt}\omega\hspace*{-1pt})$, we can define a dynamic mechanism $\{\hspace*{-1pt}\tilde{q}(\hspace*{-1pt}\tau\hspace*{-1pt},\hspace*{-1pt}\omega\hspace*{-1pt})\hspace*{-1pt},\hspace*{-1pt}\tilde{t}(\hspace*{-1pt}\tau\hspace*{-1pt},\hspace*{-1pt}\omega\hspace*{-1pt})\hspace*{-1pt},\hspace*{-1pt}\tau\hspace*{-2pt}\in\hspace*{-2pt}\mathcal{T}\hspace*{-1pt},\hspace*{-1pt}\omega\hspace*{-2pt}\in\hspace*{-2pt}[\underline{\omega}\hspace*{-1pt},\hspace*{-1pt}\overline{\omega}]\hspace*{-1pt}\}$ that is equivalent to the forward mechanism with penalty rate described above. Define,
	\begin{align} 
		&\tilde{q}(\tau,\omega):=e(\tau,\omega),\\
		&\tilde{t}(\tau,\omega):=t(\tau)-\lambda(\tau)(q(\tau)\hspace{-2pt}-\hspace{-2pt}e(\tau,\omega))\label{forward-eq}.
	\end{align}}
	\blue{The dynamic mechanism $\{\hspace*{-1pt}\tilde{q}(\hspace*{-1pt}\tau\hspace*{-1pt},\hspace*{-1pt}\omega\hspace*{-1pt})\hspace*{-1pt},\hspace*{-1pt}\tilde{t}(\hspace*{-1pt}\tau\hspace*{-1pt},\hspace*{-1pt}\omega\hspace*{-1pt})\hspace*{-1pt},\hspace*{-1pt}\tau\hspace*{-2pt}\in\hspace*{-2pt}\mathcal{T}\hspace*{-1pt},\hspace*{-1pt}\omega\hspace*{-2pt}\in\hspace*{-2pt}[\underline{\omega}\hspace*{-1pt},\hspace*{-1pt}\overline{\omega}]\hspace*{-1pt}\}$ defined above is incentive compatible and individually rational since it induces the same allocation and payment function for the seller with technology $\tau$ and wind $\omega$ as the forward mechanism with penalty rate.} 

\blue{In the equivalent dynamic mechanism $\{\hspace*{-1pt}\tilde{q}(\hspace*{-1pt}\tau\hspace*{-1pt},\hspace*{-1pt}\omega\hspace*{-1pt})\hspace*{-1pt},\hspace*{-1pt}\tilde{t}(\hspace*{-1pt}\tau\hspace*{-1pt},\hspace*{-1pt}\omega\hspace*{-1pt})\hspace*{-1pt},\hspace*{-1pt}\tau\hspace*{-2pt}\in\hspace*{-2pt}\mathcal{T}\hspace*{-1pt},\hspace*{-1pt}\omega\hspace*{-2pt}\in\hspace*{-2pt}[\underline{\omega}\hspace*{-1pt},\hspace*{-1pt}\overline{\omega}]\hspace*{-1pt}\}$ constructed above, the payment function $\tilde{t}(\hspace*{-1pt}\tau\hspace*{-1pt},\hspace*{-1pt}\omega\hspace*{-1pt})$ is linear in $\tilde{q}(\hspace*{-1pt}\tau\hspace*{-1pt},\hspace*{-1pt}\omega\hspace*{-1pt})$ (see (\ref{forward-eq})). However, in the optimal dynamic mechanism, the payment $t(\hspace*{-1pt}\tau\hspace*{-1pt},\hspace*{-1pt}\omega\hspace*{-1pt})$ is in general, nonlinear in $q(\hspace*{-1pt}\tau\hspace*{-1pt},\hspace*{-1pt}\omega\hspace*{-1pt})$ (see the example in Section \ref{sec-example}).  Therefore, even though the forward mechanism with penalty rate allows the seller to modify his generation according to  the realized wind speed $\omega$, such incorporation of $\omega$ is not as efficient as in the dynamic mechanism. }

\blue{\subsection{Dynamic Mechanism with no Penalty}}
\label{sec-penalty}

\blue{The dynamic mechanism formulated in Section \ref{sec-formulation} promises a positive expected revenue to the seller for every technology $\tau$ (see the IR constraint (\ref{IR-S-NM-P})). However, once the seller signs the agreement at $T\hspace*{-2pt}=\hspace*{-2pt}1$, he is committed to following the terms of the agreement. Similar to the forward mechanism, it is possible that in the dynamic mechanism,  the realized revenue of the seller becomes negative for very low realizations of wind speed, \textit{i.e.} $\mathcal{R}_{\tau\hspace*{-1pt},\omega}\hspace*{-1pt}<\hspace*{-1pt}0$ for small values of $\omega$ (note that  the IR constraint (\ref{IR-S-NM-P}) only guarantees $\mathcal{R}_\tau\hspace*{-1pt}=\hspace*{-1pt}  \mathbb{E}\{\mathcal{R}_{\tau\hspace*{-1pt},\omega}\}\hspace*{-1pt}\geq\hspace*{-1pt} 0$). }

\blue{In this section, we consider a variation of the dynamic mechanism that guarantees no loss for the seller for every realization of $\omega$, \textit{i.e.} $\mathcal{R}_{\tau,\omega}\geq 0$. We refer to this mechanism as \textit{the dynamic mechanism with no penalty}. The optimal dynamic mechanism  with no penalty is given by the solution to the following optimization problem:}
\blue{\begin{align}
&\blue{\hspace{50pt}\max_{q(\cdot,\cdot),t(\cdot,\cdot)}{\mathcal{W}}}\nonumber\\
\hspace{40pt}&\hspace{-35pt}\text{subject to}\nonumber\\
\hspace{40pt}&\hspace{-45pt}IC_1\hspace{-3pt}:\hspace{-2pt}\mathcal{R}_{\tau}\hspace{-4pt}\geq\hspace{-2pt}\hspace{-2pt} \mathbb{E}_\omega\hspace{-1pt}\{t(\hspace{-1pt}\hat{\tau}\hspace{-1pt},\hspace{-1pt}\sigma(\omega\hspace{-1pt})\hspace{-1pt})\hspace{-2pt}-\hspace{-2pt}C(\hspace{-1pt}q(\hspace{-1pt}\hat{\tau}\hspace{-1pt},\hspace{-1pt}\sigma(\omega\hspace{-1pt})\hspace{-1pt})\hspace{-1pt};\hspace{-1pt}\Theta(\hspace{-1pt}\tau\hspace{-1pt},\hspace{-1pt}\omega\hspace{-1pt})\hspace{-1pt})\hspace{-1pt}\}\hspace{-8pt}\quad\forall\hspace{-1pt} \tau\hspace{-1pt},\hspace{-1pt}\hat{\tau}\hspace{-1pt},\hspace{-1pt}\sigma(\hspace{-1pt}\cdot\hspace{-1pt})\hspace{-1pt},\hspace{-3pt}
\label{rent-S-NM-P-1-np} \\
\hspace{40pt}&\hspace{-45pt}IC_2\hspace{-3pt}:\hspace{-2pt}\mathcal{R}_{\tau,\omega}\hspace{-2pt}\geq\hspace{-2pt} t(\tau,\hat{\omega})\hspace{-2pt}-\hspace{-2pt}C(\hspace{-1pt}q(\tau,\hspace{-1pt}\hat{\omega});\hspace{-1pt}\Theta(\tau,\hspace{-1pt}\omega))\hspace{-1pt}\quad\forall \tau,\omega,\hat{\omega}\hspace{-1pt},\label{rent-S-NM-P-2-np} 
\\
\hspace{40pt}&\hspace{-45pt}IR\hspace{-3pt}:\hspace{-2pt}\mathcal{R}_{\tau,\omega}\hspace{-2pt}\geq\hspace{-2pt}0\quad \forall \tau, \omega. \label{IR-S-NM-P-np}
\end{align}}
\blue{The above optimization problem is similar to the optimization problem formulated for the optimal dynamic mechanism in Section \ref{sec-formulation}. The only difference is that we replace the \textit{ex-ante} IR constraint $\mathcal{R}_\tau\hspace*{-1pt}\geq\hspace*{-1pt}0$, given by (\ref{IR-S-NM-P}), with  the \textit{ex-post} IR constraint $\mathcal{R}_{\tau\hspace*{-1pt},\omega}\hspace*{-1pt}\geq\hspace*{-1pt}0$, given by (\ref{IR-S-NM-P-np}).}

\blue{It is clear that the IR constraint $\mathcal{R}_{\tau\hspace*{-1pt},\omega}\hspace*{-1pt}\geq\hspace*{-1pt}0$ is more restrictive than the \textit{ex-ante} IR constraint $\mathcal{R}_\tau\hspace*{-1pt}\geq\hspace*{-1pt}0$. Therefore, the performance of the optimal dynamic mechanism with no penalty is inferior to that of the optimal dynamic mechanism. Nevertheless, we show below that the optimal dynamic mechanism with no penalty outperforms the optimal real-time mechanism.}

\blue{\begin{theorem} \label{thm-penalty} (i) The set of IC and IR constraints for dynamic mechanism with no penalty, given by (\ref{rent-S-NM-P-1-np}-\ref{IR-S-NM-P-np}), can be reduced to the following conditions,
	\begin{align}
	q(\hspace{-1pt}\tau\hspace{-1pt},\hspace{-1pt}\omega\hspace{-1pt}) \textnormal{ is increasing in } \omega,\label{allocation-dynamic-np}
	\end{align}
\begin{align}
&\blue{\hspace{-14pt}\frac{\partial\mathcal{R}_{\tau,\omega}}{\partial \omega}=C_\theta(q(\tau,\omega);\Theta(\tau,\omega))\Theta_\omega(\tau,\omega)\geq0,}\label{IC2-dynamic-np}\\
&\hspace*{-14pt}\mathcal{R}_{\tau_i}\hspace{-2pt}-\hspace{-2pt}\mathcal{R}_{\tau_j}\hspace{-2pt}\geq\hspace{-2pt} \mathcal{R}^T\hspace*{-2pt}(\tau_j,\tau_i;q)\hspace{-2pt}+\hspace*{-2pt}\mathcal{R}^W\hspace*{-2pt}(\tau_j,\tau_i;q)\hspace{-2pt}\geq\hspace{-2pt}0,\hspace{-7pt}\quad\nonumber\\&\hspace{86pt} \forall i\hspace{-1pt},\hspace{-1pt}j\hspace{-1pt}\hspace{-2pt}\in\hspace{-2pt}\{\hspace{-1pt}1\hspace{-1pt},\hspace{-2pt}...,\hspace{-2pt}M\hspace{-1pt}\}\hspace*{-1pt},\hspace*{-1pt}i\hspace*{-3pt}>\hspace*{-3pt}j\hspace{-1pt},\hspace{-2pt}\label{ICbound-S-NM-P-np}\\
&\hspace*{-14pt}\mathcal{R}_{\tau_1}\geq \mathcal{R}^P\hspace{-1pt}(\tau_1;q).\label{IRbound-S-NM-P-np}
\end{align}	
(ii) The designer's objective, given by (\ref{eq:designerobj}), under the optimal dynamic mechanism with no penalty is higher than her objective under the optimal real-time mechanism and lower than her objective under the optimal dynamic mechanism, \textit{i.e.}
$\mathcal{W}^{\text{dynamic}}\hspace{-2pt}>\hspace{-2pt}\mathcal{W}^{\text{dynamic no penalty}}\hspace{-2pt}>\hspace{-2pt}\mathcal{W}^{\text{real-time}}$.  
\end{theorem}}

\blue{Part (i) of Theorem \ref{thm-penalty} states the analogue result of part (c) of Theorem \ref{thm-rent} for dynamic mechanisms with no penalty. We note that the set of constraints (\ref{IC2-dynamic-np},\ref{ICbound-S-NM-P-np}) for the dynamic mechanisms with no penalty is the same as the set of constraints (\ref{IC2-dynamic},\ref{ICbound-S-NM-P}) for dynamic mechanisms in part (c) of Theorem \ref{thm-rent}. This is because the IC constraints (\ref{rent-S-NM-P-1-np},\ref{rent-S-NM-P-2-np}) for dynamic mechanisms with no penalty are the same as the IC constraints  (\ref{rent-S-NM-P-1},\ref{rent-S-NM-P-2}) for dynamic mechanisms. }
\blue{However, constraint (\ref{IRbound-S-NM-P-np}) for the dynamic mechanism with no penalty is different from constraint (\ref{IRbound-S-NM-P}) for the dynamic mechanism. In the dynamic mechanism with no penalty we impose the ex-post IR constraint (\ref{IR-S-NM-P-np}) instead of the less restrictive interim IR constraint (\ref{IR-S-NM-P}) imposed for the dynamic mechanism. As a result, the designer cannot reduce the incentive payment she pays to the seller by exposing him to penalty risk for low realizations of wind speed $\omega$. The ex-post constraint requires that $\mathcal{R}_{\tau_1\hspace*{-1pt},\underline{\omega}}\hspace*{-1pt}\geq\hspace*{-1pt}0$, where $\tau_1$ and $\underline{\omega}$ denote the worst technology and lowest wind speed realization for the seller, respectively. Using constraint (\ref{IC2-dynamic-np}), which is implied by the IC constraint about $\omega$ at $T\hspace*{-1pt}=\hspace*{-1pt}2$, the expected revenue $\mathbb{E}_\omega\hspace*{-1pt}\{\hspace*{-1pt}\mathcal{R}_{\tau_1\hspace*{-1pt},\omega}\hspace*{-1pt}\}$ is strictly positive for the seller with technology $\tau_1$, and must be greater than or equal to $\mathcal{R}^{\hspace*{-1pt}P}\hspace*{-1pt}(\hspace*{-1pt}\tau_1;\hspace*{-1pt}q\hspace*{-1pt})$  (see (\ref{IRbound-S-NM-P-np})). }
	
\blue{Part (ii) of Theorem \ref{thm-penalty} follows directly from the result of part (i). First, the set of constraints (\ref{IC2-dynamic-np}-\ref{IRbound-S-NM-P-np}) for the dynamic mechanism with no penalty is more restrictive than the set of constraints  (\ref{IC2-dynamic}-\ref{IRbound-S-NM-P}) for the dynamic mechanism (of Section \ref{sec-formulation}). Therefore, the performance of the optimal dynamic mechanism with no penalty is inferior to that of the optimal dynamic mechanism.} 

\blue{Second, comparing the set of constraints (\ref{IC2-dynamic-np}-\ref{IRbound-S-NM-P-np}) for the dynamic mechanism with no penalty with the set of constraints  (\ref{IC2-real}-\ref{IRbound-R-NM}) for the real-time mechanism (in particular (\ref{ICbound-R-NM}) and (\ref{ICbound-S-NM-P-np})), it is clear that the designer's objective under the optimal dynamic mechanism with no penalty is higher than her objective under the optimal real-time mechanism. This is because under the real-time mechanism, the designer cannot distinguish between different pairs of $(\tau,\omega)$ corresponding to the same $\Theta(\hspace*{-1pt}\tau\hspace*{-1pt},\hspace*{-1pt}\omega\hspace*{-1pt})$ (see (\ref{allocation-real})), whereas, in the dynamic mechanism with no penalty, distinction among all pairs of $(\hspace*{-1pt}\tau\hspace*{-1pt},\hspace*{-1pt}\omega\hspace*{-1pt})$ is possible as the seller reports $\tau$ and $\omega$ sequentially. }
	
\blue{We note that Theorem \ref{thm-penalty} does not provide a comparison between the forward mechanism and the dynamic mechanisms with no penalty. This is because, on one hand, the forward mechanism ignores $\omega$ in its allocation function. On the other hand, in the dynamic mechanism with no penalty we impose the no penalty risk constraint, \textit{i.e.} ex-post IR (\ref{IR-S-NM-P-np}). The impact of these two constraints on the performance of the forward mechanism and the dynamic mechanism with no penalty depends on the specific parameters of the model, and thus, there is no generic ordering among them in terms of the designer's objective.}


\addtolength{\belowcaptionskip}{5pt}
\begin{figure*}[htpb!]
	\begin{subfigure}{.5\textwidth}
		\centering
		\includegraphics[width=0.92\linewidth,height=0.17\textheight]{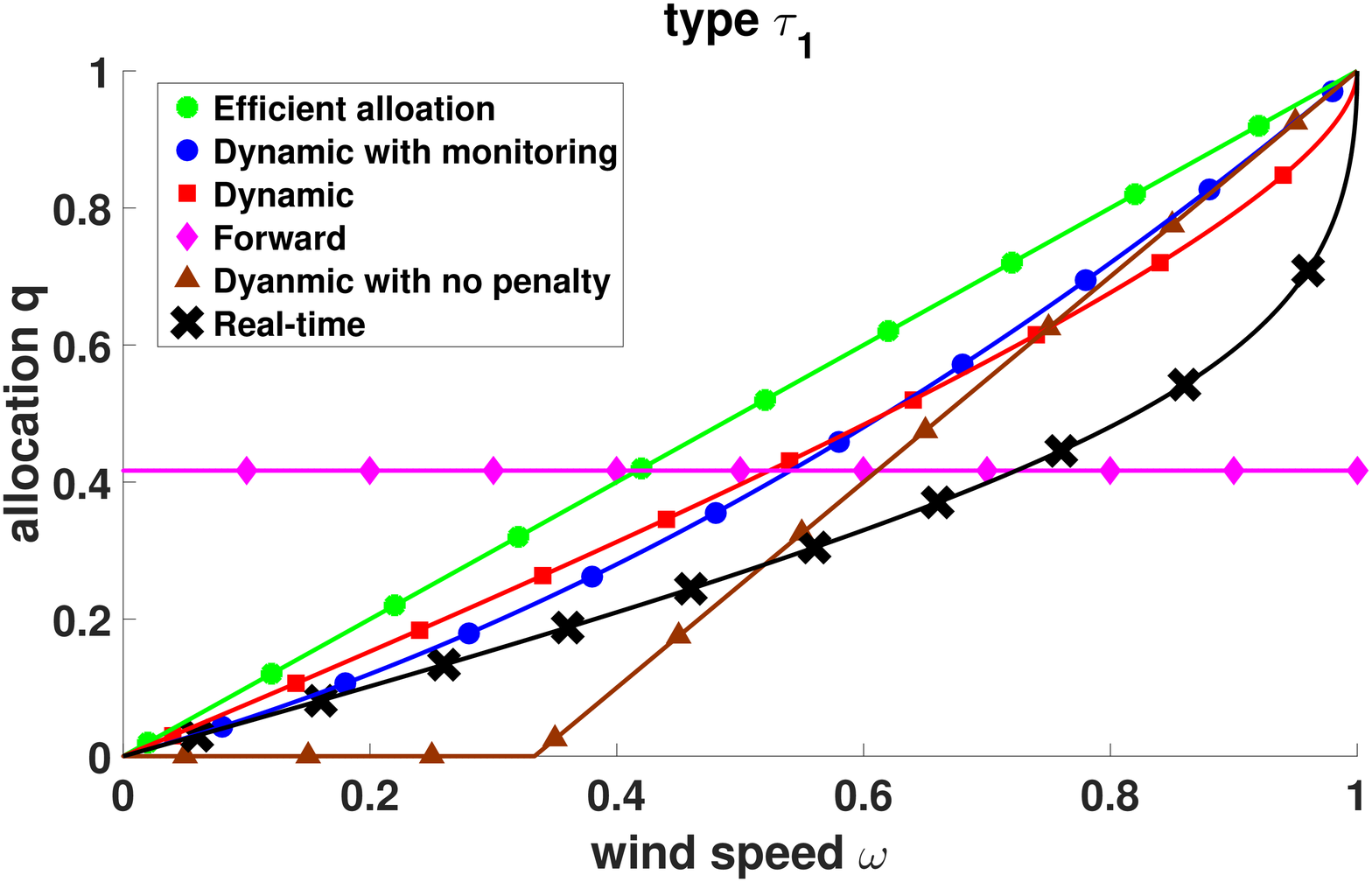}
		\vspace*{-5pt}\caption{Allocation $q(\tau_1,\omega)$}
		\label{fig:all-H}
	\end{subfigure}\hfill
	\begin{subfigure}{.5\textwidth}
		\centering
		\includegraphics[width=0.92\linewidth,height=0.17\textheight]{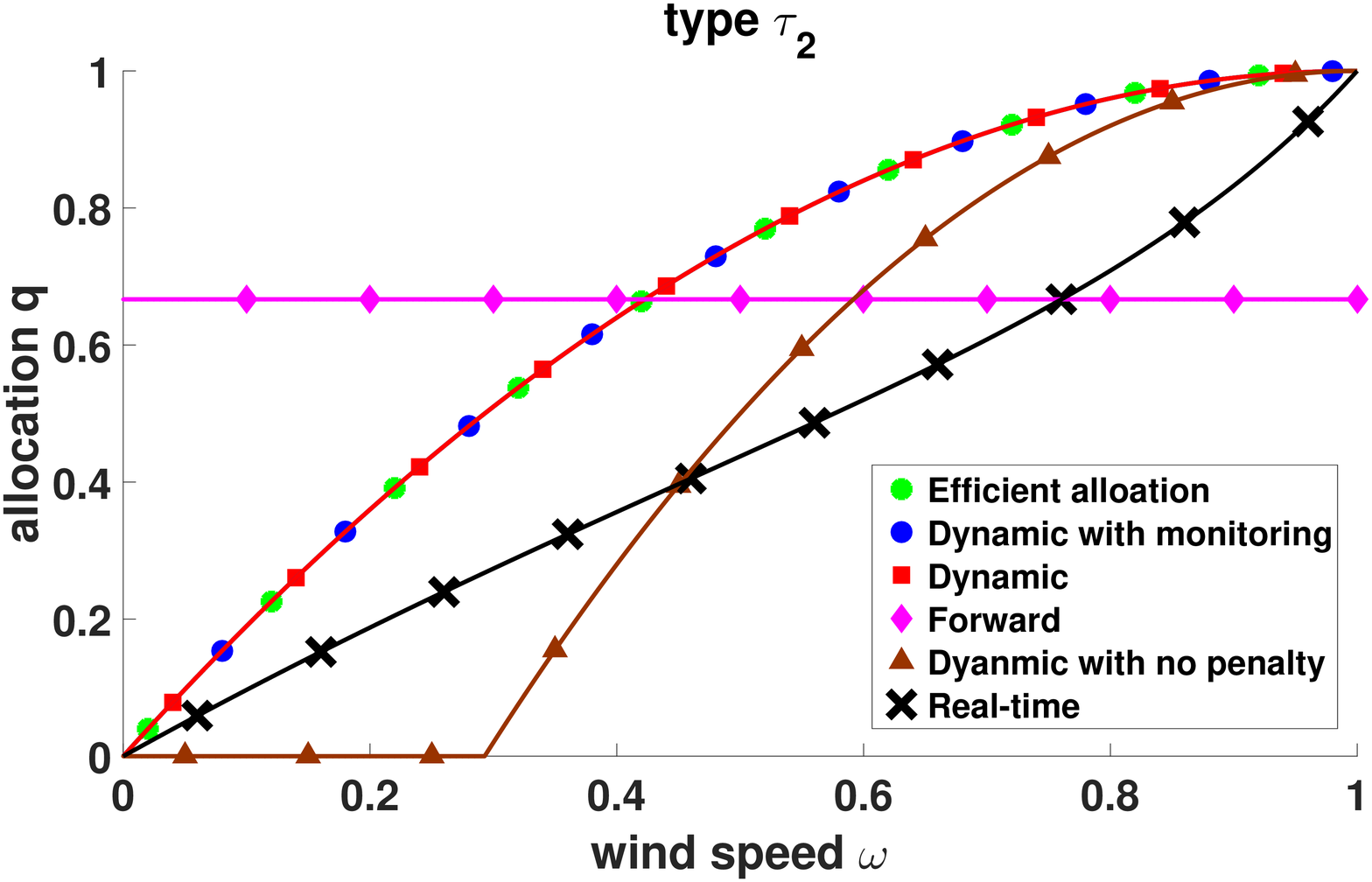}
		\vspace*{-5pt}\caption{Allocation $q(\tau_2,\omega)$}
		\label{fig:all-L}
	\end{subfigure}\\
	\begin{subfigure}{.5\textwidth}
		\centering
		\includegraphics[width=0.92\linewidth,height=0.17\textheight]{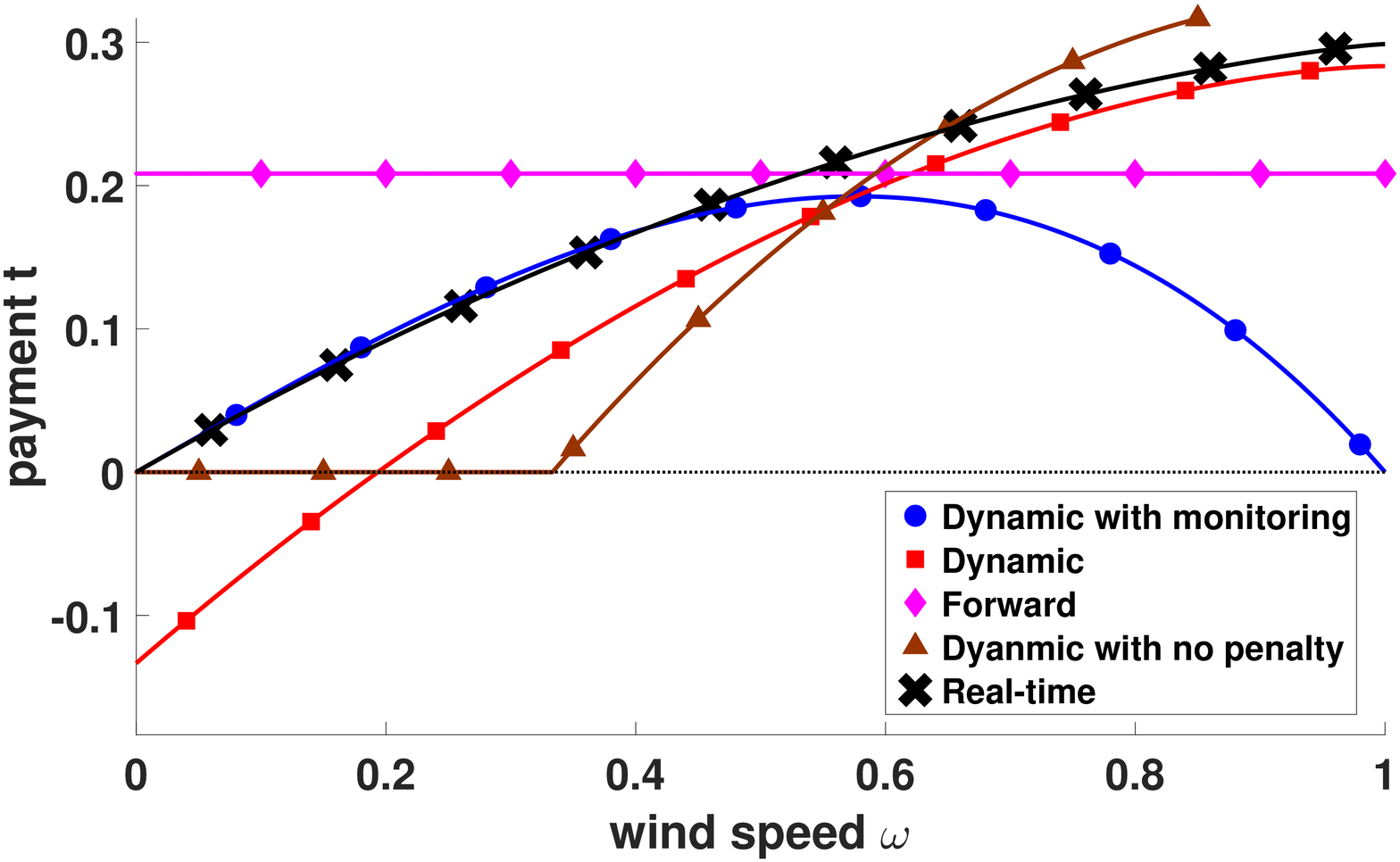}
		\vspace*{-5pt}\caption{Payment $t(\tau_1,\omega)$}
		\label{fig:pay-H}
	\end{subfigure}\hfill
	\begin{subfigure}{.5\textwidth}
		\centering
		\includegraphics[width=0.92\linewidth,height=0.17\textheight]{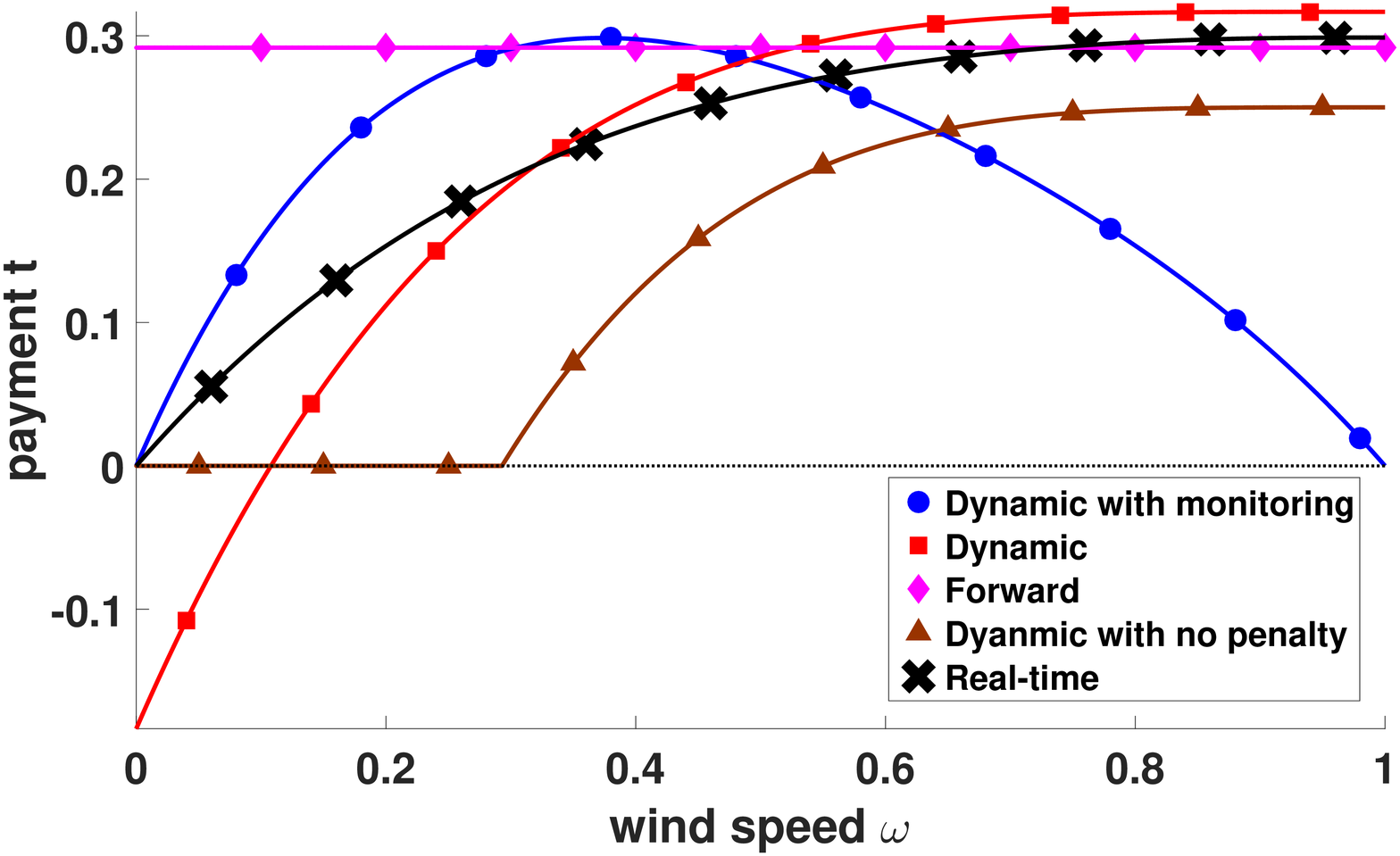}
		\vspace*{-5pt}\caption{Payment $t(\tau_2,\omega)$}
		\label{fig:pay-L}
	\end{subfigure}
	\vspace*{-3pt}\caption{Example - optimal mechanisms}
	\label{fig:example}
	\vspace*{-5pt}
\end{figure*}

\blue{\subsection{Dynamic Mechanism with Wind Monitoring}}

\blue{In the model of Section \ref{sec-model}, we use $\omega$ to denote the wind speed information that becomes available only to the seller at $T\hspace*{-1pt}=\hspace*{-1pt}2$. In this section, we consider a scenario where the designer can also monitor the realization of wind speed $\omega$.} 

\blue{In the following, we formulate the dynamic mechanism design problem under this assumption. We refer to this mechanism as \textit{the dynamic mechanism with monitoring}. Assuming that the designer monitors $\omega$, the seller is only required to reveal his private technology $\tau$ at $T\hspace{-2pt}=\hspace{-2pt}1$. Therefore, the optimal dynamic mechanism  with monitoring is given by the solution to the following optimization problem:
\begin{align}
&\max_{q(\cdot,\cdot),t(\cdot,\cdot)}{\mathcal{W}}\nonumber\\
\hspace{60pt}&\hspace{-60pt}\text{subject to}\nonumber\\
\hspace{40pt}&\hspace{-40pt}IC\hspace{-3pt}:\hspace{-1pt}\mathcal{R}_{\tau}\hspace{-3pt}\geq\hspace{-1pt}\hspace{-1pt} \mathbb{E}_\omega\hspace{-1pt}\{t(\hat{\tau},\omega)\hspace{-2pt}-\hspace{-2pt}C(\hspace{-1pt}q(\hat{\tau},\hspace{-1pt}\omega);\hspace{-1pt}\Theta(\tau,\hspace{-1pt}\omega))\}\hspace{-7pt}\quad\forall \tau\hspace{-1pt},\hspace{-1pt}\hat{\tau}\hspace{-1pt},\hspace{-1pt}
\label{rent-S-M-P}\\\hspace{40pt}&
\hspace{-40pt}IR\hspace{-3pt}:\hspace{-1pt}\mathcal{R}_{\tau}\hspace{-2pt}\geq\hspace{-2pt}0\quad \forall \tau.\label{IR-S-M-P}
\end{align} }
\blue{The above optimization problem is similar to the optimization problem for the optimal dynamic mechanism in Section \ref{sec-formulation}. However, there is no  IC constraint for $\omega$ in the dynamic mechanism with monitoring,  as the designer monitors $\omega$. }

\blue{Below, we show that wind monitoring provides two advantages to the designer. First, it improves the outcome of the dynamic mechanism. Second, it enables the designer to render the dynamic mechanism with monitoring free of any penalty risk for the seller 
	simply by modifying the payment function for different realizations of $\omega$.}

\blue{\begin{theorem} \label{thm-monitor} (i) The set of IC and IR constraints for the dynamic mechanism with monitoring, given by (\ref{rent-S-M-P},\ref{IR-S-M-P}) can be reduced to the following conditions,
	\begin{align}
	&\hspace*{-14pt}\mathcal{R}_{\tau_i}\hspace{-2pt}-\hspace{-2pt}\mathcal{R}_{\tau_j}\hspace{-2pt}=\hspace{-2pt} \mathcal{R}^T\hspace*{-2pt}(\tau_j,\tau_i;q)\hspace{-2pt}\geq\hspace{-2pt}0,\hspace{-7pt}\quad\forall i\hspace{-1pt},\hspace{-1pt}j\hspace{-1pt}\hspace{-2pt}\in\hspace{-2pt}\{\hspace{-1pt}1\hspace{-1pt},\hspace{-2pt}...,\hspace{-2pt}M\hspace{-1pt}\}\hspace*{-1pt},\hspace*{-1pt}i\hspace*{-3pt}>\hspace*{-3pt}j\hspace{-1pt},\hspace{-2pt}\label{ICbound-S-M-P}\\
	&\hspace*{-14pt}\mathcal{R}_{\tau_1}= 0.\label{IRbound-S-M-P}
	\end{align}	
	(ii) The designer's objective under the optimal dynamic mechanism with monitoring is higher than her objective under the optimal dynamic mechanism, the optimal dynamic mechanism with no penalty risk, and the optimal real-time and forward mechanisms.  \textit{i.e.}
	$\mathcal{W}^{\text{dynamic\hspace{-1pt} with\hspace{-1pt} monitoring}}\hspace{-2pt}>\hspace{-2pt}\mathcal{W}^{\text{dynamic}}\hspace{-2pt}>\hspace{-2pt}\mathcal{W}^{\text{dynamic\hspace{-1pt} no \hspace{-1pt}penalty}}>\hspace{-2pt}\mathcal{W}^{\text{real-time}}$, and  	$\mathcal{W}^{\text{dynamic with monitoring}}\hspace{-2pt}>\hspace{-2pt}\mathcal{W}^{\text{dynamic}}\hspace{-2pt}>\hspace{-2pt}\mathcal{W}^{\text{forward}}$.\\
	(iii) In the dynamic mechanism with wind monitoring, the designer can guarantee no penalty risk for the seller, without changing the mechanism outcome in terms of her objective $\mathcal{W}$, the buyer's utility $\mathcal{U}$, and the seller's revenue $\mathcal{R}$.
\end{theorem}}	
	
\blue{We summarize the results of Theorems \ref{thm-rent}-\ref{thm-monitor} on incentive payments to the seller under different market mechanisms in Table \ref{tabel-rent}. 	
For a given allocation function $q(\hspace*{-1pt}\tau\hspace*{-1pt},\hspace*{-1pt}\omega\hspace*{-1pt})$, 
define $\mathcal{R\hspace*{-1pt}}^T\hspace*{-1pt}(q)\hspace*{-2pt}:=\hspace*{-2pt}\sum_{i=2}^Mp_i\hspace*{-2pt}\sum_{j=2}^i\mathcal{R\hspace*{-1pt}}^T\hspace*{-1pt}(\hspace*{-1pt}\tau_j\hspace*{-1pt},\hspace*{-2pt}\tau_{j-1}\hspace*{-1pt};\hspace*{-1pt}q)$, $\mathcal{R\hspace*{-1pt}}^W\hspace*{-1pt}(q)\hspace*{-3pt}:=\hspace*{-3pt}\sum_{i=2}^M\hspace*{-2pt}\sum_{j=2}^ip_i\mathcal{R\hspace*{-1pt}}^T\hspace*{-2pt}(\hspace*{-1pt}\tau_j\hspace*{-1pt},\hspace*{-2pt}\tau_{j-1}\hspace*{-1pt};\hspace*{-1pt}q)$,
 and
$\mathcal{R\hspace*{-1pt}}^P\hspace*{-2pt}(\hspace*{-1pt}q\hspace*{-1pt})\hspace*{-3pt}:=\hspace*{-3pt}\mathcal{R\hspace*{-1pt}}^P\hspace*{-2pt}(\hspace*{-1pt}\tau_1\hspace*{-1pt};\hspace*{-1pt}q\hspace*{-1pt})$. 
Then, $\mathcal{R}^{\hspace*{-1pt}T}(q)$ denotes the incentive payment that the designer must pay to the seller so that he reveals truthfully his private technology $\tau$. If the designer wants to incorporate $\omega$ in energy allocation and does not monitor $\omega$, then she needs to pay the additional incentive payment $\mathcal{R}^W(q)$ so that the seller reveals $\omega$ truthfully. Without monitoring $\omega$, if the designer wants to guarantee no penalty risk for the seller, then she must pay an additional incentive payment $\mathcal{R}^P(q)$ to the seller.}

%
%
%
%
%
%
%

\renewcommand{\arraystretch}{1.35}
\begin{table}{\small \hspace{-10pt}\begin{tabular}{|c|c|c|} 
			\hline
			& \hspace{-7pt} Wind monitoring \hspace{-7pt} & No wind monitoring   \\  \hline 
			Penalty risk & $\mathcal{R}^T(q)$ & $\mathcal{R}^T(q)+\mathcal{R}^W(q)$ \\  \hline
			\hspace{-7pt} No penalty risk \hspace{-7pt} & $\mathcal{R}^T(q)$ & \hspace{-7pt} $\mathcal{R}^T(q)+\mathcal{R}^W(q)+\mathcal{R}^P(q)$  \hspace{-7pt} \\ 
			\hline
		\end{tabular}\vspace*{-5pt}}
	\caption{Incentive payment to the seller} 
	\label{tabel-rent}
\end{table} 	\vspace*{-8pt}

\section{Example}
\label{sec-example}

\blue {We consider an environment where the designer's objective is $\mathcal{W}\hspace{-1pt}=\hspace{-1pt}\mathcal{U}\hspace{-1pt}+\hspace{-1pt}0.5 \mathcal{R}$ (\textit{i.e.} $\alpha\hspace{-1pt}=\hspace{-1pt}0.5$). That is, the designer assigns more weight on the welfare of the demand than on the seller's revenue.} The buyer has utility function $\mathcal{V}(\hat{q})\hspace{-2pt}=\hspace{-2pt}\hat{q}\hspace{-2pt}-\hspace{-2pt}\frac{1}{2}\hat{q}^2$. The seller has linear cost function $C(\hat{q};\hspace{-1pt}\theta)\hspace{-2pt}=\hspace{-2pt}\theta \hat{q}$. The seller has two possible technologies; technology $\tau_1$ (inferior technology) with marginal cost $\Theta(\tau_1\hspace{-1pt},\hspace{-1pt}\omega)\hspace{-2pt}=\hspace{-2pt}1\hspace{-1pt}-\omega$, and technology $\tau_2$ (superior technology) with marginal cost $\Theta(\tau_2\hspace{-1pt},\hspace{-1pt}\omega)\hspace{-2pt}=\hspace{-2pt}(1\hspace{-1pt}-\hspace{-1pt}\omega)^2$. \blue{We assume that the designer believes that both technologies are equally likely and each has probability $0.5$.} The wind \blue{speed} $\omega$ is uniformly distributed in $[0,1]$. 

\begin{figure*}[htbp!]\vspace*{-15pt}
		\begin{floatrow}
			\floatbox{figure}[0.35\textwidth][\FBheight][t]
			{  \vspace{5pt} \hspace*{-25pt}\captionsetup{justification=centering}  \caption{ \blue{Example - different payment vs. quantity options in the dynamic mechanisms} }
				\label{fig-curves}}
			{ \hspace*{-5pt} \includegraphics[width=0.34\textwidth]{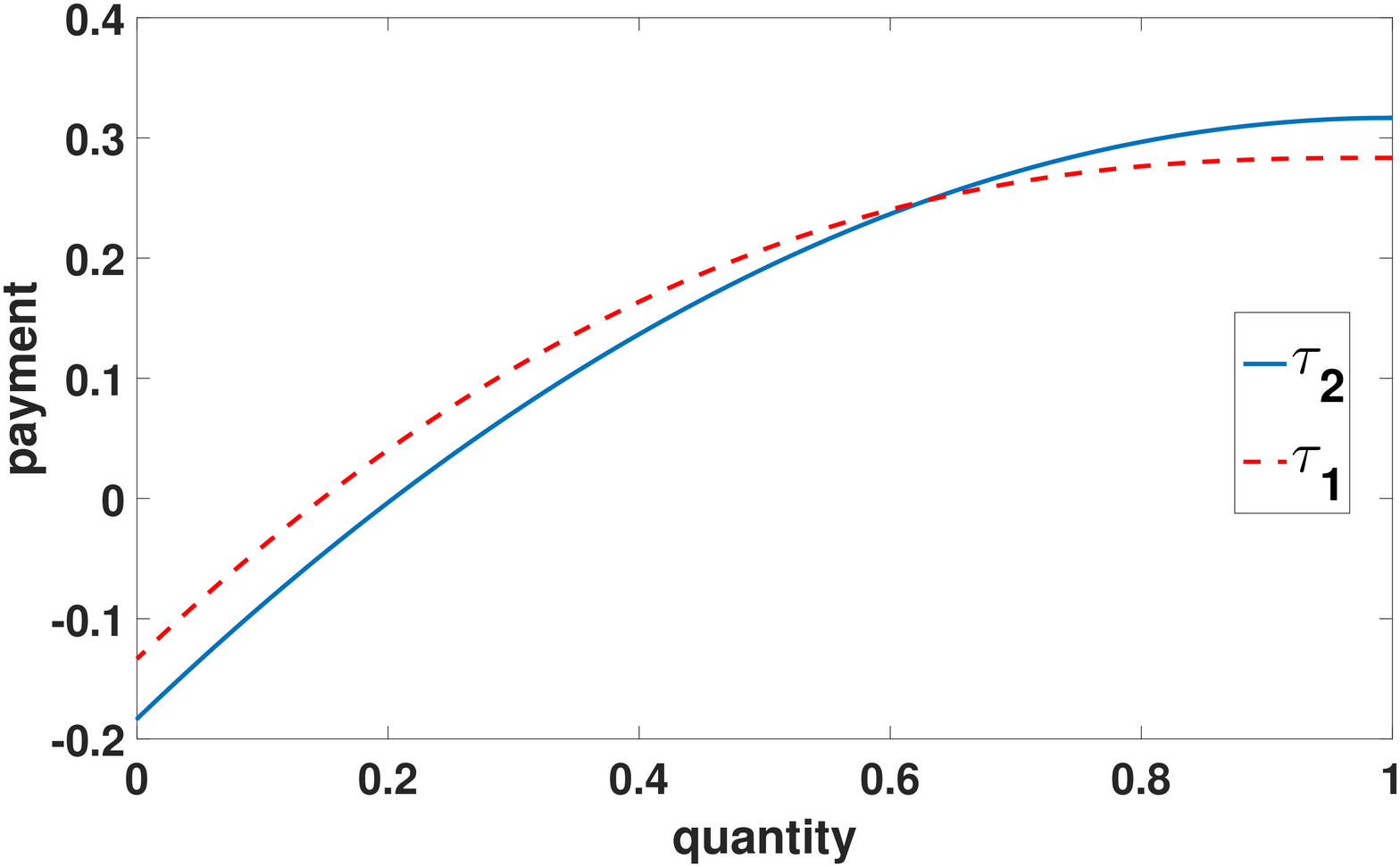}}		
			
			\hfill \hspace*{5pt}
			\floatbox{table}[0.60\textwidth][\FBheight][t]
			{ \captionsetup{justification=centering}\vspace*{5pt}
				\caption{Example - the buyer's utility $\mathcal{U}$, the information rent $\mathcal{R}$, and the designer's objective $\mathcal{W}$}
				\label{table:example}}
			{\small	\begin{tabular}{|c|c|c|c|}
					\hline
					Mechanisms &  \thead{\hspace{-7pt} Buyer's Utility \\$\mathcal{U}$} \hspace{-7pt}  &\thead{\hspace{-7pt} Seller's Revenue \\  $\mathcal{R}$} \hspace{-7pt} \vspace{-4pt} &\hspace{-7pt} \thead{Designer's Objective \\ $\mathcal{W}$} \hspace{-7pt} \\
					\hline 
					Efficient allocation & ---& --- &0.2167 \\ \hline
					Forward & 0.1372 & 0.0347 & 0.1545 \\ \hline
					Real-time & 0.0693 & 0.1210 & 0.1298\\ \hline				
					Dynamic & 0.1729 & 0.0417	 & 0.1938\\ \hline
					\hspace{-8pt} Dynamic with no penalty \hspace{-7pt} & 0.1077 & 0.0842 & 0.1498\\ \hline
					\hspace{-8pt} Dynamic with monitoring \hspace{-8pt} & 0.1813 & 0.0333 & 0.1979\\ \hline
				\end{tabular}
			}	
		\end{floatrow}\vspace*{-1pt}
	\end{figure*}

\blue{Using the results of Theorems \ref{thm-rent}-\ref{thm-monitor}, we compute the optimal forward, real-time, and dynamic mechanisms as well as the optimal dynamic mechanisms with no penalty and with monitoring. (see Appendix C \cite{companion} for the closed form solution of the general case). Figure \ref{fig:example} depicts \blue{the allocation and payment functions for each of the optimal mechanisms.} The outcome of these different mechanisms is summarized in Table \ref{table:example}. Consistent with the results of Theorems \ref{thm-buyer}-\ref{thm-monitor}, we find that $\mathcal{W}^{\text{dynamic with monitoring}}\hspace{-1pt}>\hspace{-1pt}\mathcal{W}^{\text{dynamic}}\hspace{-1pt}>\hspace{-1pt}\mathcal{W}^{\text{dynamic no penalty}}\hspace{-1pt}>\hspace{-1pt}\mathcal{W}^{\text{real-time}}$, and $\mathcal{W}^{\text{dynamic with monitoring}}\hspace{-1pt}>\hspace{-1pt}\mathcal{W}^{\text{dynamic}}\hspace{-1pt}>\hspace{-1pt}\mathcal{U}^{\text{forward}}$. We note that, as discussed in Section \ref{sec-disc}.B, there exists no general ordering between the dynamic mechanism with no penalty and the forward mechanism.}

	\begin{figure*}[htbp!]
		\centering
		\hspace{10pt}\includegraphics[width=0.45\linewidth,height=0.165\textheight]{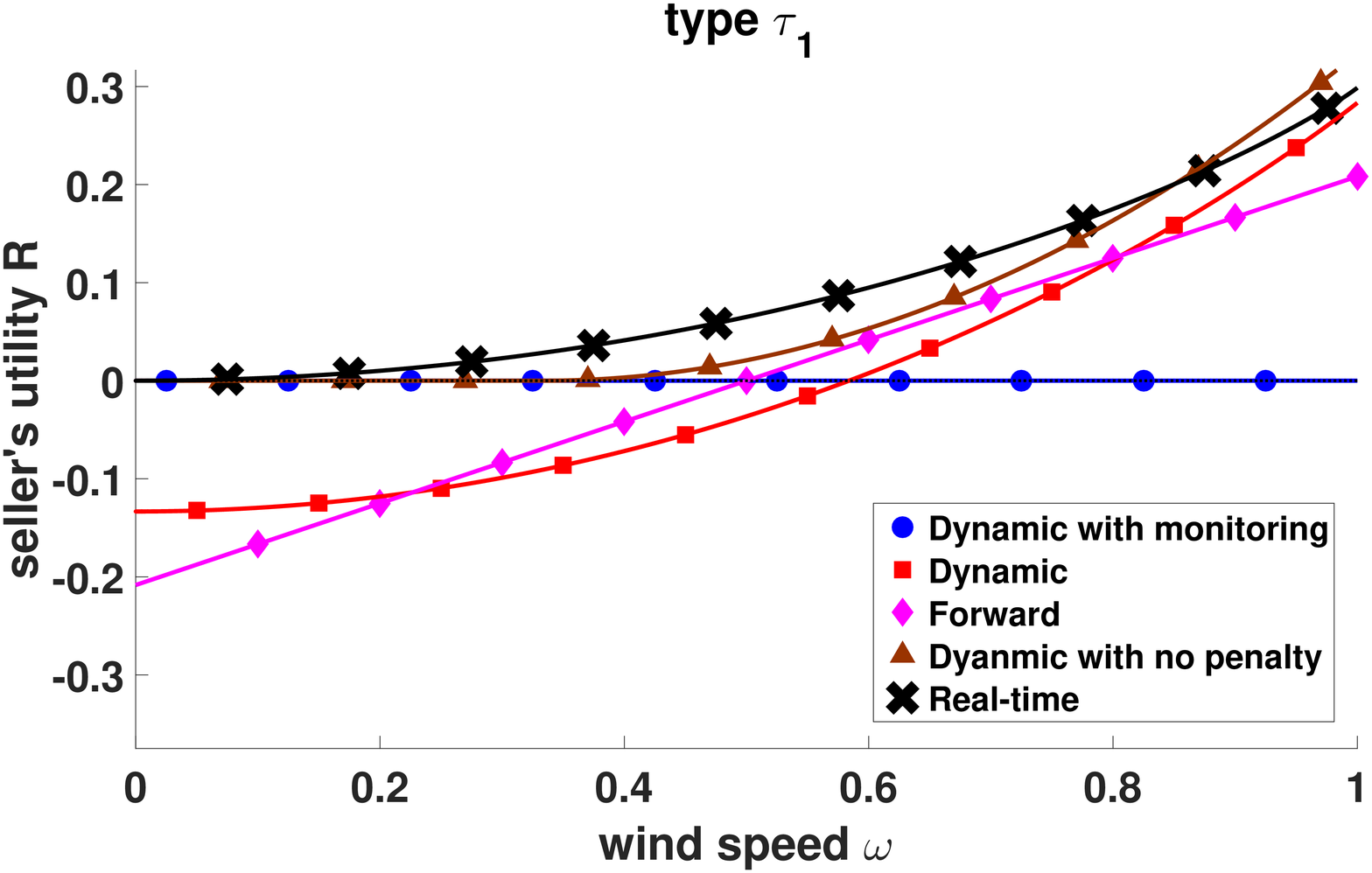}
		\hspace{-20pt}
		\includegraphics[width=0.45\linewidth,height=0.165\textheight]{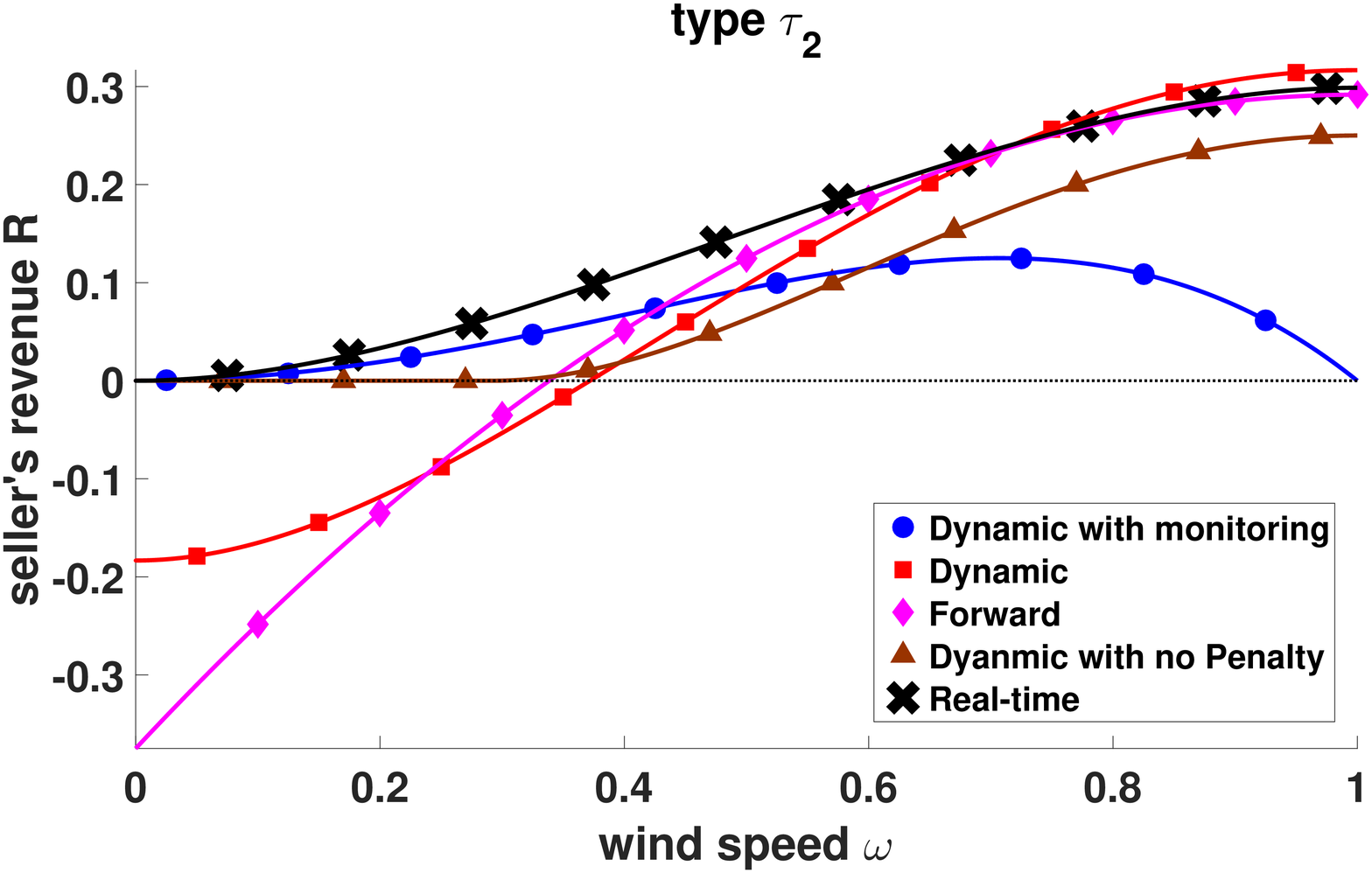}
		\vspace*{-1pt}\caption{Example - the seller's revenue}
		\label{fig:rev}
		\vspace*{-10pt}
	\end{figure*}

	\renewcommand{\arraystretch}{1.15}	

`
As we argued earlier, the seller's \blue{strategic behavior affects the efficiency of a mechanism and distorts its outcome  from} the efficient allocation \blue{(see Fig. \ref{fig:example})}. \blue{We note that for $\tau_2$,} the allocation functions in the dynamic mechanism and dynamic mechanism with monitoring are the same as the efficient allocation. \blue{Moreover, for $\tau_1$, the distortion of the allocation function from the efficient allocation is less under the dynamic mechanism than under the real-time and forward mechanisms. These observations further illustrate the advantage of the dynamic mechanism over the forward and real-time mechanisms.}

\blue{Next, we consider the seller's revenue $\mathcal{R}$. We note that the seller's revenue is the highest under the optimal real-time mechanism. This is because (i) the seller is not subject to any penalty risk under the real-time mechanism (as opposed to the forward and dynamic mechanisms); (ii) the seller reports $\tau$ and $\omega$ simultaneously, therefore, he has more power in manipulating his reports (about $\tau$ and $\omega$) to the designer than in the dynamic mechanism, where he reports $\tau$ and $\omega$ sequentially. We also note that the seller's revenue under the optimal dynamic mechanism is higher than that of the forward mechanism because the forward mechanism ignores wind speed $\omega$ whereas wind speed $\omega$ is incorporated in the dynamic mechanism, and, therefore, the seller receives additional incentive (payment) to report $\omega$ truthfully.  Next, we note that the seller's revenue under the optimal dynamic mechanism is lower than of the optimal dynamic mechanism with no penalty, and higher than of the optimal dynamic mechanism with monitoring. This is because the payment to the seller is lower in the optimal dynamic mechanism with monitoring and higher in the optimal dynamic mechanism with no penalty (see Table \ref{tabel-rent}). }

\blue{As pointed out above, the dynamic and forward mechanisms expose the seller to penalty risk (negative revenue) for low realizations of the wind speed $\omega$.} However, in the dynamic mechanism with monitoring, the dynamic mechanism with no penalty, and the real-time mechanism, the seller always receives a non-negative revenue for all realizations of $\omega$.  \blue{Figure \ref{fig:rev} shows the seller's revenue for all the mechanisms and all realizations of $\omega$.}

\blue{Next, we consider the buyer's utility $\mathcal{U}$. The buyer's utility under te dynamic mechanism is higher than that of the forward and real-time mechanisms. This is because the dynamic mechanism's efficiency is higher than that of the optimal forward and real-time mechanisms, and this is also reflected in the buyer's utility. Moreover, for a given allocation function, the incentive payment made by the buyer to the seller under the optimal dynamic mechanism is less than the incentive payments under the optimal forward and real-time mechanisms.  The buyer's utility under the optimal dynamic mechanism is higher than that of the optimal dynamic mechanism with no penalty, and lower than that of the optimal dynamic mechanism with monitoring. This is because the payment made by the buyer to the seller is lower in the optimal dynamic mechanism with monitoring and higher in the optimal dynamic mechanism with no penalty (see Table \ref{tabel-rent}).}

\blue{We note that in the forward and real-time mechanisms, which are static mechanisms, the payment the seller receives for producing a certain amount of energy $\hat{q}$ is independent of his type. However, this is not the case in the dynamic mechanism (see Fig. \ref{fig-curves}). In the dynamic mechanism the payment and  allocation depend on the seller's reports at $T=1$ and $T=2$. Therefore, depending on the seller's reports at $T=1$ and $T=2$, the designer can provide different payments to the seller for producing the same amount of energy. In Fig. \ref{fig-curves}, one can see that the seller with the superior technology $\tau_2$ receives a higher payment than the seller with the inferior technology $\tau_1$ for high quantities of energy produced. On the contrary, for low quantities of energy produced the seller with inferior technology $\tau_1$ receives a higher payment than the seller with superior technology $\tau_2$.  The difference in payments to different types of seller for the same quantity of energy produced, allows the designer to differentiate among different types of seller, thus, to increase the efficiency of the mechanism.}   

\section{\blue{Extension to Many Sellers}}

\label{sec-extension}
\blue{To demonstrate the main ideas, namely the advantage of the dynamic mechanism over the forward and real-time mechanisms, we considered a setting with only one seller in the model of Section \ref{sec-model}. However, in a general electricity market there exist many sellers competing with one another. In this section, we discuss how our results on the advantage of the dynamic mechanism over the forward and real-time mechanisms also hold for environments with many sellers.  Due to space limitation, we do not provide detailed proofs; we only provide the sketch of the proof  following steps similar to those presented for the model with one seller.}

\blue{Consider a model similar to that of Section \ref{sec-model}, with $N$ sellers. Seller $n$, $n\hspace{-2pt}\in\hspace{-2pt}\{\hspace{-1pt}1,\hspace{-1pt}...\hspace{-1pt},\hspace{-1pt}N\hspace{-1pt}\}$, has generation cost $C(\hat{q}^n;\Theta(\tau^n,\omega^n))$, where $\hat{q}^n$ denotes the amount of energy he produces, and $\tau^n$ and $\omega^n$ denotes seller $n$'s technology and wind speed, respectively. We assume that $\tau^n$ takes values in $\{\tau_1^n,...,\tau_M^n\}$ with probability $(p_1^n...,p_M^n)$. The probability distribution of $\omega^n\in[\underline{\omega},\overline{\omega}]$ is independent of $\tau_n$. The wind speeds $\omega^1,...,\omega^N$ may be correlated as the sellers can be located in geographically close locations. We assume that joint distribution of $(\hspace{-1pt}\omega^1\hspace{-2pt},\hspace{-1pt}...,\hspace{-1pt}\omega^N\hspace{-1pt})$  (resp. $(\hspace{-1pt}\tau^1\hspace{-2pt},\hspace{-1pt}...,\hspace{-1pt}\tau^N\hspace{-1pt})$), is commonly known to all sellers as well as the designer. Furthermore, Assumptions \ref{assump-FSD} and \ref{assum-nonshift} of Section \ref{sec-model}, hold for every seller $n$, $n\hspace{-2pt}\in\hspace{-2pt}\{\hspace{-1pt}1,\hspace{-1pt}...,\hspace{-1pt}N\hspace{-1pt}\}$.  }

\blue{Similar to the approach presented in Section \ref{sec-rent}, we invoke the revelation principle and restrict attention to direct revelation mechanisms that are incentive compatible and individually rational. Let $\tau^{-n}$ (resp. $\omega^{-n}$) denote the set of all sellers' technologies (resp. wind speeds) except seller $n$'s technology $\tau^n$ (resp. wind speed $\omega^n$). In the model with many sellers, seller $n$'s   allocation and payment are functions of seller $n$'s reports about $\tau^n$ and $\omega^n$ as well as all the other sellers' reports about  $\tau^{-n}$ and $\omega^{-n}$. Let $q^n(\hspace{-1pt}\tau^n\hspace{-2pt},\hspace{-1pt}\omega^n\hspace{-2pt},\hspace{-1pt}\tau^{-n}\hspace{-2pt},\hspace{-1pt}\omega^{-n}\hspace{-1pt})$, and $t^n(\hspace{-1pt}\tau^n\hspace{-2pt},\hspace{-1pt}\omega^n\hspace{-2pt},\hspace{-1pt}\tau^{-n}\hspace{-2pt},\hspace{-1pt}\omega^{-n}\hspace{-1pt})$ denote the allocation and payment functions for seller $n$, respectively.}

\blue{We assume that the designer does not reveal the reports of seller $n$ to other sellers before finalizing the payments and allocations of all sellers. Define,
	\begin{align} &\hspace{-6pt}\bar{C}^n\hspace{-1pt}(\hspace{-1pt}\hat{\tau}^n\hspace{-3pt},\hspace{-1pt}\hat{\omega}^n\hspace{-1pt};\hspace{-1pt}\tau^n\hspace{-3pt},\hspace{-1pt}\omega^n\hspace{-1pt})\hspace{-2pt}:=\hspace{-2pt}\mathbb{E}_{\tau^{-n}\hspace{-2pt},\omega^{-n}}\hspace{-1pt}\Big\{\hspace{-2pt}C(\hspace{-1pt}q^n\hspace{-1pt}(\hspace{-1pt}\hat{\tau}^n\hspace{-3pt},\hspace{-1pt}\hat{\omega}^n\hspace{-3pt},\hspace{-2pt}\tau^{-n}\hspace{-3pt},\hspace{-1pt}\omega^{-n}\hspace{-1pt});\hspace{-2pt}\Theta(\hspace{-1pt}\tau^n\hspace{-2pt},\hspace{-1pt}\omega^n\hspace{-1pt})\hspace{-1pt})\hspace{-2pt}\nonumber\\
	&\hspace{190pt}\big|\hspace{-1pt}\tau^n\hspace{-3pt},\hspace{-1pt}\omega^n\hspace{-1pt}\Big\}\hspace{-2pt},\hspace{-4pt}\label{exp-cost}\\ &\hspace{-6pt}\bar{t}^n(\hat{\tau}^n,\hat{\omega}^n)\hspace{-2pt}:=\hspace{-2pt}\mathbb{E}_{\tau^{-n},\omega^{-n}}\{t^n(\hspace{-1pt}\hat{\tau}^n\hspace{-1pt},\hspace{-1pt}\hat{\omega}^n\hspace{-1pt},\hspace{-1pt}\tau^{-n}\hspace{-1pt},\hspace{-1pt}\omega^{-n}\hspace{-1pt})\hspace{-1pt}\big|\hspace{-1pt}\tau^n\hspace{-3pt},\hspace{-1pt}\omega^n\hspace{-1pt}\}\label{exp-payment},
	\end{align} as seller $n$'s conditional expected cost and payment, respectively, when he has technology $\tau^n$ and wind $\omega^n$, and he reports $\hat{\tau}^n$ and $\hat{\omega}^n$. We note that, the expectations above are written using the fact that at equilibrium, seller $i$ believes that other sellers report their private information truthfully.}
	
	\blue{Similar to Section \ref{sec-rent}, let
	\begin{align}
 &\mathcal{R}^n_{\hspace{-1pt}\tau^n\hspace{-3pt},\omega^n\hspace{-1pt}}\hspace{-3pt}:=\hspace{-2pt}\mathbb{E}_{\tau^{-n}\hspace{-3pt},\omega^{-n}}\hspace{-2pt}\Big\{\hspace{-1pt}t^n\hspace{-1pt}(\hspace{-1pt}\tau^n\hspace{-3pt},\hspace{-1pt}\omega^n\hspace{-3pt},\hspace{-2pt}\tau^{\hspace{-1pt}-n}\hspace{-3pt},\hspace{-1pt}\omega^{\hspace{-1pt}-n}\hspace{-1pt})\hspace{-2pt}\nonumber\\&\hspace{80pt}-\hspace{-2pt}C\hspace{-2pt}\left(\hspace{-1pt}q^n\hspace{-1pt}(\hspace{-1pt}\tau^n\hspace{-3pt},\hspace{-1pt}\omega^n\hspace{-3pt},\hspace{-2pt}\tau^{\hspace{-1pt}-n}\hspace{-3pt},\hspace{-1pt}\omega^{\hspace{-1pt}-n}\hspace{-1pt});\hspace{-2pt}\Theta\hspace{-1pt}(\hspace{-1pt}\tau^n\hspace{-3pt},\hspace{-1pt}\omega^n\hspace{-1pt})\hspace{-2pt}\right)\hspace{-3pt}\Big\}\hspace{-2pt},\hspace{-1pt}\nonumber
	\end{align} denote the expected revenue of seller $n$ with technology $\tau^n$ and wind $\omega^n$. Also, define, 
	\begin{align*}
		&\mathcal{R}_{\tau^n}^n\hspace{-3pt}:=\hspace{-2pt}\mathbb{E}_{\omega^n}\{\hspace{-1pt}\mathcal{R}^n_{\tau^n,\omega^n}\hspace{-1pt}\}\nonumber,\\
		&\mathcal{R}^n\hspace{-2pt}:=\mathbb{E}_{\tau_n}\{\mathcal{R}_{\tau^n}^n\mathcal{R}_{\tau^n}^n\}.
	\end{align*}
		Moreover, let,
	\begin{align*}
		&\mathcal{U}_{\text{many}}\hspace{-3pt}:=\hspace{-2pt}\mathbb{E}_{\{\hspace{-1pt}\tau^n\hspace{-1pt},\omega^n\hspace{-1pt}\}_{n=1}^N}\hspace{-4pt}\left\{\hspace{-2pt}\mathcal{V}\hspace{-2pt}\left(\hspace{-1pt}\sum_{n=1}^N\hspace{-2pt}q^n(\hspace{-1pt}\tau^n\hspace{-2pt},\hspace{-1pt}\omega^n\hspace{-3pt},\hspace{-1pt}\tau^{-n}\hspace{-3pt},\hspace{-1pt}\omega^{-n}\hspace{-1pt})\hspace{-4pt}\right)\hspace*{-3pt}\right.\nonumber\\&\hspace{130pt}\left.-\hspace*{-3pt}\sum_{n=1}^N\hspace{-2pt}t^n(\hspace{-1pt}\tau^n\hspace{-3pt},\hspace{-1pt}\omega^n\hspace{-3pt},\hspace{-1pt}\tau^{-n}\hspace{-3pt},\omega^{-n}\hspace{-1pt})\hspace{-3pt}\right\}\hspace{-3pt},\nonumber\\
		&\mathcal{W}_{\text{many}}\hspace{-2pt}:=\hspace{-2pt}\mathcal{U}_{\text{many}}\hspace{-2pt}+\hspace{-2pt}\alpha\hspace{-2pt}\sum_{n=1}^N \mathcal{R}^n,
	\end{align*} denote the buyer's expected utility and the designer's objective, respectively.}

\blue {We discuss how the results of Theorems \ref{thm-lemma}-\ref{thm-buyer} (on the advantage of the dynamic mechanism over the forward and real-time mechanisms) continue to hold in the model with many sellers described above.} 

\blue{The key idea is the following. We write each mechanism design problem with many sellers in a form similar to the corresponding mechanism design problem with one seller (formulated in Section \ref{sec-formulation}). We write these mechanism design problems using the conditional expected cost and payment functions, defined by (\ref{exp-cost}) and (\ref{exp-payment}), respectively, instead of the cost and payment functions that appear in the problems with a single seller. We show that in each mechanism design problem and for each seller, the designer faces a set of constraints that are similar to those that arise in the case with a single seller. Therefore, the arguments used in the proofs of Theorems \ref{thm-lemma}-\ref{thm-buyer} can be directly extended to environments with many sellers.  }


\blue{We proceed with the formulation of the mechanism design problems with many sellers, and discuss how the results of Theorems \ref{thm-lemma}-\ref{thm-buyer} continue to hold in the model with many sellers.}

\blue{\textbf{Forward mechanism:} In the forward mechanism, the allocation function $q^n(\hspace{-1pt}\hat{\tau}^n\hspace{-2pt},\hspace{-1pt}\hat{\omega}^n\hspace{-2pt},\hspace{-1pt}\tau^{-n}\hspace{-2pt},\hspace{-1pt}\hat{\omega}^{-n}\hspace{-1pt})$ and payment function $t^n(\hspace{-1pt}\hat{\tau}^n\hspace{-2pt},\hspace{-1pt}\hat{\omega}^n\hspace{-2pt},\hspace{-1pt}\tau^{-n}\hspace{-2pt},\hspace{-1pt}\hat{\omega}^{-n}\hspace{-1pt})$ for seller $n$ are independent of the reported wind speeds $\hat{\omega}^n$ and $\hat{\omega}^{-n}$. Therefore, we drop the dependence on $(\hat{\omega}^n,\hat{\omega}^{-n})$, and denote the conditional expected cost and payment functions by $\bar{C}^n(\hat{\tau}^n;\tau^{n},\omega^n)$ and $\bar{t}^n(\hat{\tau}^n)$, respectively. 
	The optimal forward mechanism with many sellers is given by the solution to the following optimization problem:
	\begin{align}
	&\max_{\{q^n(.),t^n(.)\}_{n=1}^N} \mathcal{W}_{\text{many}}\nonumber\\
	\hspace{15pt}&\hspace{-15pt}\text{subject to}\nonumber\\
	&IC^n\hspace{-3pt}:\mathcal{R}^n_{\tau^n}\hspace{-2pt}\geq\hspace{-2pt} \bar{t}^n(\hat{\tau^n})\hspace{-2pt}-\hspace{-2pt}\mathbb{E}_\omega\{\bar{C}^n(\hspace{-1pt}\hat{\tau};\hspace{-1pt}\tau,\hspace{-1pt}\omega^n)\hspace{-1pt}\}\hspace{-1pt}\quad\forall \tau^n,\hat{\tau}^n,n\hspace{-1pt}
	\label{rent-forward-multi} \\
	&IR^n\hspace{-3pt}: \mathcal{R}^n_{\tau^n}\geq 0\quad \forall \tau^n,n.\label{IR-F-multi}
	\end{align}}
\blue{We note that in the above optimization problem, the designer faces a set of constraints, given by (\ref{rent-forward-multi},\ref{IR-F-multi}), for every seller $n$. These constraints are similar to those given by (\ref{rent-forward},\ref{IR-F}) for the forward mechanism with a single seller. Therefore, we can show that seller's expected revenue $\mathcal{R}_{\tau^n}^n$ for every seller $i$ must satisfy a set of constraints similar to those given by (\ref{ICbound-monitoring},\ref{IRbound-monitoring}) in part (a) of Theorem \ref{thm-rent} in terms of conditional expected cost $\bar{C}^n\hspace{-1pt}(\hat{\tau}^n\hspace{-1pt};\hspace{-1pt}\tau^{n}\hspace{-2pt},\hspace{-1pt}\omega^n)$.}    
\vspace{5pt}

\blue{\textbf{Real-time mechanism:} The real-time mechanism design problem with many sellers can be formulated as follows. 
	\begin{align}
	&\hspace{70pt}\max_{\{q^n(.),t^n(.)\}_{n=1}^N}{\mathcal{W}_{\text{many}}}\nonumber\\
	&\hspace{-3pt}\text{subject to}\nonumber\\
	&IC^n\hspace{-3pt}:\hspace{-1pt}\mathcal{R}_{\tau^n,\omega^n}^n\hspace{-3pt}\geq\hspace{-3pt} \bar{t}^n\hspace{-1pt}(\hspace{-1pt}\hat{\tau}^n\hspace{-1pt},\hspace{-1pt}\hat{\omega}^n\hspace{-1pt})\hspace{-2pt}-\hspace{-2pt}\bar{C}^n\hspace{-1pt}(\hspace{-1pt}\hspace{-1pt}\hat{\tau}^n\hspace{-1pt},\hspace{-1pt}\hat{\omega}^n\hspace{-1pt};\hspace{-1pt}\hspace{-1pt}\tau^n\hspace{-1pt},\hspace{-1pt}\omega^n\hspace{-1pt})\quad\nonumber\\&\hspace{150pt}\forall \tau^n\hspace{-1pt},\hspace{-1pt}\omega^n\hspace{-1pt},\hspace{-1pt}\hat{\tau}^n\hspace{-1pt},\hspace{-1pt}\hat{\omega}^n\hspace{-1pt}, n,
	\label{rent-R-NM-multi}\\
	&IR^n\hspace{-3pt}:\mathcal{R}_{\tau^n,\omega^n}^n\geq 0\quad \forall \tau^n,\omega^n,n. \label{IR-R-NM-multi}
	\end{align} }
\blue{Note that the set of constraints (\ref{rent-R-NM-multi},\ref{IR-R-NM-multi}), for every seller $n$, is similar to the set of constraints (\ref{rent-R-NM},\ref{IR-R-NM}) in the real-time mechanism with a single seller. Therefore, we can show that expected revenue $\mathcal{R}_{\tau^n\hspace{-2pt},\omega^n}^n$ must satisfy a set of constraints similar to those given by Theorem \ref{thm-lemma}. Consequently, we can show that the set of constraints that the designer faces for every seller $n$ can be reduced to a set of constraints that are similar to those given by (\ref{IC2-real}-\ref{IRbound-R-NM}) in part (b) of Theorem \ref{thm-rent} in terms of expected seller's revenue  $\mathcal{R}_{\tau^n\hspace{-2pt},\omega^n}^n$ and $\mathcal{R}_{\tau^n}^n$, and the conditional expected cost $\bar{C}^n\hspace{-1pt}(\hat{\tau}^n\hspace{-2pt},\hspace{-1pt}\hat{\omega}^n\hspace{-1pt};\hspace{-1pt}\tau^{n}\hspace{-2pt},\hspace{-1pt}\omega^n)$. We can also show that $q(\tau^n\hspace{-1pt},\hspace{-1pt}\omega^n\hspace{-1pt},\hspace{-1pt}\tau^{-n}\hspace{-1pt},\hspace{-1pt}\omega^{-n})$ must only be a function of $\Theta^n\hspace{-1pt}(\tau^n\hspace{-2pt},\hspace{-1pt}\omega^n)$ and $\Theta^m\hspace{-1pt}(\tau^m\hspace{-3pt},\hspace{-1pt}\omega^m)$, $m\hspace{-2pt}\neq\hspace{-2pt} n$, since all sellers report simultaneously about their own technologies and wind speeds (see (\ref{allocation-real})). Using an argument similar to the one in Theorem \ref{thm-rent} for seller $n$, we can also show that $\mathbb{E}_{\tau^{-n}\hspace{-1pt},\omega^{-n}}\hspace{-1pt}\{q^n(\tau^n\hspace{-2pt},\hspace{-1pt}\omega^n\hspace{-2pt},\hspace{-1pt}\tau^{-n}\hspace{-1pt},\hspace{-1pt}\omega^{-n})\hspace{-1pt}\}$ must be increasing in $\omega^n$.}    
\vspace{5pt}

\blue{\textbf{Dynamic mechanism:} The optimal dynamic mechanism with many sellers is given by the solution to the following optimization problem:
\begin{align}
&\blue{\hspace{50pt}\max_{\{q^n(.),t^n(.)\}_{n=1}^N}{\mathcal{W}_{\text{many}}}}\nonumber\\
\hspace{40pt}&\hspace{-35pt}\text{subject to}\nonumber\\
\hspace{40pt}&\hspace{-45pt}IC_1^n\hspace{-3pt}:\hspace{-2pt}\mathcal{R}_{\tau^n}^n\hspace{-4pt}\geq\hspace{-2pt} \mathbb{E}_{\omega^n}\hspace{-1pt}\{\bar{t}^n\hspace{-1pt}(\hspace{-1pt}\hat{\tau}^n\hspace{-1pt},\hspace{-1pt}\sigma^n(\omega^n\hspace{-1pt})\hspace{-1pt})\hspace{-2pt}-\hspace{-2pt}\bar{C}^n\hspace{-1pt}(\hspace{-1pt}\hat{\tau}^n\hspace{-2pt},\hspace{-1pt}\sigma^n\hspace{-1pt}(\omega^n\hspace{-1pt})\hspace{-1pt};\hspace{-1pt}\tau^n\hspace{-2pt},\hspace{-1pt}\omega^n\hspace{-1pt})\hspace{-1pt}\}\hspace{-8pt}\quad\nonumber\\&\hspace{123pt}\forall\hspace{-1pt} \tau^n\hspace{-1pt},\hspace{-1pt}\hat{\tau}^n\hspace{-1pt},\hspace{-1pt}\sigma^n(\hspace{-1pt}\cdot\hspace{-1pt})\hspace{-1pt},\hspace{-1pt}n,\hspace{-3pt}
\label{rent-S-NM-P-1-multi} \\
\hspace{40pt}&\hspace{-45pt}IC_2^n\hspace{-3pt}:\hspace{-2pt}\mathcal{R}_{\tau^n\hspace{-1pt},\omega^n}^n\hspace{-2pt}\geq\hspace{-2pt} \bar{t}^n\hspace{-1pt}(\hspace{-1pt}\tau^n\hspace{-2pt},\hspace{-1pt}\hat{\omega}^n\hspace{-1pt})\hspace{-2pt}-\hspace{-2pt}\bar{C}^n\hspace{-1pt}(\hspace{-1pt}\tau^n\hspace{-2pt},\hspace{-1pt}\hat{\omega}^n;\hspace{-1pt}\tau^n\hspace{-3pt},\hspace{-1pt}\omega^n\hspace{-1pt})\hspace{-1pt}\quad\hspace{-5pt}\forall \tau^n\hspace{-3pt},\hspace{-1pt}\omega^n\hspace{-2pt},\hspace{-1pt}\hat{\omega}^n\hspace{-2pt},\hspace{-1pt}n,\hspace{-4pt}\label{rent-S-NM-P-2-multi} 
\\
\hspace{40pt}&\hspace{-45pt}IR^n\hspace{-1pt}:\hspace{-2pt}\mathcal{R}_{\tau^n}^n\hspace{-2pt}\geq\hspace{-2pt}0\quad \forall \tau,n. \label{IR-S-NM-P-multi}
\end{align}
As we mentioned above, we assume that the designer does not reveal other sellers' report about $\tau^{-n}$ at $T=1$. Therefore, seller $n$'s IC constraint about $\omega$ at $T=2$ is given by (\ref{rent-S-NM-P-2-multi}), where seller $n$ assumes that other sellers report   truthfully $\tau^{-n}$ and $\omega^{-n}$. We note that  the set of constraints (\ref{rent-S-NM-P-1-multi}-\ref{IR-S-NM-P-multi}), for every seller $n$,  is similar to the set of constraints (\ref{rent-S-NM-P-1}-\ref{IR-S-NM-P}) in the dynamic mechanism with a single seller. Therefore, we can show that the expected seller's revenue $\mathcal{R}_{\tau^n\hspace{-1pt},\omega^n}^n$ must satisfy a set of constraints similar to those given by Theorem \ref{thm-lemma}. Furthermore, we can show that the expected seller's revenue $\mathcal{R}_{\tau^n\hspace{-1pt},\omega^n}^n$ and $\mathcal{R}_{\tau^n}^n$ for every seller $n$ must satisfy a set of constraints that are similar to the ones given by ((\ref{IC2-dynamic}-\ref{IRbound-S-NM-P}) in part (c) of Theorem \ref{thm-rent}. Using an argument similar to Theorem \ref{thm-rent} for seller $n$, we can also show that $\mathbb{E}_{\tau^{-n}\hspace{-1pt},\omega^{-n}}\{q^n(\tau^n\hspace{-2pt},\hspace{-1pt}\omega^n\hspace{-2pt},\hspace{-1pt}\tau^{-n}\hspace{-2pt},\hspace{-1pt}\omega^{-n})\hspace{-1pt}\}$ must be increasing in $\omega^n$ as in (\ref{allocation-dynamic}) in part (c) of Theorem \ref{thm-rent}.}

\blue{The above arguments show that all the mechanism design problems with many sellers considered in this section, have sets of constraints for each seller that are similar to those that arise in the corresponding problems with a single seller. Thus, we can show, by arguments similar to those given in the proof of Theorem \ref{thm-buyer}, that the set of constraints for the dynamic mechanism with many sellers is less restrictive than the set of constraints for the forward and real-time mechanisms with many sellers. Therefore, we can establish that the dynamic mechanism with many sellers outperforms the forward and real-time mechanism with many sellers. }

\begin{remark}\blue{
The optimal forward and real-time mechanism with many sellers are standard static multi-unit auctions (see \cite{krishna2009auction}). However, the  optimal dynamic mechanism with many sellers is a form of \textit{handicap auction}, first introduced in \cite{esHo2007optimal}. }
	
	\blue{In a standard auction, seller $n$ bids his generation cost function simultaneously and the auctioneer determines the outcomes based on all sellers' bids. }
	
		\blue{In a handicap auction, at $T\hspace{-2pt}=\hspace{-2pt}1$, seller $n$ bids  his current information about his generation cost (\textit{i.e.} $\tau^n$), so as to pick  a menu of payment-quantity curve, from which he can select his generation at $T\hspace{-1pt}=\hspace{-1pt}2$. Then, at $T\hspace{-2pt}=\hspace{-2pt}2$, seller $n$ observes the realized wind speed $\omega^n$, and competes with other sellers for generation based on the payment-quantity curve  he won at $T\hspace{-2pt}=\hspace{-2pt}1$. Sellers with  better technologies anticipate to have higher generations in real-time. Therefore, in a handicap auction, at $T\hspace{-2pt}=\hspace{-2pt}1$, they bid for payment-quantity curves that give them high marginal prices for high generation quantities (which are more likely for them), but low marginal prices for low generation quantities (which are less likely for them). On the other hand, sellers with worse technologies bid for  payment-quantity curves that give them high marginal prices for low generation quantities (which are more likely for them), but low marginal prices for high generation quantities (which are less likely for them); see Fig. \ref{fig-curves} in the example of Section \ref{sec-example}.}
\end{remark} 

\section{Conclusion}
\blue{We investigated mechanism design problems for wind energy.} We proposed  a dynamic market mechanism that  couples the outcomes of the real-time and forward mechanisms. We showed that the proposed dynamic mechanism  outperforms the forward and real-time mechanisms with respect to the integration of wind energy into the grid. \blue{In the dynamic mechanism, the seller must sequentially reveal his private information to the designer, and refine his commitment accordingly.}  

We also investigated the effects of wind monitoring and penalty risk exposure on the market outcome.      


\label{sec-con}

\small
\bibliographystyle{abbrv}
\bibliography{bib}

\newpage
\appendices
\section{Proofs of the Main Results} \label{sec:appA}
The proofs of Theorems \ref{thm-lemma}-\blue{\ref{thm-monitor}} are based on Lemmas \ref{lemma-IC2}-\ref{lemma-S-M} that we state below. The proofs of Lemmas \ref{lemma-IC2}-\ref{lemma-S-M} are given in Appendix B.

Let $g(\omega)$ denote the corresponding  probability density function for $\omega$, \textit{i.e.} $g(\omega):=\frac{\partial G(\omega)}{\partial \omega}$.

First, we prove the following sufficient and necessary condition for the $IC_2$ constraint (reporting the true $\omega$) for \blue{the dynamic mechanism}.    
\begin{lemma}[revenue equivalence]
	If \blue{the dynamic mechanism} is incentive compatible, and the allocation rule $q(\tau,\omega)$ is continuous in $\omega$, then, for all $\omega$ and $\acute{\omega}$,
	\begin{eqnarray}
	\mathcal{R}_{\tau,\omega}=\mathcal{R}_{\tau,\acute{\omega}}-\int_{\acute{\omega}}^{\omega} {C_\theta(q(\tau,\hat{\omega});\Theta(\tau,\hat{\omega}))\Theta_\omega(\tau,\hat{\omega})d\hat{\omega}},\label{RE-2}
	\end{eqnarray}
	and $q(\tau,\omega)$ is increasing in $\omega$.
	Moreover, if (\ref{RE-2}) holds and $q(\tau,\omega)$ is increasing in $\omega$, then $IC_2$ is satisfied.
	\label{lemma-IC2}
\end{lemma}

We can now provide the proof for Theorem \ref{thm-lemma} using the result of Lemma \ref{lemma-IC2}.

\begin{proof}[\textbf{Proof of Theorem \ref{thm-lemma}}]

	We note that the IC constraint (\ref{rent-R-NM}) for \blue{the real-time mechanisms} implies the $IC_2$ constraint for the dynamic mechanism (given by (\ref{rent-S-NM-P-2}), by setting $\hat{\tau}\hspace{-1pt}=\hspace{-1pt}\tau$.  By Lemma \ref{lemma-IC2}, the $IC_2$ constraint for the dynamic mechanism is satisfied only if equation (\ref{rent-IC2}) holds.

\end{proof}	

Lemma \ref{lemma-IC2} characterizes the seller's \blue{revenue} \blue{in the dynamic mechanism} given that he tells the truth at $T\hspace{-2pt}=\hspace{-2pt}1$ about his technology $\tau$. To complete the characterization of the seller's optimal strategy at $T\hspace{-2pt}=\hspace{-2pt}2$, we show, via Lemma \ref{lemma-off-report} below, that if the seller misreports his technology at $T\hspace{-2pt}=\hspace{-2pt}1$ (off-equilibrium path), he later corrects his lie at $T\hspace{-2pt}=\hspace{-2pt}2$.

\begin{lemma}
	Consider \blue{the dynamic mechanism} that satisfies the IC constraints for $\omega$ and $\tau$. If a seller with technology $\tau$ misreports $\hat{\tau}$, $\hat{\tau}\neq\tau$, at $T=1$, then, for every wind realization $\omega$ at $T=2$, he corrects his lie by reporting $\hat{\omega}=\sigma^*(\hat{\tau};\tau,\omega)$ such that,
	\begin{align}
	\Theta(\tau,\omega)=\Theta(\hat{\tau},\hat{\omega}).\label{eq-off-report}
	\end{align}
	\label{lemma-off-report}
\end{lemma}
\textbf{Remark:} Note that by Assumption \ref{assum-nonshift} on non-shifting support, for any $\tau,\hat{\tau},\omega$, there exists a unique $\hat{\omega}$ (given the strict monotonicity of $\Theta(\tau,\omega)$ in $\omega$) that satisfies equation (\ref{eq-off-report}), and $\sigma^*(\hat{\tau};\tau,\omega)$ is well-defined.

Using the results of Lemmas \ref{lemma-IC2} and \ref{lemma-off-report}, we characterize below the seller's expected gain by misreporting his technology at $T\hspace{-2pt}=\hspace{-2pt}1$ \blue{in the dynamic mechanism}.

\begin{lemma}
	For \blue{the dynamic mechanism} that satisfies the set of IC constraints for $\omega$ and $\tau$, the maximum utility of the seller with technology $\tau$ reporting $\hat{\tau}$ at $T\vspace{-2pt}=\vspace{-2pt}1$ is given by,
	\begin{align}
	&\mathbb{E}_\omega\{t(\hat{\tau},\sigma^*(\hat{\tau};\tau,\omega))-C(q(\hat{\tau},\sigma^*(\hat{\tau};\tau,\omega));\Theta(\tau,\omega))\}\nonumber\\
	&=\mathcal{R}_{\hat{\tau}}\hspace{-2pt}-\hspace{-4pt}\int\hspace{-5pt}\int_{\omega}^{\sigma^*(\hat{\tau};\tau,\omega)}{\hspace{-20pt}C_\theta(q(\hat{\tau},\hat{\omega});\Theta(\hat{\tau},\hat{\omega}))\Theta_\omega(\hat{\tau},\hat{\omega})d\hat{\omega}dG(\omega)}\label{eq-misreport-rent1}
	\end{align}
	\label{lemma-misreport}
\end{lemma}

\blue{The following result, for the dynamic mechanism, characterizes the payments that incentivize the seller to report truthfully his technology $\tau$ and wind speed $\omega$. 
}

\begin{lemma}\label{lemma-S-NM}
	\blue{For the dynamic mechanism}, the set of IC constraints for $\tau$, given by (\ref{rent-S-NM-P-1}), can be replaced by the following inequality constraints,
	\begin{align}
	&\mathcal{R}_{\tau_{i}}\hspace{-2pt}-\hspace{-2pt}\mathcal{R}_{\tau_{j}}\hspace{-2pt}\geq\hspace{-2pt} \mathcal{R}^T\hspace*{-1pt}(\tau_{j},\tau_{i};q)\hspace{-2pt}+\hspace{-2pt}\mathcal{R}^W\hspace*{-1pt}(\tau_{j},\tau_{i};q)\quad \forall  i,j\hspace*{-2pt}\in\hspace*{-2pt}\{2,..,M\}, i\hspace*{-2pt}>\hspace*{-2pt}j.\nonumber
	\end{align}
\end{lemma}

\vspace{10pt}

To provide the proof for Theorems \ref{thm-rent}-\blue{\ref{thm-buyer}}, 
we need the following results \blue{for the forward and the real-time mechanisms. }

\begin{lemma}\label{lemma-F}
		\blue{For} the forward mechanism, the set of IC constraints and IR constraints, given by (\ref{rent-forward}) and (\ref{IR-F}), respectively, can be replaced by the following constraints,
		\begin{align}
		&\text{$q(\tau)$  only depends on $\tau$}\nonumber\\
		&\mathcal{R}_{\tau_{i}}\hspace{-2pt}-\hspace{-2pt}\mathcal{R}_{\tau_{i-1}}\hspace{-2pt}=\hspace{-2pt} \mathcal{R}^T\hspace*{-1pt}(\tau_{i-1},\tau_{i};q)\hspace{-2pt}\quad \forall i\hspace*{-2pt}\in\hspace*{-2pt}\{2,..,M\},\nonumber\\
		&\mathcal{R}_{\tau_1}\blue{\geq}0.\nonumber
		\end{align}


\end{lemma}


\begin{lemma}	\label{lemma-R-NM}
	\blue{For the real-time mechanism}, the set of IC and IR constraints given by (\ref{rent-R-NM}) and (\ref{IR-R-NM}), respectively, can be replaced by the following constraints,
	\begin{align}
	&\text{$q(\tau,\omega)$ only depends on $\Theta(\tau,\omega)$}\nonumber\\
	&\mathcal{R}_{\tau_{i}}\hspace{-2pt}-\hspace{-2pt}\mathcal{R}_{\tau_{i-1}}\hspace{-2pt}=\hspace{-2pt} \mathcal{R}^T\hspace*{-1pt}(\tau_{i-1},\tau_{i};q)\hspace{-2pt}+\hspace{-2pt}\mathcal{R}^W\hspace*{-1pt}(\tau_{i-1},\tau_{i};q)\quad \forall i\hspace*{-2pt}\in\hspace*{-2pt}\{2,..,M-1\},\nonumber\\
	&R_{\tau_1}\blue{\geq}\mathcal{R}^P(\tau_1;q)\nonumber,
	\end{align}
	and the seller's revenue satisfies (\ref{rent-IC2}).
	
	
\end{lemma}

\vspace{10pt}

Using the result of Lemmas \blue{\ref{lemma-S-NM}-\ref{lemma-R-NM}}, we first provide the proof for Theorem \blue{\ref{thm-rent}}.
%
%
\vspace{10pt}
%

\begin{proof}[\textbf{Proof of Theorem \ref{thm-rent}}]
	
	We first show that $R^T\hspace*{-2pt}(\hspace{-1pt}\tau_j\hspace{-1pt},\hspace*{-3pt}\tau_i;\hspace*{-1pt}q\hspace{-1pt})\hspace*{-3pt}\geq\hspace*{-2pt}0$  and $R^W\hspace*{-1pt}(\hspace{-1pt}\tau_j\hspace{-1pt},\hspace{-2pt}\tau_i\hspace{-1pt};\hspace{-1pt}q\hspace{-1pt})\hspace*{-2pt}\geq \hspace*{-2pt}0$  for $i,j\in\{1,...,M\},i>j$ (strict if $q(\hspace{-1pt}\tau_j\hspace{-1pt},\hspace{-1pt}\omega\hspace{-1pt})\hspace*{-2pt}\neq\hspace*{-2pt} 0$ for some $\omega$). 
	
	First, note that by Assumption \ref{assump-FSD} we have $\Theta(\tau_i,\omega)<\Theta(\tau_j,\omega)$ for $i>j$. Thus, $C(q(\tau_j,\omega);\Theta(\tau_j,\omega))\hspace*{-2pt}\geq\hspace*{-2pt} C(q(\tau_j,\omega);\Theta(\tau_i,\omega))$ with strict inequality if $q(\tau_j,\omega)> 0$. Therefore,
	\begin{align*}
	\mathcal{R}^{\hspace{-1pt}T}\hspace{-2pt}(\hspace{-1pt}\tau_j\hspace{-1pt},\hspace{-2pt}\tau_i;\hspace{-1pt}q)\hspace{-3pt}:=\hspace{-5pt}\int{\hspace{-5pt}\left(\hspace{-1pt}C(\hspace{-1pt}q(\hspace{-1pt}\tau_j\hspace{-1pt},\hspace{-1pt}\omega\hspace{-1pt});\hspace{-2pt}\Theta(\hspace{-1pt}\tau_j\hspace{-1pt},\hspace{-1pt}\omega\hspace{-1pt})\hspace{-1pt})\hspace{-2pt}-\hspace{-2pt}C(\hspace{-1pt}q(\hspace{-1pt}\tau_j\hspace{-1pt},\hspace{-1pt}\omega\hspace{-1pt});\hspace{-2pt}\Theta(\hspace{-1pt}\tau_i\hspace{-1pt},\hspace{-1pt}\omega\hspace{-1pt})\hspace{-1pt})\hspace{-1pt}\right)\hspace{-2pt}dG(\omega)}\hspace{-2pt}\geq\hspace{-2pt} 0,
	\end{align*}
	with strict inequality if $q(\tau_j,\omega)>0$ for a set of $\omega$'s with positive probability.
	
	Second, we have $C_\theta(q(\tau_j,\omega);\Theta(\tau_j,\hat{\omega}))\hspace*{-2pt}\leq\hspace*{-2pt} C_\theta(q(\tau_j,\hat{\omega});\Theta(\tau_j,\hat{\omega}))$ for $\omega \leq \hat{\omega}$ since $q(\tau;\hat{\omega})$ is increasing in $\hat{\omega}$ by Lemma \ref{lemma-IC2} and $C(q,\theta)$ is increasing in $q$ by Assumption \ref{assump-FSD}. Moreover, by the result of Lemma \ref{lemma-off-report}, we have $\sigma^*(\tau_j;\tau_i,\omega)\geq \omega$ for $i>j$. Therefore, we have
	\begin{align*} 
	&\mathcal{R}^{\hspace{-1pt}W}\hspace{-2pt}(\hspace{-1pt}\tau_j\hspace{-1pt},\hspace{-2pt}\tau_i;\hspace{-1pt}q)\hspace{-3pt}:=\hspace{-5pt}\int\hspace{-8pt}\int_{\omega}^{\sigma^*\hspace{-1pt}(\tau_j;\tau_i,\omega)}\hspace{-35pt}\left(C_\theta(q(\hspace{-1pt}\tau_j\hspace{-1pt},\hspace{-1pt}\omega\hspace{-1pt});\hspace{-1pt}\Theta(\hspace{-1pt}\tau_j\hspace{-1pt},\hspace{-1pt}\hat{\omega}\hspace{-1pt})\hspace{-1pt})\hspace{-2pt}-\hspace{-2pt}C_\theta(q(\hspace{-1pt}\tau_j\hspace{-2pt},\hspace{-1pt}\hat{\omega}\hspace{-1pt});\hspace{-1pt}\Theta(\hspace{-1pt}\tau_j\hspace{-1pt},\hspace{-1pt}\hat{\omega}\hspace{-1pt})\hspace{-1pt})\hspace{-1pt}\right)\nonumber\\
	&\hspace{163pt}\Theta_\omega\hspace{-1pt}(\hspace{-1pt}\tau_j\hspace{-1pt},\hspace{-1pt}\hat{\omega}\hspace{-1pt})\hspace{-1pt}d\hat{\omega}dG(\omega)\hspace*{-2pt}\geq \hspace*{-2pt}0,
	\end{align*}
	since $\Theta_\omega\hspace{-1pt}(\hspace{-1pt}\tau_j\hspace{-1pt},\hspace{-1pt}\hat{\omega}\hspace{-1pt})\hspace{-2pt}< \hspace{-2pt}0$;
the inequality is strict if $q(\tau_j,\omega)>0$ for a set of $\omega$'s with positive probability.

\vspace{10pt}

In the following, we provide the proof for each part of Theorem \ref{thm-rent} separately.
	\begin{enumerate}[a)]
%
%
		
		\item \blue{The proof for part (a) directly follows from the result of Lemma \ref{lemma-F}.}
		
		
		\item \blue{The proof for part (b) directly follows from the result of Lemma \ref{lemma-R-NM}.}
		
		\item \blue{By the result of Theorem \ref{thm-lemma}, the set of IC constraints (\ref{rent-S-NM-P-2}) can be replaced by (\ref{allocation-dynamic},\ref{IC2-dynamic}). Moreover,} by the result of Lemma \ref{lemma-S-NM}, we can reduce the set of IC constraints (\ref{rent-S-NM-P-1}) to the set of inequality constraints (\ref{ICbound-S-NM-P}). 
		
		Furthermore, set of inequality constraints  (\ref{ICbound-S-NM-P}) implies that $\mathcal{R}_{\tau_1}\leq\mathcal{R}_{\tau_2}\leq...\leq \mathcal{R}_{\tau_M}$ since $R^T(\tau_{i-1},\tau_i;q)\geq0$ and $R^W(\tau_{i-1},\tau_i;q)\geq0$. Thus, \blue{the set of IR constraints (\ref{IR-S-NM-P})} can be reduced to $\mathcal{R}_{\tau_1}\blue{\geq}0$.

		\end{enumerate}

\end{proof}

\vspace{10pt}

\blue{Next}, we provide the proof for Theorem \ref{thm-buyer}.

\vspace{10pt}

\begin{proof}[\textbf{Proof of Theorem \ref{thm-buyer}}]
	The proof directly follows from the result of Theorem \ref{thm-rent}. We note that the objective functions in all the mechanism design problems are \blue{$\mathcal{W}$}, and the problems differ only in the set of constraints they have to satisfy.
	\begin{itemize}
		\item The set of constraints for \blue{the real-time} mechanism, given by part (b) of Theorem \ref{thm-rent}, is more restrictive than the set of constraints for \blue{the dynamic mechanism} given by part (c) of Theorem \ref{thm-rent}. \blue{Therefore, the designer's objective $\mathcal{W}^{\text{dyanmic}}$ is higher than his objective $\mathcal{W}^{\text{real-time}}$.} 
		\item 
		The set of constraints for the forward mechanism,  given by part (a) of Theorem \ref{thm-rent}, is more restrictive than the set of constraints for \blue{the dynamic mechanisms}, given by part (c) of Theorem \ref{thm-rent}. \blue{Therefore, the designer's objective $\mathcal{W}^{\text{dyanmic}}$ is higher than his objective $\mathcal{W}^{\text{forward}}$.}
	\end{itemize}
\end{proof}

\vspace{10pt}

\blue{We now provide the proof of Theorem \ref{thm-penalty} on the dynamic mechanism with no penalty.}

\vspace{10pt}

\begin{proof}[\blue{\textbf{Proof of Theorem \ref{thm-penalty}}}]\\
\blue{\textbf{(i)} We note that the set of IC constraints (\ref{rent-S-NM-P-1-np},\ref{rent-S-NM-P-2-np}) for the dynamic mechanism with no penalty is identical to the set of IC constraints (\ref{rent-S-NM-P-1-np},\ref{rent-S-NM-P-2-np}) for dynamic mechanism. Therefore, constraints (\ref{allocation-dynamic-np}-\ref{ICbound-S-NM-P-np}) directly follow from the result of part (c) of Theorem \ref{thm-rent}.}

	Next, we show that the set of IR constraints (\ref{IR-S-NM-P-np}) is satisfied if and only if $\mathcal{R}_{\tau_i,\underline{\omega}}\geq \mathcal{R}^P(\tau_i;q)$ and the seller's revenue satisfies (\ref{rent-IC2}). First, by part (i) of Theorem \ref{thm-lemma}, we have  $\mathcal{R}_{\tau,\omega}$ is increasing in $\omega$. Hence,  the set of IR constraints (\ref{IR-S-NM-P-np}) is satisfied if and only if $\mathcal{R}_{\tau_i,\underline{\omega}}\geq 0$, for all $i\in\{1,...,M\}$. 
			Second, using  (\ref{rent-IC2}) along with $\mathcal{R}_{\tau_i,\underline{\omega}}\geq 0$, $i\in\{1,...,M\}$, we have, 
			\begin{align*}
			\mathcal{R}_{\tau_i} &\geq \int_{\underline{\omega}}^{\overline{\omega}}\hspace{-4pt}\int_{\underline{\omega}}^\omega C_\theta(q(\tau_i,\hat{\omega});\Theta(\tau,\hat{\omega}))\Theta_\omega(\tau_i,\hat{\omega})d\hat{\omega}dG(\omega)\\
			&=\int_{\underline{\omega}}^{\overline{\omega}}\hspace{-4pt}\int_{\hat{\omega}}^{\overline{\omega}} C_\theta(q(\tau_i,\hat{\omega});\Theta(\tau_i,\hat{\omega}))\Theta_\omega(\tau_i,\hat{\omega})dG(\omega)d\hat{\omega}\\
			&=\int_{\underline{\omega}}^{\overline{\omega}}\left[1-G(\hat{\omega})\right] C_\theta(q(\tau_i,\hat{\omega});\Theta(\tau_i,\hat{\omega}))\Theta_\omega(\tau_i,\hat{\omega})d\hat{\omega}\\
			&= \mathcal{R}^P(\tau_i;q).\nonumber
			\end{align*}
			Thus, the set of IR constraints (\ref{IR-S-NM-P-np}) is satisfied if and only if (\ref{IRbound-S-NM-P-np}) is satisfied.
		
		\blue{\textbf{(ii)}  We note that the objective functions in all the mechanism design problems are \blue{$\mathcal{W}$}, and the problems differ only in the set of constraints they have to satisfy. The set of constraints for the real-time mechanism,  given by part (b) of Theorem \ref{thm-rent}, is more restrictive than the set of constraints for the dynamic mechanisms with no penalty, given by part (i) above. Therefore, the designer's objective $\mathcal{W}^{\text{dyanmic no penalty}}$ is higher than his objective $\mathcal{W}^{\text{real-time}}$. Moreover, the set of constraints for the dynamic mechanism,  given by part (c) of Theorem \ref{thm-rent}, is less restrictive than the set of constraints for the dynamic mechanisms with no penalty, given by part (i) above. Therefore, the designer's objective $\mathcal{W}^{\text{dyanmic no penalty}}$ is lower than his objective $\mathcal{W}^{\text{dynamic}}$.}
\end{proof}

\blue{To provide the proof for Theorem \ref{thm-monitor}, we need the following result for the dynamic mechanism with monitoring.}

\begin{lemma}\label{lemma-S-M}
	For \blue{the dynamic mechanism with monitoring}, the set of IC and IR constraints, given by (\ref{rent-S-M-P}) and (\ref{IR-S-M-P}), respectively, can be replaced by the following constraints,
	\begin{align}
	&\mathcal{R}_{\tau_{i}}\hspace{-2pt}-\hspace{-2pt}\mathcal{R}_{\tau_{i-1}}\hspace{-2pt}=\hspace{-2pt} \mathcal{R}^T\hspace*{-1pt}(\tau_{i-1},\tau_{i};q)\hspace{-2pt}\quad \forall i\hspace*{-2pt}\in\hspace*{-2pt}\{2,..,M-1\},\nonumber\\
	&R_{\tau_1}\geq 0\nonumber.
	\end{align}
\end{lemma}

\blue{Using the result of Lemma \ref{lemma-S-M}, we provide the proof of Theorem \ref{thm-monitor} below.}

\vspace{10pt}

\begin{proof}[\blue{\textbf{Proof of Theorem \ref{thm-monitor}}}]\\
	\blue{\textbf{(i)} The proof for part (i) directly follows from the result of Lemma \ref{lemma-S-M}.}
	
	\blue{\textbf{(ii)} The objective functions in all the mechanism design problems are \blue{$\mathcal{W}$}, and the problems differ only in the set of constraints they have to satisfy. The set of constraints for the dynamic mechanism,  given by part (c) of Theorem \ref{thm-rent}, is more restrictive than the set of constraints for the dynamic mechanisms with monitoring, given by part (i) above. Therefore, the designer's objective $\mathcal{W}^{\text{dyanmic with monitoring}}$ is higher than $\mathcal{W}^{\text{dynamic}}$. Moreover, by the result of Theorems \ref{thm-buyer} and \ref{thm-penalty}, we have $\mathcal{W}^{\text{dynamic}}\geq\mathcal{W}^{\text{dynamic no penalty}}\geq \mathcal{W}^{\text{real-time}}$ and $\mathcal{W}^{\text{dynamic}}\geq \mathcal{W}^{\text{forward}}$.}
	
	\blue{\textbf{(iii)}} Define the following modified payment function:
	\begin{align*}
	\hat{t}(\hspace{-1pt}\tau\hspace{-1pt},\hspace{-1pt}\omega\hspace{-1pt})\hspace{-2pt}:=\hspace{-2pt}\mathbb{E}_\omega\{\hspace{-1pt}t(\hspace{-1pt}\tau\hspace{-1pt},\hspace{-1pt}\omega\hspace{-1pt})\hspace{-1pt}\}\hspace{-2pt}+\hspace{-2pt}\left(C(q(\hspace{-1pt}\tau\hspace{-1pt},\hspace{-1pt}\omega\hspace{-1pt});\hspace{-1pt}\Theta(\hspace{-1pt}\tau\hspace{-1pt},\hspace{-1pt}\omega\hspace{-1pt})\hspace{-1pt})\hspace{-2pt}-\hspace{-2pt}\mathbb{E}_\omega\hspace{-2pt}\left\{\hspace{-1pt}C(q(\hspace{-1pt}\tau\hspace{-1pt},\hspace{-1pt}\omega\hspace{-1pt});\hspace{-1pt}\Theta(\hspace{-1pt}\tau\hspace{-1pt},\hspace{-1pt}\omega\hspace{-1pt})\hspace{-1pt})\hspace{-1pt}\right\}\hspace{-1pt}\right).
	\end{align*} 
	We have $\mathbb{E}_\omega\{\hat{t}(\tau,\omega)\}=\mathbb{E}_\omega\{t(\tau,\omega)\}$. Thus, the seller's strategic report for $\tau$ at $T\hspace{-2pt}=\hspace{-2pt}1$ does not change. Consider a modified mechanism with the modified payment function $\hat{t}(\tau,\omega)$ and the original allocation function $q(\tau,\omega)$. This modified mechanism satisfies the set of IC constraint for $\tau$ and it satisfies the ex-post IR constraint, \textit{i.e.},
	\begin{align*}
	\mathcal{R}_{\tau,\omega}=\mathbb{E}_\omega\{t(\tau,\omega)\}-\mathbb{E}_\omega\left\{C(q(\tau,\omega);\Theta(\tau,\omega))\right\}=\mathcal{R}_\tau\geq0.
	\end{align*} 
	We note that with the monitoring of $\omega$, the modified mechanism keeps the same allocation function, and therefore, results in the same \blue{ designer's objective  $\mathcal{W}$}, seller's revenue $\mathcal{R}$, and buyer's utility $\mathcal{U}$.
\end{proof}

\vspace{40pt}

\section*{Appendix B - Proof of Lemmas}

\label{sec:appB}

\begin{proof}[\textbf{Proof of Lemma \ref{lemma-IC2}}]
	Assume that the IC for $\omega$ and $\breve{\omega}$ is satisfied. Then,
	\begin{align*}
	t(\tau,\omega)-C(q(\tau,\omega);\Theta(\tau,\omega))&\geq t(\tau,\breve{\omega})-C(q(\tau,\breve{\omega});\Theta(\tau,\omega)),\nonumber\\
	t(\tau,\breve{\omega})-C(q(\tau,\breve{\omega};\Theta(\tau,\breve{\omega}))&\geq t(\tau,\omega)-C(q(\tau,\omega);\Theta(\tau,\breve{\omega})).\nonumber
	\end{align*}
	Therefore, 
	\begin{align}
	C(q(\tau,\breve{\omega});\Theta(\tau,\breve{\omega}))\hspace*{-2pt}-\hspace*{-2pt}C(q(\tau,\breve{\omega});\Theta(\tau,\omega))
	\hspace*{-2pt}
	\leq\hspace*{-2pt} \mathcal{R}_{\tau,\omega}\hspace*{-2pt}-\hspace*{-2pt}\mathcal{R}_{\tau,\breve{\omega}}&\nonumber\\ \leq\hspace*{-2pt} C(q(\tau,\omega);\Theta(\tau,\breve{\omega}))\hspace*{-2pt}-\hspace*{-2pt}C(q(\tau,\omega);\Theta(\tau,\omega))&.\label{epi-lemma1-eq1}
	\end{align}
	Set $\breve{\omega}\hspace{-1pt}=\hspace{-1pt}\omega\hspace{-1pt}-\hspace{-1pt}\epsilon$ where $\epsilon\rightarrow 0$. We have $\frac{\partial \mathcal{R}_{\tau,\omega}}{\partial \omega}\hspace{-1pt}=\hspace{-1pt}C_\theta(q(\tau,\omega);\Theta(\tau,\omega))\Theta_\omega(\tau,\omega)$.
	
	Moreover, (\ref{epi-lemma1-eq1}) implies that $q(\tau,\breve{\omega})\hspace{-2pt}<\hspace{-2pt} q(\tau,\omega)$ for $\breve{\omega}\hspace{-2pt}<\hspace{-2pt} \omega$, since by assumption $\Theta(\tau,\omega)$ is decreasing in $\omega$ and $C(\hat{q};\theta)$ is convex and increasing in $\hat{q}$.
	
	To prove the converse, assume that (\ref{RE-2}) holds and $q(\tau,\omega)$ is increasing in $\omega$. Then,
	for any $\tau,\omega,\breve{\omega}$, we have
	\begin{align}
	&\mathcal{R}_{\tau,\omega}\hspace*{-2pt}-\hspace*{-2pt}\left[t(\tau,\breve{\omega})-C(q(\tau,\breve{\omega});\Theta(\tau,\omega))\right]\hspace*{-2pt}=\\&\left[\mathcal{R}_{\tau,\omega}-\mathcal{R}_{\tau,\breve{\omega}}\right]\hspace{-2pt}-\hspace*{-2pt}\left[C(q(\tau,\breve{\omega});\Theta(\tau,\breve{\omega}))\hspace*{-2pt}-\hspace*{-2pt}C(q(\tau,\breve{\omega});\Theta(\tau,\omega))\right]\nonumber\\
	&=\hspace*{-4pt}\int_{\omega}^{\breve{\omega}}{\hspace*{-4pt}C_\theta(q(\tau,\hat{\omega});\Theta(\tau,\hat{\omega}))\Theta_\omega(\tau,\hat{\omega})d\hat{\omega}}\hspace*{-2pt}\nonumber\\&\hspace*{8pt}-\hspace*{-2pt}\left[C(q(\tau,\breve{\omega});\Theta(\tau,\breve{\omega}))\hspace*{-2pt}-\hspace*{-2pt}C(q(\tau,\breve{\omega});\Theta(\tau,\omega))\right]\geq 0, \nonumber
	\end{align}
	where the last last inequality is true since $C_\theta(q;\theta)$ is increasing in $\theta$, $\Theta(\tau,\omega)$ is  decreasing in $\omega$, and $q(\tau,\omega)$ is increasing in $\omega$.
\end{proof}

\vspace{5pt}

\begin{proof}[\textbf{Proof of Lemma \ref{lemma-off-report}}] We first note that by Assumption \ref{assum-nonshift} and the monotonicity of $\Theta(\hat{\tau},\hat{\omega})$ in $\omega$, there exists a unique $\hat{\omega}$ such that $\Theta(\tau,\omega)=\Theta(\hat{\tau},\hat{\omega})$, \textit{i.e.} $\sigma^*(\hat{\tau};\tau,\omega)$ is well-defined.
	
	Now, consider a seller with technology $\hat{\tau}$ and wind realization $\hat{\omega}$. Then, the $IC_2$ constraint requires
	\begin{align*}
	t(\hat{\tau},\hat{\omega})\hspace{-2pt}-\hspace{-2pt}C(q(\hat{\tau},\hat{\omega});\hspace{-1pt}\Theta(\hat{\tau},\hat{\omega})\hspace{-1pt})\hspace{-2pt}\geq\hspace{-2pt} t(\hat{\tau},\omega')\hspace{-2pt}-\hspace{-2pt}C(q(\hat{\tau},\omega');\hspace{-1pt}\Theta(\hat{\tau},\hat{\omega})\hspace{-1pt})\quad \hspace{-8pt}\forall \omega'\hspace{-1pt}.
	\end{align*}
	Replacing $\Theta(\hat{\tau},\hat{\omega})$ by $\Theta(\tau,\omega)$ in the above equation, we get
	\begin{align*}
	t(\hat{\tau},\hat{\omega})\hspace{-2pt}-\hspace{-2pt}C(q(\hat{\tau},\hat{\omega});\hspace{-1pt}\Theta(\tau,\omega)\hspace{-1pt})\hspace{-2pt}\geq\hspace{-2pt} t(\hat{\tau},\omega')\hspace{-2pt}-\hspace{-2pt}C(q(\hat{\tau},\omega');\hspace{-1pt}\Theta(\tau,\omega)\hspace{-1pt})\quad \hspace{-8pt}\forall \omega'\hspace{-1pt}.
	\end{align*}
	Now, consider a seller with technology $\tau$ that misreported $\hat{\tau}$ at $T\hspace{-2pt}=\hspace{-2pt}1$, and has a wind realization $\omega$ at $T\hspace{-2pt}=\hspace{-2pt}2$. The above inequality asserts that it is optimal for him to report $\hat{\omega}$ at $T\hspace{-2pt}=\hspace{-2pt}2$ so that $\Theta(\tau,\omega)=\Theta(\hat{\tau},\hat{\omega})$.
\end{proof}

\vspace{5pt}

\begin{proof}[\textbf{Proof of Lemma \ref{lemma-misreport}}]
	We have,
	\begin{align}
	&\mathbb{E}_\omega\{t(\hat{\tau},\sigma^*(\hat{\tau};\tau,\omega))-C(\hat{\tau},\sigma^*(\hat{\tau};\tau,\omega);\Theta(\tau,\omega))\}\nonumber\\
	&=\int{\hspace{-3pt}\left[t(\hspace{-1pt}\hat{\tau}\hspace{-1pt},\hspace{-1pt}\sigma^*(\hspace{-1pt}\hat{\tau}\hspace{-1pt};\hspace{-1pt}\tau\hspace{-1pt},\hspace{-1pt}\omega\hspace{-1pt})\hspace{-1pt})\hspace{-2pt}-\hspace{-2pt}C(\hspace{-1pt}q(\hspace{-1pt}\hat{\tau}\hspace{-1pt},\hspace{-1pt}\sigma^*(\hspace{-1pt}\hat{\tau}\hspace{-1pt};\hspace{-1pt}\tau\hspace{-1pt},\hspace{-1pt}\omega\hspace{-1pt})\hspace{-1pt})\hspace{-1pt};\hspace{-1pt}\Theta(\hspace{-1pt}\tau\hspace{-1pt},\hspace{-1pt}\omega\hspace{-1pt})\hspace{-1pt})\hspace{-1pt}\right]dG(\omega)}. \label{appB-eq1}
	\end{align}
	Using Lemma \ref{lemma-off-report}, we get
	\begin{align}
	&\int{\left[t(\hat{\tau},\sigma^*(\hat{\tau};\tau,\omega))-C(q(\hat{\tau},\sigma^*(\hat{\tau};\tau,\omega));\Theta(\tau,\omega))\right]dG(\omega)}
	\nonumber\\
	&=\hspace{-4pt}\int{\hspace{-4pt}\left[t(\hspace{-1pt}\hat{\tau}\hspace{-1pt},\hspace{-1pt}\sigma^*(\hspace{-1pt}\hat{\tau}\hspace{-1pt};\hspace{-1pt}\tau\hspace{-1pt},\hspace{-1pt}\omega\hspace{-1pt})\hspace{-1pt})\hspace{-2pt}-\hspace{-2pt}C(\hspace{-1pt}q(\hspace{-1pt}\hat{\tau}\hspace{-1pt},\hspace{-1pt}\sigma^*(\hspace{-1pt}\hat{\tau}\hspace{-1pt};\hspace{-1pt}\tau\hspace{-1pt},\hspace{-1pt}\omega\hspace{-1pt})\hspace{-1pt})\hspace{-1pt};\hspace{-1pt}\Theta(\hspace{-1pt}\hat{\tau}\hspace{-1pt},\hspace{-1pt}\sigma^*(\hspace{-1pt}\hat{\tau}\hspace{-1pt};\hspace{-1pt}\tau\hspace{-1pt},\hspace{-1pt}\omega\hspace{-1pt})\hspace{-1pt})\hspace{-1pt})\right]\hspace{-1pt}dG(\omega)}\nonumber
	\\&=\int{\mathcal{R}_{\hat{\tau},\sigma^*(\hat{\tau};\tau,\omega)}dG(\omega)}. \label{appB-eq2}
	\end{align}
	Then, by Lemma \ref{lemma-IC2}, 
	\begin{align}
	&\int{\mathcal{R}_{\hat{\tau},\sigma^*(\hat{\tau};\tau,\omega)}dG(\omega)}\hspace{-2pt}=\nonumber\\
	&\hspace{-4pt}\int{\hspace{-3pt}\left[\hspace{-1pt}\mathcal{R}_{\hat{\tau},\omega}\hspace{-2pt}-\hspace{-4pt}\int_{\omega}^{\sigma^*(\hat{\tau};\tau,\omega)}{\hspace{-20pt}C_\theta(q(\hat{\tau},\hat{\omega});\Theta(\hat{\tau},\hat{\omega}))\Theta_\omega(\hat{\tau},\hat{\omega})d\hat{\omega}}\right]dG(\omega)}. \label{lemma-eq-epi1}
	\end{align}
	Furthermore, $\mathcal{R}_{\hat{\tau}}=\int{\mathcal{R}_{\hat{\tau},\omega}dG(\omega)}$; thus, the RHS of (\ref{lemma-eq-epi1}) can be rewritten as,
	\begin{align}
	\mathcal{R}_{\hat{\tau}}\hspace{-2pt}-\hspace{-4pt}\int\hspace{-6pt}\int_{\omega}^{\sigma^*(\hat{\tau};\tau,\omega)}{\hspace{-15pt}C_\theta(q(\hat{\tau},\hat{\omega});\Theta(\hat{\tau},\hat{\omega}))\Theta_\omega(\hat{\tau},\hat{\omega})d\hat{\omega}dG(\omega)}. \label{appB-eq3}
	\end{align}
	
	The assertion of Lemma \ref{lemma-misreport} follows from (\ref{appB-eq1})-(\ref{appB-eq3}). 
\end{proof}

\vspace{5pt}

\begin{proof}[\textbf{Proof of Lemma \ref{lemma-S-NM}}]
	We first prove that the set of IC constraints for $\tau$ given by (\ref{rent-S-NM-P-1}) can be reduced to $\mathcal{R}_{\tau_i}\hspace{-2pt}-\hspace{-2pt}\mathcal{R}_{\tau_j}\hspace{-2pt}\geq\hspace{-2pt} \mathcal{R}^T\hspace*{-1pt}(\tau_{j},\tau_{i};q)\hspace{-2pt}+\hspace{-2pt}\mathcal{R}^W$ for all $i,j\hspace*{-2pt}\in\hspace*{-2pt}\{2,..,M\}$. Next, we show that the set of IC constraints for $\tau$ can be further reduced to $\mathcal{R}_{\tau_{i}}\hspace{-2pt}-\hspace{-2pt}\mathcal{R}_{\tau_{j}}\hspace{-2pt}\geq\hspace{-2pt} \mathcal{R}^T\hspace*{-1pt}(\tau_{j},\tau_{i};q)\hspace{-2pt}+\hspace{-2pt}\mathcal{R}^W\hspace*{-1pt}(\tau_{j},\tau_{i};q)$ for only $i\hspace*{-2pt}>\hspace*{-2pt}j$, $i,j\hspace*{-2pt}\in\hspace*{-2pt}\{2,..,M\}$.
	
	Using Lemma \ref{lemma-misreport}, we can rewrite the set of $IC_1$ constraints , given by (\ref{rent-S-NM-P-1}), as follows, 
	\begin{align}
	\hspace{-5pt}\mathcal{R}_{\tau_i}\hspace*{-2pt}-\hspace*{-2pt}\mathcal{R}_{\tau_j}\hspace*{-2pt}\geq\hspace*{-2pt} -\hspace*{-4pt}\int{\hspace*{-6pt}\int_{\omega}^{\sigma^*(\tau_j;\tau_i,\omega)}\hspace*{-25pt}C_\theta(q(\hspace{-1pt}\tau_j\hspace{-1pt},\hspace{-1pt}\hat{\omega}\hspace{-1pt});\Theta(\hspace{-1pt}\tau_j\hspace{-1pt},\hspace{-1pt}\hat{\omega}))\Theta_\omega(\hspace{-1pt}\tau_j\hspace{-1pt},\hspace{-1pt}\hat{\omega}\hspace{-1pt})d\hat{\omega}dG(\hspace{-1pt}\omega\hspace{-1pt})}. \label{thm-rent-c-eq1}
	\end{align}
	Below, we prove that the RHS of (\ref{thm-rent-c-eq1}) is equal to $\mathcal{R}^T\hspace{-1pt}(\hspace{-1pt}\tau_j\hspace{-1pt},\hspace{-1pt}\tau_i;\hspace{-1pt}q\hspace{-1pt})\hspace{-2pt}+\hspace{-2pt}\mathcal{R}^W\hspace{-1pt}(\hspace{-1pt}\tau_j\hspace{-1pt},\hspace{-1pt}\tau_i;\hspace{-1pt}q\hspace{-1pt})$. We have,
	\begin{align}
	\hspace{-4pt}\mathcal{R}^{T}\hspace{-2pt}(\hspace{-1pt}\tau_j\hspace{-1pt},\hspace{-1pt}\tau_i;\hspace{-1pt}q)&\hspace{-2pt}=\hspace{-4pt}\int{\hspace{-4pt}\left[C(q(\tau_j\hspace{-1pt},\hspace{-1pt}\omega);\hspace{-1pt}\Theta(\tau_j\hspace{-1pt},\hspace{-1pt}\omega))\hspace{-2pt}-\hspace{-2pt}C(q(\tau_j\hspace{-1pt},\hspace{-1pt}\omega);\hspace{-1pt}\Theta(\tau_i\hspace{-1pt},\hspace{-1pt}\omega)\hspace{-1pt})\hspace{-1pt}\right]\hspace{-1pt}dG(\omega)}\nonumber\\
	&\hspace{-2pt}=\hspace{-4pt}\int{\hspace*{-6pt}\int_{\Theta(\tau_i,\omega)}^{\Theta(\tau_j,\omega)}\hspace{-16pt}C_\theta(q(\tau_j,\omega);\hat{\theta})d\hat{\theta}dG(\omega)}\nonumber\\
	&\hspace{-2pt}=\hspace{-4pt}\int{\hspace*{-6pt}\int_{\sigma^*(\tau_j;\tau_i,\omega)}^{\omega}\hspace{-24pt}C_\theta(q(\tau_j,\omega);\Theta(\tau_j,\hat{\omega}))\Theta_\omega(\tau_j,\hat{\omega})d\hat{\omega}dG(\omega)}\nonumber\\
	&\hspace{-2pt}=\hspace{-2pt}-\hspace{-4pt}\int{\hspace*{-6pt}\int_{\omega}^{\sigma^*(\tau_j;\tau_i,\omega)}\hspace{-35pt}C_\theta(q(\tau_j,\omega);\Theta(\tau_j,\hat{\omega}))\Theta_\omega(\tau_j,\hat{\omega})d\hat{\omega}dG(\omega)}\hspace{-1pt},\hspace{-5pt} \label{thm-rent-c-eq2}
	\end{align}
	where the third equality results from a change of variable from $\hat{\theta}$ to $\hat{\omega}$ as $\hat{\theta}:=\Theta(\hat{\omega},\tau_j)$. Note that $\Theta(\tau_i,\omega)=\Theta(\tau_j,\sigma^*(\tau_j;\tau_i,\omega))$ by Lemma \ref{lemma-off-report}, thus, the new boundaries of integration with respect to $\hat{\omega}$ in (\ref{thm-rent-c-eq2}) are given by $\sigma^*(\tau_j;\tau_i,\omega)$ and  $\omega$ .
	
	Adding $R^{T}\hspace{-1pt}(\tau_j,\tau_i;q)$, given by (\ref{thm-rent-c-eq2}), to $R^W(\tau_j,\tau_i;q)$, we obtain,
	\begin{align}
	&\mathcal{R}^{T}(\tau_j,\tau_i;q)\hspace{-2pt}+\hspace{-2pt}\mathcal{R}^W(\tau_j,\tau_i;q)\hspace{-2pt}=\hspace*{-2pt}\nonumber\\
	&-\hspace*{-4pt}\int{\hspace*{-6pt}\int_{\omega}^{\sigma^*(\tau_j;\tau_i,\omega)}\hspace*{-15pt}C_\theta(q(\tau_j,\hat{\omega});\Theta(\tau_j,\hat{\omega}))\Theta_\omega(\tau_j,\hat{\omega})d\hat{\omega}dG(\omega)},\label{thm-rent-c-eq3}
	\end{align}
	which is equal to RHS of (\ref{thm-rent-c-eq1}).
	
	Now we show that the set of IC constraints for $\tau$ can be further reduced to  $\mathcal{R}_{\tau_{i}}\hspace{-2pt}-\hspace{-2pt}\mathcal{R}_{\tau_{j}}\hspace{-2pt}\geq\hspace{-2pt} \mathcal{R}^T\hspace*{-1pt}(\tau_{j},\tau_{i};q)\hspace{-2pt}+\hspace{-2pt}\mathcal{R}^W\hspace*{-1pt}(\tau_{j},\tau_{i};q)$ for only $i\hspace*{-2pt}>\hspace*{-2pt}j$, $i,j\hspace*{-2pt}\in\hspace*{-2pt}\{1,..,M\}$.
	
	Using (\ref{thm-rent-c-eq3}), we can write the set of inequality constraints (\ref{thm-rent-c-eq1}) as,
	\begin{align}
	\hspace{-2pt}\mathcal{R}^T(\tau_j,\tau_i;q)\hspace{-2pt}&+\hspace{-2pt}\mathcal{R}^W(\tau_j,\tau_i;q)\nonumber\\
	\hspace{30pt}\leq \hspace{-2pt}
	\mathcal{R}_{\tau_i}\hspace{-2pt}&-\hspace{-2pt}\mathcal{R}_{\tau_j}\hspace{-2pt} \leq \hspace{-2pt}\nonumber\\
	-\mathcal{R}^T(\tau_i,\tau_j;q)\hspace{-2pt}&-\hspace{-2pt}\mathcal{R}^W(\tau_i,\tau_j;q) \quad \forall i,j\hspace{-2pt}\in\hspace{-2pt}\{1,...,M\}, i\hspace{-2pt}>\hspace{-2pt}j,
	\label{eq:thm-down-1}
	\end{align}
	where the lower bound is from $IC_1$ constraint that ensures $\tau_i$ does not report $\tau_j$, and the upper bound is from the $IC_1$ constraint that ensures $\tau_j$ does not report $\tau_i$. We note that allocation $q(\tau,\omega)$ is implementable if and only if the set of constraints described by (\ref{eq:thm-down-1}) has a feasible solution for $\mathcal{R}_{\tau_i}$, $i\hspace{-2pt}\in\hspace{-2pt}\{1,2,..,M\}$. Given an allocation rule $q(\tau,\omega)$, the set of constraints given by (\ref{eq:thm-down-1}), has a feasible solutions only if for any arbitrary increasing sequence $(\hspace{-1pt}k_1\hspace{-1pt},\hspace{-1pt}k_2\hspace{-1pt}),\hspace{-1pt}(\hspace{-1pt}k_2\hspace{-1pt},\hspace{-1pt}k_3\hspace{-1pt}),\hspace{-1pt}\cdots\hspace{-1pt},\hspace{-1pt}(\hspace{-1pt}k_{m-1}\hspace{-1pt},\hspace{-1pt}k_m\hspace{-1pt})$, where $\tau_{k_{r-1}}\hspace{-2pt}<\hspace{-2pt}\tau_{k_r}$ for $2\hspace{-1pt}\leq\hspace{-2pt} r\hspace{-2pt}\leq \hspace{-2pt}m$, 
	\begin{align}
	&\hspace{-1pt}\mathcal{R}^T\hspace{-2pt}(\tau_{k_1}\hspace{-1pt},\hspace{-1pt}\tau_{k_2};\hspace{-1pt}q)\hspace{-2pt}+\hspace{-2pt}\mathcal{R}^W\hspace{-2pt}(\tau_{k_1}\hspace{-1pt},\hspace{-1pt}\tau_{k_2};\hspace{-1pt}q)\nonumber\\
	+&\mathcal{R}^T\hspace{-2pt}(\tau_{k_2}\hspace{-1pt},\hspace{-1pt}\tau_{k_3};\hspace{-1pt}q)\hspace{-2pt}+\hspace{-2pt}\mathcal{R}^W\hspace{-2pt}(\tau_{k_2}\hspace{-1pt},\hspace{-1pt}\tau_{k_3};\hspace{-1pt}q)\nonumber\\
	&\vdots\nonumber\\
	+&\mathcal{R}^T\hspace{-2pt}(\tau_{k_{m-1}}\hspace{-1pt},\hspace{-1pt}\tau_{k_m};\hspace{-1pt}q)\hspace{-2pt}+\mathcal{R}^W\hspace{-2pt}(\tau_{k_{m-1}}\hspace{-1pt},\hspace{-1pt}\tau_{k_m};\hspace{-1pt}q)\hspace{-2pt}\nonumber\\
	&\leq \sum_{r=2}^m \Big[\mathcal{R}_{\tau_{k_r}}\hspace{-2pt}-\hspace{-2pt}\mathcal{R}_{\tau_{k_{r-1}}}\Big]\hspace{-2pt}\nonumber\\
	&=\mathcal{R}_{\tau_{k_m}}\hspace{-2pt}-\hspace{-2pt}\mathcal{R}_{\tau_{k_{1}}}\nonumber\\ &\leq-\mathcal{R}^T\hspace{-2pt}(\tau_{k_m}\hspace{-1pt},\hspace{-1pt}\tau_{k_1};\hspace{-1pt}q)\hspace{-2pt}-\hspace{-2pt}\mathcal{R}^W\hspace{-2pt}(\tau_{k_m}\hspace{-1pt},\hspace{-1pt}\tau_{k_1};\hspace{-1pt}q).\label{thm-rent-c-eq4}
	\end{align}
	
	In the following, we show that for a given implementable allocation $q(\tau,\omega)$, the set of constraint  (\ref{eq:thm-down-1}) can be reduced to
	\begin{align}
	\hspace{-6pt}\mathcal{R}_{\tau_i}\hspace{-2pt}-\hspace{-2pt}\mathcal{R}_{\tau_j}\hspace{-2pt}
	\geq\hspace{-2pt} \mathcal{R}^T\hspace{-1pt}(\tau_j,\hspace{-1pt}\tau_i;\hspace{-1pt}q) \hspace{-2pt}+\hspace{-2pt}\mathcal{R}^W\hspace{-1pt}(\tau_j,\hspace{-1pt}\tau_i;\hspace{-1pt}q) \quad \hspace{-2pt}\forall i,\hspace{-1pt}j\hspace{-2pt}\in\hspace{-2pt}\{\hspace{-1pt}1\hspace{-1pt},\hspace{-1pt}...,\hspace{-1pt}M\hspace{-1pt}\}, i\hspace{-2pt}>\hspace{-2pt}j. \label{eq:thm-down-2}\hspace{-7pt}
	\end{align}

	\blue{We note that $\mathcal{W}=\mathcal{S}-(1-\alpha)\mathcal{R}$, for $\alpha\in[0,1]$. }	
	Therefore, \blue{for a given allocation function $q(\tau,\omega)$,} the designer wants to minimize $\mathcal{R}$. Let $\mathcal{R}^*_{\tau}$ denote the optimal seller's revenue that satisfies (\ref{eq:thm-down-2}) for a given implementable allocation function $q(\tau,\omega)$. Construct a graph with $M$ nodes, where there is an edge between between node $i$ and $j$ if (\ref{eq:thm-down-2}) is binding for $i$ and $j$, $i>j$. 
	
	First, we note that the resulting graph must be connected. If not, then there exist at least two unconnected clusters of nodes. Consider a cluster that does not include node $1$. Then one can  reduce the value of $\mathcal{R}^*_{\tau}$ by $\epsilon>0$ for all the nodes in that cluster without violating any of the constraints (\ref{eq:thm-down-2}), and improve the outcome of the mechanism by reducing the seller's revenue $\mathcal{R}$.
	
	Now, assume that the optimal seller's revenue $\mathcal{R}^*_\tau$ that is determined by only considering the set of constraints (\ref{eq:thm-down-2}), does not satisfy the set of constraints (\ref{eq:thm-down-1}). Then, there exists $i>j$ so that the following constraint that is violated,
	\begin{align*}
	\mathcal{R}_{\tau_i}^*-\mathcal{R}^*_{\tau_j}\nleq -\mathcal{R}^T(\tau_i,\tau_j;q)-\mathcal{R}^W(\tau_i,\tau_j;q).
	\end{align*}
	Let $C=\{(i,k_1),(k_1,k_2),\cdots,(k_m,j)\}$ denote a path between node $i$ and node $j$ in the connected graph constructed above. Then,
	\begin{align}
	\mathcal{R}_{\tau_i}^*-\mathcal{R}^*_{\tau_j}&=\sum_{(\hat{k},k)\in C} \mathcal{R}^T(\tau_{\hat{k}},\tau_k;q)+\mathcal{R}^W(\tau_{\hat{k}},\tau_k;q)\nonumber\\
	&\nleq -\mathcal{R}^T(\tau_i,\tau_j;q)-\mathcal{R}^W(\tau_i,\tau_j;q),
	\end{align}
	which contradicts (\ref{thm-rent-c-eq4}). 
	
\end{proof}


\begin{proof}[\textbf{Proof of Lemma \ref{lemma-F}}] 
	
	\blue {We can rewrite the set of IC constraints (\ref{rent-forward}) as follows,
	\begin{align*}
	\mathcal{R}_{\tau_{i}}\hspace{-2pt}\hspace{25pt}&\hspace{-25pt}\geq\hspace{-2pt}t(\hspace{-1pt}\tau_{j}\hspace{-1pt})\hspace{-2pt}-\hspace{-2pt}\mathbb{E}_\omega\{ C(\hspace{-1pt}q(\hspace{-1pt}\tau_{j}\hspace{-1pt}\hspace{-1pt});\hspace{-1pt}\Theta(\hspace{-1pt}\tau_{i}\hspace{-1pt},\hspace{-1pt}\omega\hspace{-1pt})\hspace{-1pt})\}\hspace{-2pt}\\
	&\hspace{-28pt}=\hspace*{-2pt}\mathcal{R}_{\tau_{j}}\hspace{-4pt}+\hspace{-3pt}\Big[\hspace{-1pt}\mathbb{E}_\omega\{\hspace{-1pt}C(\hspace{-1pt}q(\hspace{-1pt}\tau_{j}\hspace{-1pt});\hspace{-1pt}\Theta(\hspace{-1pt}\tau_{j}\hspace{-1pt},\hspace{-1pt}\omega\hspace{-1pt})\hspace{-1pt})\hspace{-1pt}\}\hspace{-2pt}-\hspace{-2pt}\mathbb{E}_\omega\{\hspace{-1pt}C(\hspace{-1pt}q(\hspace{-1pt}\tau_{j}\hspace{-1pt});\hspace{-1pt}\Theta(\hspace{-1pt}\tau_{i}\hspace{-1pt},\hspace{-1pt}\omega\hspace{-1pt})\hspace{-1pt})\hspace{-1pt}\}\hspace{-2pt}\Big]\hspace*{-2pt},\nonumber\\
	&\hspace{-28pt}=\hspace*{-2pt}\mathcal{R}_{\tau_{j}}\hspace{-2pt}+\hspace{-2pt}\mathcal{R}^T(\hspace{-1pt}\tau_j\hspace{-1pt},\hspace{-1pt}\tau_i\hspace{-1pt};q\hspace{-1pt})\quad\forall i,j\in\{1,...,M\},
	\end{align*}
	where the last equality holds by definition.}
	We proceed as follows. We first consider a relaxed version of the forward mechanism design problem. We determine the solution to the relaxed problem, and show it is also a feasible solution for the original forward mechanism design problem.
	 
	Consider the following relaxed version of the forward mechanism design problem,
	\begin{align}
	&\max \mathcal{\blue{\mathcal{W}}}\nonumber\\
	\hspace{-1pt}\text{subject to}&\nonumber\\
	\hspace{-20pt}\mathcal{R}_{\tau_{i}}\hspace{-2pt}\hspace{25pt}&\hspace{-25pt}\geq\hspace{-2pt}t(\hspace{-1pt}\tau_{i-1}\hspace{-1pt})\hspace{-2pt}-\hspace{-2pt}\mathbb{E}_\omega\{ C(\hspace{-1pt}q(\hspace{-1pt}\tau_{i-1}\hspace{-1pt}\hspace{-1pt});\hspace{-1pt}\Theta(\hspace{-1pt}\tau_{i}\hspace{-1pt},\hspace{-1pt}\omega\hspace{-1pt})\hspace{-1pt})\}\hspace{-2pt}\nonumber\\
	&\hspace{-28pt}=\hspace*{-2pt}\mathcal{R}_{\tau_{i-1}}\hspace{-4pt}+\hspace{-3pt}\Big[\hspace{-1pt}\mathbb{E}_\omega\{\hspace{-1pt}C(\hspace{-1pt}q(\hspace{-1pt}\tau_{i-1}\hspace{-1pt});\hspace{-1pt}\Theta(\hspace{-1pt}\tau_{i-1}\hspace{-1pt},\hspace{-1pt}\omega\hspace{-1pt})\hspace{-1pt})\hspace{-1pt}\}\hspace{-2pt}-\hspace{-2pt}\mathbb{E}_\omega\{\hspace{-1pt}C(\hspace{-1pt}q(\hspace{-1pt}\tau_{i-1}\hspace{-1pt});\hspace{-1pt}\Theta(\hspace{-1pt}\tau_{i}\hspace{-1pt},\hspace{-1pt}\omega\hspace{-1pt})\hspace{-1pt})\hspace{-1pt}\}\hspace{-2pt}\Big]\hspace*{-2pt},\nonumber\\
	&\hspace{140pt}i\hspace{-2pt}\in\hspace{-2pt}\{2,...,M\},\hspace{-5pt}\label{lemma-F-eq1}\\
	&\hspace{-22pt}\mathcal{R}_{\tau_i}\geq0,i\hspace{-2pt}\in\hspace{-2pt}\{1,...,M\},
	\end{align}
	where we only include the set of IC constraints that ensures a seller with type $\tau_{i}$ does not report type $\tau_{i-1}$, and the other IC constraints are omitted.

	Note that $\mathbb{E}_\omega\{\hspace{-1pt}C(\hspace{-1pt}q(\hspace{-1pt}\tau_{i-1}\hspace{-1pt});\hspace{-1pt}\Theta(\hspace{-1pt}\tau_{i-1}\hspace{-1pt},\hspace{-1pt}\omega\hspace{-1pt})\hspace{-1pt})\hspace{-2pt}\}\geq\mathbb{E}_\omega\{\hspace{-1pt}C(\hspace{-1pt}q(\hspace{-1pt}\tau_{i-1}\hspace{-1pt});\hspace{-1pt}\Theta(\hspace{-1pt}\tau_{i}\hspace{-1pt},\hspace{-1pt}\omega\hspace{-1pt})\hspace{-1pt})\hspace{-1pt}\}$ since $\Theta(\hspace{-1pt}\tau_{i-1}\hspace{-1pt},\hspace{-1pt}\omega\hspace{-1pt})\geq \Theta(\hspace{-1pt}\tau_{i}\hspace{-1pt},\hspace{-1pt}\omega\hspace{-1pt})$ by Assumption \ref{assump-FSD}. Thus, from (\ref{lemma-F-eq1}) we have $\mathcal{R}_{\tau_i}\geq \mathcal{R}_{\tau_j}$ for $i>j$. \blue{The designer's objective is $\mathcal{W}=\mathcal{S}-(1-\alpha)\mathcal{R}$. Therefore, for a given allocation function $q(\tau,\omega)$, the designer} wants to minimize  $\mathcal{R}_{\tau_i}$ $i\in\{1,2,...,M\}$. Therefore, at the optimal solution to the relaxed problem, 
	\begin{align*}
	&\mathcal{R}_{\tau_{i}}\hspace{-2pt}=\hspace{-2pt}\mathcal{R}_{\tau_{i-1}}\hspace{-3pt}+\hspace{-3pt}\Big[\hspace{-1pt}\mathbb{E}_\omega\{\hspace{-1pt}C(\hspace{-1pt}q(\hspace{-1pt}\tau_{i-1}\hspace{-1pt});\hspace{-1pt}\Theta(\hspace{-1pt}\tau_{i-1}\hspace{-1pt},\hspace{-1pt}\omega\hspace{-1pt})\hspace{-1pt})\hspace{-1pt}\}\hspace{-2pt}-\hspace{-2pt}\mathbb{E}_\omega\{\hspace{-1pt}C(\hspace{-1pt}q(\hspace{-1pt}\tau_{i-1}\hspace{-1pt});\hspace{-1pt}\Theta(\hspace{-1pt}\tau_{i}\hspace{-1pt},\hspace{-1pt}\omega\hspace{-1pt})\hspace{-1pt})\hspace{-1pt}\}\hspace{-2pt}\Big]\hspace{-2pt},\\
	&\mathcal{R}_{\tau_1}=0.
	\end{align*}
	 
	Therefore, we can write,
	\begin{align}
	\mathcal{R}_{\tau_{i}}\hspace{-2pt}=\sum_{j=1}^{i-1}\hspace{-2pt}\Big[\hspace{-1pt}\mathbb{E}_\omega\{\hspace{-1pt}C(\hspace{-1pt}q(\hspace{-1pt}\tau_j\hspace{-1pt});\hspace{-1pt}\Theta(\hspace{-1pt}\tau_{j}\hspace{-1pt},\hspace{-1pt}\omega\hspace{-1pt})\hspace{-1pt})\hspace{-1pt}\}\hspace{-2pt}-\hspace{-2pt}\mathbb{E}_\omega\{\hspace{-1pt}C(\hspace{-1pt}q(\hspace{-1pt}\tau_j\hspace{-1pt});\hspace{-1pt}\Theta(\hspace{-1pt}\tau_{j+1}\hspace{-1pt},\hspace{-1pt}\omega\hspace{-1pt})\hspace{-1pt})\hspace{-1pt}\}\hspace{-1pt}\Big]\hspace{-1pt}. \label{optimalforward-rent}
	\end{align}
	Substituting $\mathcal{R}_{\tau_i}$, $1\hspace{-2pt}\leq i\hspace{-2pt}\leq\hspace{-2pt} M$, we get
	\begin{align*}
	&\mathcal{S}-\blue{(1-\alpha)}\sum_{i=1}^M p_i\mathcal{R}_{\tau_i}\hspace{-2pt}\\
	&\hspace{51pt}=\hspace{-2pt}\sum_{i=1}^np_i\left[\mathcal{V}(q(\tau_i))\hspace{-2pt}-\hspace{-2pt}\mathbb{E}_\omega\{\hspace{-1pt}C(q(\tau_i);\Theta(\tau_i,\omega))\hspace{-1pt}\}\right]\\
	&\hspace{-1pt}-\hspace{-2pt}\blue{(\hspace{-1pt}1\hspace{-2pt}-\hspace{-2pt}\alpha\hspace{-1pt})}\hspace{-2pt}\sum_{i=1}^{M}\hspace{-1pt}p_i\hspace{-3pt}\sum_{j=1}^{i-1}\hspace{-2pt}\Big[\hspace{-1pt}\mathbb{E}_\omega\hspace{-1pt}\{\hspace{-1pt}C(\hspace{-1pt}q(\hspace{-1pt}\tau_j\hspace{-1pt});\hspace{-1pt}\Theta(\hspace{-1pt}\tau_{j}\hspace{-1pt},\hspace{-1pt}\omega\hspace{-1pt})\hspace{-1pt})\hspace{-1pt}\}\hspace{-2pt}-\hspace{-2pt}\mathbb{E}_\omega\hspace{-1pt}\{\hspace{-1pt}C(\hspace{-1pt}q(\hspace{-1pt}\tau_j\hspace{-1pt});\hspace{-1pt}\Theta(\hspace{-1pt}\tau_{j+1}\hspace{-1pt},\hspace{-1pt}\omega\hspace{-1pt})\hspace{-1pt})\hspace{-1pt}\}\hspace{-2pt}\Big]\\
	&\hspace{51pt}=\hspace{-2pt}\sum_{i=1}^M p_i\left[\mathcal{V}(q(\tau_i))\hspace{-2pt}-\hspace{-2pt}\mathbb{E}_\omega\{\hspace{-1pt}C(q(\tau_i);\Theta(\tau_i,\omega))\hspace{-1pt}\}\right]-\\
	&\hspace{-1pt}\hspace{-2pt}\blue{(\hspace{-1pt}1\hspace{-2pt}-\hspace{-2pt}\alpha\hspace{-1pt})}\hspace{-4pt}\sum_{j=1}^{M-1}\hspace{-5pt}\left(\hspace{-1pt}\sum_{i=j+1}^{M}\hspace{-4pt}p_i\hspace{-3pt}\right)\hspace{-4pt}\Big[\hspace{-1pt}\mathbb{E}_\omega\hspace{-1pt}\{\hspace{-1pt}C(\hspace{-1pt}q(\hspace{-1pt}\tau_j\hspace{-1pt});\hspace{-1pt}\Theta(\hspace{-1pt}\tau_{j}\hspace{-1pt},\hspace{-1pt}\omega\hspace{-1pt})\hspace{-1pt})\hspace{-1pt}\}\hspace{-2pt}-\hspace{-2pt}\mathbb{E}_\omega\hspace{-1pt}\{\hspace{-1pt}C(\hspace{-1pt}q(\hspace{-1pt}\tau_j\hspace{-1pt});\hspace{-1pt}\Theta(\hspace{-1pt}\tau_{j+1}\hspace{-1pt},\hspace{-1pt}\omega\hspace{-1pt})\hspace{-1pt})\hspace{-1pt}\}\hspace{-2pt}\Big]\\
	&\hspace{51pt}=\hspace{-2pt}\sum_{i=1}^M p_i\Big(\mathcal{V}(q(\tau_i))\hspace{-2pt}-\hspace{-2pt}\mathbb{E}_\omega\{\hspace{-1pt}C(q(\tau_i);\Theta(\tau_i,\omega))\hspace{-1pt}\}-\\
	&\hspace{-1pt}\hspace{-3pt}\blue{(\hspace{-1pt}1\hspace{-2pt}-\hspace{-2pt}\alpha\hspace{-1pt})}\hspace{-3pt}\left(\sum_{j=i+1}^{M}\hspace{-3pt}\frac{p_j}{p_i}\hspace{-2pt}\right)\hspace{-4pt}\Big[\hspace{-1pt}\mathbb{E}_\omega\hspace{-1pt}\{\hspace{-1pt}C(\hspace{-1pt}q(\hspace{-1pt}\tau_j\hspace{-1pt});\hspace{-1pt}\Theta(\hspace{-1pt}\tau_{j}\hspace{-1pt},\hspace{-1pt}\omega\hspace{-1pt})\hspace{-1pt})\hspace{-1pt}\}\hspace{-2pt}-\hspace{-2pt}\mathbb{E}_\omega\hspace{-1pt}\{\hspace{-1pt}C(\hspace{-1pt}q(\hspace{-1pt}\tau_j\hspace{-1pt});\hspace{-1pt}\Theta(\hspace{-1pt}\tau_{j+1}\hspace{-1pt},\hspace{-1pt}\omega\hspace{-1pt})\hspace{-1pt})\hspace{-1pt}\}\hspace{-2pt}\Big]\hspace{-1pt}\Big)\hspace{-1pt}.\\
\end{align*} 
	By maximizing the above expression with respect to $q(\tau_i)$ and using the first-order condition, we find that the optimal $q(\tau_i)$ is determined by the following equation
	\begin{align}
	\hspace{-6pt}v(q)\hspace{-2pt}=&\mathbb{E}_\omega\hspace{-1pt}\{\hspace{-1pt}c(q;\hspace{-1pt}\Theta(\tau_i,\hspace{-1pt}\omega)\hspace{-1pt})\hspace{-1pt}\}\nonumber\\&\hspace{-2pt}+\hspace{-2pt}\blue{(\hspace{-1pt}1\hspace{-2pt}-\hspace{-2pt}\alpha\hspace{-1pt})}\hspace{-5pt}\sum_{j=i+1}^M\hspace{-2pt}\frac{p_i}{p_j}\Big[\mathbb{E}_\omega\hspace{-1pt}\{c(\hspace{-1pt}q;\hspace{-1pt}\Theta(\hspace{-1pt}\tau_{i}\hspace{-1pt},\hspace{-1pt}\omega\hspace{-1pt})\hspace{-2pt})\}\hspace{-2pt}-\hspace{-2pt}\mathbb{E}_\omega\{c(\hspace{-1pt}q;\hspace{-1pt}\Theta(\hspace{-1pt}\tau_{i+1}\hspace{-1pt},\hspace{-1pt}\omega\hspace{-1pt})\hspace{-2pt})\hspace{-1pt}\}\hspace{-1pt}\Big]\hspace{-1pt}.\label{optimalforward-q}\hspace{-4pt}
	\end{align}
	The above equation has a unique solution since the LHS is decreasing in $q$ by the concavity of $\mathcal{V}(q)$, and the RHS is increasing in $q$ by Assumption \ref{assump-FSD}. Moreover, note that the RHS is decreasing in $i$ by Assumption \ref{assump-FSD}, therefore, $q(\tau_i)>q(\tau_j)$ for $i>j$.
	
	Now, we show that the optimal solution to the relaxed problem satisfies the omitted constraints. Consider the IC constraint for type $\tau_i$ reporting $\tau_j$. We need to show that the the utility of a seller with technology $\tau_i$ is higher when he reports truthfully than his utility from misreporting $\tau_j$. That is, 
	\begin{align*}
	\mathcal{R}_{\tau_i}-\left[t(\tau_j)-\mathbb{E}_\omega\{C(q(\tau_j);\Theta(\tau_i,\omega))\}\right]=&\mathcal{R}_{\tau_i}\hspace{-2pt}-\hspace{-2pt}\mathcal{R}_{\tau_j}\hspace{-2pt}\\&\hspace*{-130pt}+\hspace{-2pt}\left[\mathbb{E}_\omega\{C(q(\tau_j);\Theta(\tau_i,\omega))\}\hspace{-2pt}-\hspace{-2pt}\mathbb{E}_\omega\{C(q(\tau_j);\Theta(\tau_j,\omega))\}\right]\hspace{-2pt}\geq\hspace{-2pt} 0.
	\end{align*}
	For $i>j$, we can write,
	\begin{align*}
	&\mathcal{R}_{\tau_i}\hspace{-2pt}-\hspace{-2pt}\mathcal{R}_{\tau_j}\hspace{-2pt}+\hspace{-2pt}\left[\mathbb{E}_\omega\{C(q(\tau_j);\Theta(\tau_i,\omega)\hspace{-1pt})\hspace{-1pt}\}\hspace{-2pt}-\hspace{-2pt}\mathbb{E}_\omega\{C(q(\tau_j,\omega);\Theta(\tau_j,\omega))\hspace{-1pt}\}\hspace{-1pt}\right]\\
	=&\sum_{k=j}^{i-1}\hspace{-2pt}\Big[\hspace{-1pt}\mathbb{E}_\omega\{C(\hspace{-1pt}q(\hspace{-1pt}\tau_k\hspace{-1pt}\hspace{-1pt});\hspace{-1pt}\Theta(\hspace{-1pt}\tau_{k}\hspace{-1pt},\hspace{-1pt}\omega\hspace{-1pt})\hspace{-1pt})\}\hspace{-2pt}-\hspace{-2pt}\mathbb{E}_\omega\{C(\hspace{-1pt}q(\hspace{-1pt}\tau_k\hspace{-1pt});\hspace{-1pt}\Theta(\hspace{-1pt}\tau_{k+1}\hspace{-1pt},\hspace{-1pt}\omega\hspace{-1pt})\hspace{-1pt})\}\hspace{-1pt}\Big]\\&+\hspace{-2pt}\left[\mathbb{E}_\omega\{C(q(\tau_j);\Theta(\tau_i,\omega)\hspace{-1pt})\hspace{-1pt}\}\hspace{-2pt}-\hspace{-2pt}\mathbb{E}_\omega\{C(q(\tau_j);\Theta(\tau_j,\omega))\hspace{-1pt}\}\hspace{-1pt}\right]\\
	=&\sum_{k=j}^{i-1}\hspace{-2pt}\Big[\hspace{-1pt}\mathbb{E}_\omega\{C(\hspace{-1pt}q(\hspace{-1pt}\tau_k\hspace{-1pt}\hspace{-1pt});\hspace{-1pt}\Theta(\hspace{-1pt}\tau_{k}\hspace{-1pt},\hspace{-1pt}\omega\hspace{-1pt})\hspace{-1pt})\}\hspace{-2pt}-\hspace{-2pt}\mathbb{E}_\omega\{C(\hspace{-1pt}q(\hspace{-1pt}\tau_k\hspace{-1pt});\hspace{-1pt}\Theta(\hspace{-1pt}\tau_{k+1}\hspace{-1pt},\hspace{-1pt}\omega\hspace{-1pt})\hspace{-1pt})\}\hspace{-1pt}\Big]\\	&\hspace{2pt}-\sum_{k=j}^{i-1}\hspace{-2pt}\Big[\hspace{-1pt}\mathbb{E}_\omega\{C(\hspace{-1pt}q(\hspace{-1pt}\tau_j\hspace{-1pt});\hspace{-1pt}\Theta(\hspace{-1pt}\tau_{k}\hspace{-1pt},\hspace{-1pt}\omega\hspace{-1pt})\hspace{-1pt})\hspace{-1pt}\}\hspace{-2pt}-\hspace{-2pt}\mathbb{E}_\omega\{C(\hspace{-1pt}q(\hspace{-1pt}\tau_j\hspace{-1pt});\hspace{-1pt}\Theta(\hspace{-1pt}\tau_{k+1}\hspace{-1pt},\hspace{-1pt}\omega\hspace{-1pt})\hspace{-1pt})\hspace{-1pt}\}\hspace{-1pt}\Big]\geq 0,
	\end{align*}
	since $q(\tau_k)\geq q(\tau_j)$ and $C(\hat{q};\Theta(\tau_k,\omega))-C(\hat{q};\Theta(\tau_{k+1},\omega))$ is increasing in $\hat{q}$ by Assumption \ref{assump-FSD}.

	Similarly for $i<j$, we have, 
	\begin{align*}
	&\mathcal{R}_{\tau_i}\hspace{-2pt}-\hspace{-2pt}\mathcal{R}_{\tau_j}\hspace{-2pt}+\hspace{-2pt}\left[\mathbb{E}_\omega\{C(q(\tau_j);\Theta(\tau_i,\omega)\hspace{-1pt})\hspace{-1pt}\}\hspace{-2pt}-\hspace{-2pt}\mathbb{E}_\omega\{C(q(\tau_j,\omega);\Theta(\tau_j,\omega))\hspace{-1pt}\}\hspace{-1pt}\right]\\
	=&-\sum_{k=i}^{j-1}\hspace{-2pt}\Big[\hspace{-1pt}\mathbb{E}_\omega\{C(\hspace{-1pt}q(\hspace{-1pt}\tau_k\hspace{-1pt});\hspace{-1pt}\Theta(\hspace{-1pt}\tau_{k}\hspace{-1pt},\hspace{-1pt}\omega\hspace{-1pt})\hspace{-1pt})\}\hspace{-2pt}-\hspace{-2pt}\mathbb{E}_\omega\{C(\hspace{-1pt}q(\hspace{-1pt}\tau_k\hspace{-1pt});\hspace{-1pt}\Theta(\hspace{-1pt}\tau_{k+1}\hspace{-1pt},\hspace{-1pt}\omega\hspace{-1pt})\hspace{-1pt})\}\hspace{-1pt}\Big]\\&+\Big[\mathbb{E}_\omega\{C(q(\tau_j);\Theta(\tau_i,\omega))\}-\mathbb{E}_\omega\{C(q(\tau_j);\Theta(\tau_j,\omega))\}\Big]\\
	=&-\sum_{k=i}^{j-1}\hspace{-2pt}\Big[\hspace{-1pt}\mathbb{E}_\omega\{C(\hspace{-1pt}q(\hspace{-1pt}\tau_k\hspace{-1pt});\hspace{-1pt}\Theta(\hspace{-1pt}\tau_{k}\hspace{-1pt},\hspace{-1pt}\omega\hspace{-1pt})\hspace{-1pt})\}\hspace{-2pt}-\hspace{-2pt}\mathbb{E}_\omega\{C(\hspace{-1pt}q(\hspace{-1pt}\tau_k\hspace{-1pt});\hspace{-1pt}\Theta(\hspace{-1pt}\tau_{k+1}\hspace{-1pt},\hspace{-1pt}\omega\hspace{-1pt})\hspace{-1pt})\}\hspace{-1pt}\Big]\\
	&\hspace{2pt}+\sum_{k=j}^{i-1}\hspace{-2pt}\Big[\hspace{-1pt}\mathbb{E}_\omega\{C(\hspace{-1pt}q(\hspace{-1pt}\tau_j\hspace{-1pt});\hspace{-1pt}\Theta(\hspace{-1pt}\tau_{k}\hspace{-1pt},\hspace{-1pt}\omega\hspace{-1pt})\hspace{-1pt})\}\hspace{-2pt}-\hspace{-2pt}\mathbb{E}_\omega\{C(\hspace{-1pt}q(\hspace{-1pt}\tau_j\hspace{-1pt});\hspace{-1pt}\Theta(\hspace{-1pt}\tau_{k+1}\hspace{-1pt},\hspace{-1pt}\omega\hspace{-1pt})\hspace{-1pt})\}\hspace{-1pt}\Big]\geq 0,
	\end{align*}
	where the last inequality is true since $q(\tau_k)\hspace{-2pt}\leq\hspace{-2pt} q(\tau_j)$ for $k\hspace{-2pt}\leq \hspace{-2pt}j$, and $C(\hat{q}\hspace{-1pt};\hspace{-1pt}\Theta(\tau_k,\hspace{-1pt}\omega)\hspace{-1pt})\hspace{-2pt}-\hspace{-2pt}C(\hat{q}\hspace{-1pt};\hspace{-1pt}\Theta(\tau_{k+1},\hspace{-1pt}\omega)\hspace{-1pt})$ is increasing in $\hat{q}$ by Assumption \ref{assump-FSD}. 
	
\end{proof}


\vspace{5pt}

\begin{proof}[\textbf{Proof of Lemma \ref{lemma-R-NM}}]
	First we note that in the \blue{real-time} mechanism the seller reports $\tau$ and $\omega$ simultaneously, and his cost only depends $\Theta(\tau,\omega)$. Therefore, allocation function $q(\tau,\omega)$ and payment function $t(\tau,\omega)$ must only depend on $\Theta(\tau,\omega)$ rather than exact values of $\tau$ and $\omega$. 
	Therefore, the mechanism design problem can be written only in terms of $\theta$, where the buyer designs $\{q(\theta),t(\theta)\}$, and asks the seller to report $\theta$ instead of $(\tau,\omega)$. The reformulation of the \blue{real-time} mechanism in terms of $\theta$ can be written as follows,
	\begin{align}
	&\max_{q(\cdot)}{\blue{\mathcal{W}}}\label{R-NM-max}\\
	&\hspace{-15pt}\text{subject to}\nonumber\\
	&IC:\mathcal{R}_\theta\geq t(\hat{\theta})-C(q(\hat{\theta});\theta),\quad \forall \theta,\label{IC-R-NM-ref}\\
	&IR: \mathcal{R}_\theta\geq 0,\quad \forall \theta\label{R-NM-mon}.
	\end{align}
	
	\begin{claim} The sets of IC constraints (\ref{IC-R-NM-ref}) and IR constraints  (\ref{R-NM-mon}) are satisfied if and only if  
		\begin{align}
		&\frac{\partial \mathcal{R}_{\theta}}{\partial \theta}=-C_\theta(q(\theta);\theta),	\label{lemma-R-NM-eq1}\\
		&\mathcal{R}_{\overline{\theta}}\geq0,
		\label{lemma-R-NM-eq3}
		\end{align}
		and $q(\theta)$ is decreasing in $\theta$. \label{claim-R-NM}
	\end{claim}
	
	We prove the result of Claim \ref{claim-R-NM} below. Assume (\ref{IC-R-NM-ref}) is true. Consider the following two IC constraints, where the first one requires that a seller with type $\theta$ does not gain by misreporting $\hat{\theta}$, and the second one requires that a seller with true type $\hat{\theta}$ does not gain by misreporting $\theta$.
	\begin{align*}
	t(\hat{\theta})-C(q(\hat{\theta});\hat{\theta})&\geq t(\theta)-C(q(\theta);\hat{\theta}),\nonumber\\
	t(\theta)-C(q(\theta);\theta)&\geq t(\hat{\theta})-C(q(\hat{\theta});\theta).\nonumber
	\end{align*}
	By subtracting the above two IC constraints, we obtain,
	\begin{align}
	C(q(\hat{\theta});\hat{\theta})\hspace*{-2pt}-\hspace*{-2pt}C(q(\hat{\theta});\theta)
	\hspace*{-2pt}
	\leq\hspace*{-2pt} \mathcal{R}_{\theta}\hspace*{-2pt}-\hspace*{-2pt}\mathcal{R}_{\hat{\theta}}\hspace{-2pt}\leq\hspace*{-2pt} C(q(\theta);\hat{\theta})\hspace*{-2pt}-\hspace*{-2pt}C(q(\theta);\theta).\label{lemma-R-NM-eq2}
	\end{align}
	Let $\hat{\theta}=\theta+\epsilon$ and take $\epsilon\rightarrow 0$. Then $\frac{\partial \mathcal{R}_{\theta}}{\partial \theta}=-C_\theta(q(\theta);\theta)$. 
	Moreover, by (\ref{lemma-R-NM-eq2}),
	\begin{align*} C(q(\hat{\theta});\hat{\theta})\hspace*{-2pt}-\hspace*{-2pt}C(q(\hat{\theta});\theta)
	\hspace*{-2pt}
	\leq\hspace*{-2pt} C(q(\theta);\hat{\theta})\hspace*{-2pt}-\hspace*{-2pt}C(q(\theta);\theta),\end{align*}
	which along with Assumption \ref{assump-FSD}, implies that $q(\theta)\hspace{-2pt}>\hspace{-2pt} q(\hat{\theta})$  
	for $\theta\hspace{-2pt}<\hspace{-2pt} \hat{\theta}$. 
	
	Furthermore,  by Assumption \ref{assump-FSD} $\frac{\partial \mathcal{R}_{\theta}}{\partial \theta}\hspace*{-2pt}=\hspace*{-2pt}-C_\theta(\hspace*{-1pt}q(\theta);\hspace*{-1pt}\theta)\hspace*{-2pt}\leq\hspace*{-2pt}0$. Therefore, (\ref{R-NM-mon}) is satisfied if and only if $\mathcal{R}_{\overline{\theta}}\hspace*{-2pt}\geq\hspace*{-2pt}0$.

	To prove the converse, assume that (\ref{lemma-R-NM-eq1}) and (\ref{lemma-R-NM-eq3}) hold and $q(\theta)$ is decreasing in $\theta$. First, we show that the IC constraints (\ref{IC-R-NM-ref}) is satisfied.	
	For any $\theta$ and $\hat{\theta}$, we have
	\begin{align}
	&\mathcal{R}_{\theta}\hspace*{-2pt}-\hspace*{-2pt}\left[t(\hat{\theta})-C(q(\hat{\theta});\theta)\right]\hspace*{-2pt}=\hspace{-2pt}\left[\mathcal{R}_{\theta}-\mathcal{R}_{\hat{\theta}}\right]\hspace{-2pt}-\hspace*{-2pt}\left[C(q(\hat{\theta});\hat{\theta})\hspace*{-2pt}-\hspace*{-2pt}C(q(\hat{\theta});\theta)\right]\nonumber\\
	&=\hspace*{-4pt}\int_{\theta}^{\hat{\theta}}{\hspace*{-6pt}C_\theta(q(\breve{\theta});\breve{\theta})d\breve{\theta}}\hspace*{-2pt}-\hspace*{-2pt}\left[C(q(\hat{\theta});\hat{\theta})\hspace*{-2pt}-\hspace*{-2pt}C(q(\hat{\theta});\theta))\right]\hspace{-2pt}\geq\hspace{-2pt} 0, \nonumber
	\end{align}
	where the second equality follows from (\ref{lemma-R-NM-eq1}) and the last inequality is true since $C_\theta(	\hat{q};\theta)$ is decreasing in $\theta$ by Assumption \ref{assump-FSD}, and $q(\theta)$ is decreasing in $\theta$. 
	
	Moreover, we have $\mathcal{R}_\theta=\mathcal{R}_{\overline{\theta}}-\int_{\theta}^{\overline{\theta}}C_\theta(\hspace*{-1pt}q(\hat{\theta});\hspace*{-1pt}\hat{\theta})d\hat{\theta}\hspace{-2pt}\geq\hspace{-2pt}0$. This completes the proof of Claim 1.

	We note that $q(\tau,\omega)$ only depends on $\Theta(\tau,\omega)$. Thus, the sets of constraints given by (\ref{lemma-R-NM-eq1}) and (\ref{rent-IC2}) are equivalent.

	Next, we show that we can replace the set of IR constraints (\ref{IR-R-NM}) by $\mathcal{R}_{\tau_1}=R^P(\tau_{1};q)$. Equation (\ref{lemma-R-NM-eq2}) implies that $\mathcal{R}_{\theta}\geq\mathcal{R}_{\hat{\theta}} $ for $\theta\hspace{-2pt}<\hspace{-2pt} \hat{\theta}$, since $q(\theta)$ is decreasing in $\theta$. That is, $\mathcal{R}_{\Theta(\tau_1,\underline{\omega})} \leq \mathcal{R}_{\hat{\theta}}$ for all $\hat{\theta}$. Thus,  \blue{the set of IR constraints (\ref{IR-R-NM}) is satisfied if and only if} $\mathcal{R}_{\tau_1,\underline{\omega}}\blue{\geq} 0$. Therefore, by (\ref{lemma-R-NM-eq1}), we can write,
	\begin{align*}
	\mathcal{R}_{\tau_1}&\hspace{-2pt}=\hspace{-2pt}\mathbb{E}_\omega\{\mathcal{R}_{\Theta(\tau_1,\omega)}\}\blue{+\mathcal{R}_{\tau_1,\underline{\omega}}}\\&\blue{\hspace{-2pt}\geq\hspace{-2pt}\mathbb{E}_\omega\{\mathcal{R}_{\Theta(\tau_1,\omega)}\}}\\
	&\hspace{-2pt}=\hspace{-2pt}\int_{\underline{\omega}}^{\omega}\hspace{-4pt}\int_{\underline{\omega}}^{\hat{\omega}}\hspace{-4pt} C_\theta(q(\tau_1,\hat{\omega});\Theta(\tau_1,\hat{\omega})\hspace{-1pt})\Theta_\omega(\tau_1,\hat{\omega})d\hat{\omega}dG(\omega)\\		&=\int_{\underline{\omega}}^{\overline{\omega}}\hspace{-4pt}\int_{\hat{\omega}}^{\overline{\omega}} C_\theta(q(\tau_1,\hat{\omega});\Theta(\tau_1,\hat{\omega}))\Theta_\omega(\tau_1,\hat{\omega})dG(\omega)d\hat{\omega}\\
	&=\int_{\underline{\omega}}^{\overline{\omega}}\left[1-G(\hat{\omega})\right] C_\theta(q(\tau_1,\hat{\omega});\Theta(\tau_1,\hat{\omega}))\Theta_\omega(\tau_1,\hat{\omega})d\hat{\omega}\\
	&=\mathcal{R}^P(\tau_1;q).
	\end{align*}

	Finally, we show that $\mathcal{R}_{\tau_i}\hspace{-1pt}-\hspace{-1pt}\mathcal{R}_{\tau_{i-1}}\hspace{-1pt}=\hspace{-1pt}\mathcal{R}^T(\tau_{i-1}\hspace{-1pt},\hspace{-1pt}\tau_{i};q) +\mathcal{R}^W(\tau_{i-1}\hspace{-1pt},\hspace{-1pt}\tau_{i};q)$. Using (\ref{lemma-R-NM-eq1}), we have
	\begin{align*}
	\mathcal{R}_{\tau_{i}}\hspace{-2pt}-\hspace{-2pt}\mathcal{R}_{\tau_{i-1}}\hspace{-2pt}&=\hspace{-2pt}\mathbb{E}_{\hat{\omega}}\{\mathcal{R}_{\tau_{i-1},\hat{\omega}}\}\hspace{-2pt}-\hspace{-2pt}\mathbb{E}_{\hat{\omega}}\{\mathcal{R}_{\tau_{i-1},\hat{\omega}}\}\\&=\hspace{-2pt}\mathbb{E}_{\hat{\omega}}\{\mathcal{R}_{\Theta(\tau_{i},\hat{\omega})}\hspace{-2pt}-\hspace{-2pt}\mathcal{R}_{\Theta(\tau_{i-1},\hat{\omega})}\}\\&=\hspace{-2pt}-\hspace{-4pt}\int\hspace{-5pt}\int_{\Theta(\tau_{i-1},\hat{\omega})}^{\Theta(\tau_{i},\hat{\omega})}\hspace{-4pt}C_\theta(q(\breve{\theta});\breve{\theta})d\breve{\theta}dG(\hat{\omega}).
	\end{align*}
	By the change of variable $\breve{\theta}=\Theta(\tau_{i-1},\hat{\omega})$, we obtain,
	\begin{align*}
	&\int\hspace{-6pt}\int_{\Theta(\tau_{i-1},\omega)}^{\Theta(\tau_{i},\omega)}\hspace{-20pt}C_\theta(q(\breve{\theta});\breve{\theta})d\breve{\theta}dG(\omega)\\
	&=-\int\hspace{-6pt}\int_{\omega}^{\sigma^*(\tau_{i-1};\tau_{i},\omega)}\hspace{-28pt}C_\theta(q(\tau_{i-1},\hat{\omega});\Theta(\tau_{i-1},\hat{\omega}))\Theta_\omega(\tau_{i-1},\hat{\omega})d\hat{\omega}dG(\omega)\\
	&=\int\hspace{-6pt}\int_{\sigma^*(\hat{\tau};\tau,\omega)}^{\omega}\hspace{-15pt}C_\theta(q(\hat{\tau},\hat{\omega});\Theta(\hat{\tau},\hat{\omega}))\Theta_\omega(\hat{\tau},\hat{\omega})d\hat{\omega}dG(\omega)
	\end{align*}
	which is the same as the RHS of (\ref{thm-rent-c-eq1}). Thus, from the proof of Lemma \ref{lemma-S-NM}, we have $\mathcal{R}_{\tau_{i}}-\mathcal{R}_{\tau_{i-1}}=\mathcal{R}^T(\tau_{i-1},\tau_{i};q) +\mathcal{R}^W(\tau_{i-1},\tau_{i};q)$.

\end{proof} 


\vspace{5pt}

\begin{proof}[\textbf{Proof of Lemma \ref{lemma-S-M}}]

	The IC constraint (\ref{rent-S-M-P}) for the \blue{dynamic mechanism with monitoring} is given by,
	\begin{align*}
	\mathbb{E}_\omega\hspace{-1pt}\{\hspace{-1pt}\mathcal{R}_{\tau_i,\omega}\hspace{-1pt}\}\hspace{-2pt}\hspace{-1pt}\geq\hspace{-1pt}\hspace{-1pt} \mathbb{E}_\omega\hspace{-1pt}\{t(\tau_j,\omega)\hspace{-2pt}-\hspace{-2pt}C(\hspace{-1pt}q(\tau_j,\hspace{-1pt}\omega);\hspace{-1pt}\Theta(\tau_i,\hspace{-1pt}\omega))\}\hspace{-5pt}\quad\forall \tau\hspace{-1pt},\hspace{-1pt}\tau_j\hspace{-1pt}.\hspace{-1pt}
	\end{align*}
	We can rewrite the above IC constraint as,
	\begin{align*}
	\mathcal{R}_{\tau_i}\hspace{-1pt}-\hspace{-1pt}\mathcal{R}_{\tau_j}&\geq \mathbb{E}_\omega\{t(\tau_j\hspace{-1pt},\hspace{-1pt}\omega)-C(q(\tau_j\hspace{-1pt},\hspace{-1pt}\omega);\hspace{-1pt}\Theta(\tau_i\hspace{-1pt},\hspace{-1pt}\omega))\}\hspace{-1pt}-\hspace{-1pt}\mathcal{R}_{\tau_j}\\
	&=\hspace{-1pt}\int{\left[C(q(\tau_j\hspace{-1pt},\hspace{-1pt}\omega);\hspace{-1pt}\Theta(\tau_j\hspace{-1pt},\hspace{-1pt}\omega))\hspace{-1pt}-\hspace{-1pt}C(q(\tau_j\hspace{-1pt},\hspace{-1pt}\omega);\hspace{-1pt}\Theta(\tau_i\hspace{-1pt},\hspace{-1pt}\omega)\hspace{-1pt})\hspace{-1pt}\right]dG(\omega)}\\&=\mathcal{R}^T(\tau_j,\tau_i;q)\hspace{-5pt}\quad\forall \tau_i\hspace{-1pt},\hspace{-1pt}\tau_j\hspace{-1pt}.
	\end{align*}
	Therefore, we can replace the set of IC constraints (\ref{rent-S-M-P}) by $\mathcal{R}_{\tau_i}-\mathcal{R}_{\tau_i}\geq R^T(\tau_j,\tau_i;q)$ for all $i,j\in[1,...,M]$. 
	
	Using an argument similar to that of the proof of Lemma \ref{lemma-S-NM}, we can further reduce the set of IC constraints to $\mathcal{R}_{\tau_{i}}\hspace{-2pt}-\hspace{-2pt}\mathcal{R}_{\tau_{j}}\hspace{-2pt}\geq\hspace{-2pt} \mathcal{R}^T\hspace*{-1pt}(\tau_{j},\tau_{i};q)\hspace{-2pt}$ for only $i\hspace*{-2pt}>\hspace*{-2pt}j$, $i,j\hspace*{-2pt}\in\hspace*{-2pt}\{1,..,M\}$.
\end{proof}

\vspace*{90pt} \section*{Appendix C - Closed Form Solutions}
\label{sec:appC}
In this appendix, we provide closed form solutions for the mechanism design problems formulated in Sections \ref{sec-formulation} and \ref{sec-disc}. 

\vspace{10pt}

\subsection{The forward mechanism} 

\vspace{5pt}

\blue{Using} the results of Lemmas \ref{lemma-F}, the optimal allocation $q(\tau)$ is given by the unique solution to the following equation,

\begin{align*}
v(q)=&\mathbb{E}_\omega\{c(q;\Theta(\tau_i,\omega))\}\nonumber\\&+\blue{(\hspace{-1pt}1\hspace{-2pt}-\hspace{-2pt}\alpha\hspace{-1pt})}\hspace{-5pt}\sum_{j=i+1}^i\frac{p_j}{p_i}\Big[\mathbb{E}_\omega\{c(\hspace{-1pt}q;\hspace{-1pt}\Theta(\hspace{-1pt}\tau_{i}\hspace{-1pt},\hspace{-1pt}\omega\hspace{-1pt})\hspace{-2pt})\}\hspace{-2pt}-\hspace{-2pt}\mathbb{E}_\omega\{c(\hspace{-1pt}q;\hspace{-1pt}\Theta(\hspace{-1pt}\tau_{i+1}\hspace{-1pt},\hspace{-1pt}\omega\hspace{-1pt})\hspace{-2pt})\}\Big];
\end{align*}
the optimal payment function $t(\tau_i,\omega)$, $i\hspace{-2pt}\in\hspace{-1pt}\{\hspace{-1pt}1,2,...,M\hspace{-1pt}\}$ is given by,
\begin{align*}
t(\tau_{i})\hspace{-2pt}&=\hspace{-1pt}\mathbb{E}_\omega\{C(q(\tau_{i});\Theta(\tau_{i},\omega))\hspace{-1pt}\}\hspace{-2pt}+\hspace{-2pt}\mathcal{R}_{\tau_{i}}\hspace{-2pt}\\
&=\hspace{-1pt}\mathbb{E}_\omega\{C(q(\tau_i);\Theta(\tau_i,\omega))\hspace{-1pt}\}\hspace{-2pt}\\&\hspace{8pt}+\hspace{-2pt}\sum_{j=1}^{i+1}\hspace{-2pt}\Big[\hspace{-1pt}\mathbb{E}_\omega\hspace{-1pt}\{\hspace{-1pt}C(\hspace{-1pt}q(\hspace{-1pt}\tau_j\hspace{-1pt},\hspace{-1pt}\omega\hspace{-1pt});\hspace{-1pt}\Theta(\hspace{-1pt}\tau_{j}\hspace{-1pt},\hspace{-1pt}\omega\hspace{-1pt})\hspace{-1pt})\hspace{-1pt}\}\hspace{-2pt}-\hspace{-2pt}\mathbb{E}_\omega\hspace{-1pt}\{\hspace{-1pt}C(\hspace{-1pt}q(\hspace{-1pt}\tau_j\hspace{-1pt},\hspace{-1pt}\omega\hspace{-1pt});\hspace{-1pt}\Theta(\hspace{-1pt}\tau_{j+1}\hspace{-1pt},\hspace{-1pt}\omega\hspace{-1pt})\hspace{-1pt})\hspace{-1pt}\}\hspace{-1pt}\Big]\hspace{-1pt}.
\end{align*}

\vspace{5pt}

\blue{\begin{proof}[\textbf{Proof.}]
The proof  follows directly from equations (\ref{optimalforward-rent},\ref{optimalforward-q}) (in the proof of Lemma \ref{lemma-F}), which determine the optimal allocation function $q(\tau)$ and the seller's revenue $\mathcal{R}_\tau$.
\end{proof}}

\vspace{10pt}

\subsection{The \blue{real-time} mechanism} 

\vspace{5pt}

\blue{Using} the result Lemma \ref{lemma-R-NM}, the optimal allocation $q(\tau,\omega)$ is given by the unique solution to the following equation,\footnote{We assume that $c(q;\theta)+\blue{(\hspace{-1pt}1\hspace{-2pt}-\hspace{-2pt}\alpha\hspace{-1pt})}\frac{F(\theta)}{f(\theta)}c_\theta(q;\theta)$ is increasing in $q$ for all $\theta$.}
\begin{align*}
v(q)=c(q;\Theta(\tau,\omega))+\blue{(1-\alpha)}\frac{F(\Theta(\tau,\omega))}{f(\Theta(\tau,\omega))}c_\theta(q;\Theta(\tau,\omega));
\end{align*}
the optimal payment function $t(\tau,\omega)$ is given by,
\begin{align*}
t(\tau,\omega)\hspace{-2pt}&=\hspace{-1pt}C(q(\tau,\omega);\Theta(\tau,\omega))\hspace{-2pt}+\hspace{-2pt}\mathcal{R}_{\tau,\omega}\hspace{-2pt}\\&
=\hspace{-1pt}C(q(\tau,\omega);\Theta(\tau,\omega))\hspace{-2pt}+\hspace{-2pt}\int_{\Theta(\tau,\omega)}^{\overline{\theta}}C_\theta(q(\theta);\theta)d\theta.
\end{align*}

\vspace{5pt}

\begin{proof}[\textbf{Proof.}]
Using (\ref{lemma-R-NM-eq1}) \blue{in the proof of Lemma \ref{lemma-R-NM}}, we can write,
\begin{align*}
\mathcal{R}_\theta=\int_{\theta}^{\overline{\theta}}C_\theta(q(\theta);\theta)d\theta,
\end{align*}
by setting $\mathcal{R}_{\overline{\theta}}=\mathcal{R}_{\tau_1,\underline{\omega}}=0$. Let $F(\theta):=\sum_{i=1}^Mp_i F_i(\theta)$ and $f(\theta):=\sum_{i=1}^Mp_if_i(\theta)$. Then, we can write,
\begin{align*}
\blue{\mathcal{W}}\hspace{-1pt}&=\hspace{-1pt}\blue{\mathcal{S}-(1-\alpha)\mathcal{R}}\\
&=\hspace{-2pt}\int\hspace{-3pt}\left[ \mathcal{V}(q(\theta)\hspace{-1pt})\hspace{-2pt}-\hspace{-2pt}C(q(\theta);\hspace{-1pt}\theta)\right]\hspace{-1pt}dF(\theta)\hspace{-2pt}-\hspace{-2pt}\blue{(\hspace{-1pt}1\hspace{-2pt}-\hspace{-2pt}\alpha\hspace{-1pt})}\hspace{-4pt}\int\hspace{-6pt}\int_{\hat{\theta}}^{\overline{\theta}}\hspace{-5pt}C_\theta(q(\theta);\hspace{-1pt}\theta)d\theta dF(\hat{\theta})\\
&=\hspace{-2pt}\int\hspace{-3pt}\left[ \mathcal{V}(q(\theta)\hspace{-1pt})\hspace{-2pt}-\hspace{-2pt}C(q(\theta);\hspace{-1pt}\theta)\right]\hspace{-2pt}dF(\theta)\hspace{-2pt}-\hspace{-2pt}\blue{(\hspace{-1pt}1\hspace{-2pt}-\hspace{-2pt}\alpha\hspace{-1pt})}\hspace{-4pt}\int\hspace{-6pt}\int^{\theta}_{\underline{\theta}}\hspace{-5pt}C_\theta(q(\theta);\hspace{-1pt}\theta)dF(\hat{\theta})d\theta\\
&=\hspace{-2pt}\int\hspace{-3pt}\left[ \mathcal{V}(q(\theta)\hspace{-2pt}\hspace{-1pt})\hspace{-2pt}-\hspace{-2pt}C(q(\theta);\hspace{-1pt}\theta)\right]\hspace{-1pt}dF(\theta)\hspace{-2pt}-\blue{(\hspace{-1pt}1\hspace{-2pt}-\hspace{-2pt}\alpha\hspace{-1pt})}\hspace{-4pt}\int\hspace{-3pt}C_\theta(q(\theta);\hspace{-1pt}\theta)F(\theta)d\theta\\
&=\hspace{-2pt}\int\hspace{-3pt}\left[ \mathcal{V}(q(\theta))-C(q(\theta);\theta)-\blue{(\hspace{-1pt}1\hspace{-2pt}-\hspace{-2pt}\alpha\hspace{-1pt})}\frac{F(\theta)}{f(\theta)}C_\theta(q(\theta);\theta)\right]dF(\theta).
\end{align*} Hence, using Claim \ref{claim-R-NM} \blue{in the proof of Lemma \ref{lemma-R-NM}}, we can rewrite the optimization problem (\ref{R-NM-max}) as follows,
\begin{align}
&\max_{q(\cdot)}{\int\hspace{-3pt}\left[ \mathcal{V}(q(\theta)\hspace{-1pt})\hspace{-2pt}-\hspace{-2pt}C(q(\theta);\hspace{-1pt}\theta)\hspace{-2pt}-\hspace{-2pt}\blue{(\hspace{-1pt}1\hspace{-2pt}-\hspace{-2pt}\alpha\hspace{-1pt})}\frac{F(\theta)}{f(\theta)}C_\theta(q(\theta);\hspace{-1pt}\theta)\right]\hspace{-2pt}dF(\theta)}\hspace{-2pt}\\
&\hspace{-15pt}\text{subject to}\label{R-NM-max2}\\
&q(\theta) \text{ is decreasing in }\theta.\nonumber
\end{align}   
Consider a relaxed version of the above optimization problem by ignoring the monotonicity constraint on $q(\theta)$. The optimal solution $q(\theta)$ to the relaxed problem is determined by maximizing the integrand  in  (\ref{R-NM-max2}) for each $\theta$, and is given by unique solution to the following equation,
\begin{align}
v(q)=c(q;\theta)+\blue{(\hspace{-1pt}1\hspace{-2pt}-\hspace{-2pt}\alpha\hspace{-1pt})}\frac{F(\theta)}{f(\theta)}c_\theta(q;\theta).
\end{align}
We note that if $c(q;\theta)+\blue{(\hspace{-1pt}1\hspace{-2pt}-\hspace{-2pt}\alpha\hspace{-1pt})}\frac{F(\theta)}{f(\theta)}c_\theta(q;\theta)$ is increasing in $\theta$, then the optimal $q(\theta)$ determined  above is decreasing in $\theta$, and thus, automatically satisfies the ignored monotonicity condition on $q(\theta)$.
\end{proof}

\vspace{10pt}

\subsection{The \blue{dynamic mechanism with monitoring}} 

\vspace{5pt}

\blue{Using} the results of Lemma \ref{lemma-S-M}, the optimal allocation $q(\tau,\omega)$ is given by the unique solution to the following equation,
\begin{align*}
v(q)\hspace{-1pt}=\hspace{-1pt}c(q;\hspace{-1pt}\Theta(\tau_i,\hspace{-1pt}\omega)\hspace{-1pt})\hspace{-2pt}+\hspace{-2pt}\blue{(\hspace{-1pt}1\hspace{-2pt}-\hspace{-2pt}\alpha\hspace{-1pt})}\hspace{-4pt}\sum_{j=i+1}^M\hspace{-2pt}\frac{p_j}{p_i}\Big[c(\hspace{-1pt}q;\hspace{-1pt}\Theta(\hspace{-1pt}\tau_{i}\hspace{-1pt},\hspace{-1pt}\omega\hspace{-1pt})\hspace{-2pt})\hspace{-2pt}-\hspace{-2pt}c(\hspace{-1pt}q;\hspace{-1pt}\Theta(\hspace{-1pt}\tau_{i+1}\hspace{-1pt},\hspace{-1pt}\omega\hspace{-1pt})\hspace{-2pt})\hspace{-1pt}\Big]\hspace{-1pt};
\end{align*}
the optimal payment function $t(\tau_i,\omega)$, $i\hspace{-2pt}\in\hspace{-1pt}\{\hspace{-1pt}1,2,...,M\hspace{-1pt}\}$ is given by,
\begin{align*}
t(\tau_{i},\omega)\hspace{-2pt}&=\hspace{-1pt}C(q(\tau_{i},\omega);\Theta(\tau_{i},\omega))\hspace{-2pt}+\hspace{-2pt}\mathcal{R}_{\tau_{i},\omega}\hspace{-2pt}\\
&=\hspace{-1pt}C(q(\tau_i,\omega);\Theta(\tau_i,\omega))\hspace{-2pt}\\&\hspace{10pt}+\hspace{-2pt}\sum_{j=1}^{i+1}\hspace{-2pt}\Big[\hspace{-1pt}C(\hspace{-1pt}q(\hspace{-1pt}\tau_j\hspace{-1pt},\hspace{-1pt}\omega\hspace{-1pt});\hspace{-1pt}\Theta(\hspace{-1pt}\tau_{j}\hspace{-1pt},\hspace{-1pt}\omega\hspace{-1pt})\hspace{-1pt})\hspace{-2pt}-\hspace{-2pt}C(\hspace{-1pt}q(\hspace{-1pt}\tau_j\hspace{-1pt},\hspace{-1pt}\omega\hspace{-1pt});\hspace{-1pt}\Theta(\hspace{-1pt}\tau_{j+1}\hspace{-1pt},\hspace{-1pt}\omega\hspace{-1pt})\hspace{-1pt})\hspace{-1pt}\Big].
\end{align*}

\vspace{5pt}

\begin{proof}[\textbf{Proof.}]
	\blue{Using the result of Lemma \ref{lemma-S-M}, \blue{the dynamic mechanism with monitoring} is given by the solution to the following optimization problem,
	\begin{align*}
	&\max \;\;\blue{\mathcal{W}}\\
	\text{subject to}&\\
	&\mathcal{R}_{\tau_{i}}-\mathcal{R}_{\tau_{i-1}}= R^T(\tau_{i-1},\tau_{i};q),i\hspace{-2pt}\in\hspace{-2pt}\{2,...,M\},\\
	&\mathcal{R}_{\tau_{1}}\geq 0.
	\end{align*}}
	
	
	\blue{Note that $\mathcal{W}\hspace{-1pt}=\hspace{-1pt}\mathcal{S}\hspace{-1pt}-\hspace{-1pt}(\hspace{-1pt}1\hspace{-1pt}-\hspace{-1pt}\alpha\hspace{-1pt})\mathcal{R}$.} As a result, at the optimal solution of the relaxed problem we have,
	\begin{align*}
	&\mathcal{R}_{\tau_{i}}=\mathcal{R}_{\tau_{i-1}}+R^T(\tau_{i-1},\tau_{i};q),i\hspace{-2pt}\in\hspace{-2pt}\{2,...,M\},\\
	&\mathcal{R}_{\tau_1}=0.
	\end{align*}
	
	Substituting $R_{\tau_i}$, $1\hspace{-2pt}\leq\hspace{-2pt} i\hspace{-2pt}\leq \hspace{-2pt}M$, we obtain,
	\begin{align*}
	\blue{\mathcal{W}}\hspace{-1pt}&\hspace{-1pt}=\hspace{-1pt}\mathcal{S}\hspace{-1pt}-\blue{(\hspace{-1pt}1\hspace{-2pt}-\hspace{-2pt}\alpha\hspace{-1pt})}\sum_{j=1}^np_j\mathcal{R}_{\tau_j}&\\
	&=\hspace{-2pt}\mathcal{S}-\blue{(\hspace{-1pt}1\hspace{-2pt}-\hspace{-2pt}\alpha\hspace{-1pt})}\sum_{j=1}^{M}\left(p_{j}\sum_{i=1}^{j-1}R^T(\tau_{i},\tau_{i+1};q)\right)\\&\hspace{-2pt}=\hspace{-2pt}\mathcal{S}-\blue{(\hspace{-1pt}1\hspace{-2pt}-\hspace{-2pt}\alpha\hspace{-1pt})}\sum_{i=1}^{M-1}\left(\sum_{j=i+1}^{M}p_{j}\right)R^T(\tau_{i},\tau_{i+1};q)\\
	&=\hspace{-2pt}\sum_{i=1}^Mp_i\hspace{-2pt}\int \hspace{-4pt}\left[\mathcal{V}(q(\tau_i,\omega))\hspace{-2pt}-\hspace{-2pt}C(q(\tau_i,\omega);\Theta(\tau_i,\omega))\right]\hspace{-2pt}dG(\omega)-\blue{(\hspace{-1pt}1\hspace{-2pt}-\hspace{-2pt}\alpha\hspace{-1pt})}\Bigg[\\	&\hspace{-2pt}\sum_{i=1}^{M-1}\hspace{-3pt}\left(\hspace{-1pt}\sum_{j=i+1}^{M}\hspace{-1pt}p_{j}\hspace{-3pt}\right)\hspace{-4pt}\int{\hspace{-5pt}\left[C(\hspace{-1pt}q(\hspace{-1pt}\tau_{i}\hspace{-1pt},\hspace{-1pt}\omega\hspace{-1pt});\hspace{-1pt}\Theta(\hspace{-1pt}\tau_{i}\hspace{-1pt},\hspace{-1pt}\omega\hspace{-1pt})\hspace{-2pt})\hspace{-2pt}-\hspace{-2pt}C(\hspace{-1pt}q(\hspace{-1pt}\tau_{i}\hspace{-1pt},\hspace{-1pt}\omega);\hspace{-1pt}\Theta(\hspace{-1pt}\tau_{i+1}\hspace{-1pt},\hspace{-1pt}\omega\hspace{-1pt})\hspace{-2pt})\hspace{-1pt}\right]\hspace{-2pt}dG(\omega)}\hspace{-1pt}\Bigg]\\
	&=\hspace{-2pt}\sum_{i=1}^M p_i\hspace{-2pt}\int \hspace{-4pt}\bigg[\mathcal{V}(q(\tau_i,\omega))\hspace{-2pt}-\hspace{-2pt}C(q(\tau_i,\omega);\Theta(\tau_i,\omega))-\blue{(\hspace{-1pt}1\hspace{-2pt}-\hspace{-2pt}\alpha\hspace{-1pt})}\Bigg[\\
	&\sum_{j=i+1}^M\hspace{-2pt}\frac{p_j}{p_i}\Big[C(\hspace{-1pt}q(\hspace{-1pt}\tau_{i}\hspace{-1pt},\hspace{-1pt}\omega\hspace{-1pt});\hspace{-1pt}\Theta(\hspace{-1pt}\tau_{i}\hspace{-1pt},\hspace{-1pt}\omega\hspace{-1pt})\hspace{-2pt})\hspace{-2pt}-\hspace{-2pt}C(\hspace{-1pt}q(\hspace{-1pt}\tau_{i}\hspace{-1pt},\hspace{-1pt}\omega);\hspace{-1pt}\Theta(\hspace{-1pt}\tau_{i+1}\hspace{-1pt},\hspace{-1pt}\omega\hspace{-1pt})\hspace{-2pt})\Big]\hspace{-1pt}\bigg]\hspace{-2pt}dG(\omega)\Bigg].
	\end{align*}
	By maximizing the integrand point-wise with respect to $q(\tau_i,\omega)$ and using the first-order condition, we find that the optimal value of $q(\tau_i,\omega)$ is determined by the following equation,
	\begin{align*}
	v(q)\hspace{-1pt}=\hspace{-1pt}c(q;\hspace{-1pt}\Theta(\tau_i,\hspace{-1pt}\omega)\hspace{-1pt})\hspace{-1pt}+\hspace{-1pt}\blue{(\hspace{-1pt}1\hspace{-2pt}-\hspace{-2pt}\alpha\hspace{-1pt})}\hspace{-4pt}\sum_{j=i+1}^M\hspace{-2pt}\frac{p_i}{p_j}\Big[c(\hspace{-1pt}q;\hspace{-1pt}\Theta(\hspace{-1pt}\tau_{i}\hspace{-1pt},\hspace{-1pt}\omega\hspace{-1pt})\hspace{-2pt})\hspace{-2pt}-\hspace{-2pt}c(\hspace{-1pt}q;\hspace{-1pt}\Theta(\hspace{-1pt}\tau_{i+1}\hspace{-1pt},\hspace{-1pt}\omega\hspace{-1pt})\hspace{-2pt})\hspace{-1pt}\Big]\hspace{-1pt}.
	\end{align*}
	The above equation has a unique solution since the LHS is decreasing in $q$ by the concavity of $\mathcal{V}(q)$, and the RHS is increasing in $q$ by Assumption \ref{assump-FSD}.

\end{proof}


\vspace{10pt}


\subsection{The \blue{dynamic} mechanism} 

\vspace{5pt}

For the \blue{dynamic} mechanism there exists no closed form solution for \blue{ arbitrary $M$ and parameters of the model}, since the set of binding constraints from the inequality constraints given by (\ref{ICbound-S-NM-P}) cannot be determined a priori, and depends on the allocation function $q(\tau,\omega)$ \blue{(see \cite{battaglini2015optimal} for more discussion)}. However, for the case with two possible technologies, \textit{i.e.} $M=2$, we provide the closed form solutions for the dynamic mechanism \blue {in the following}. 

	If $c(\hat{q};\Theta(\tau_1,\omega))\hspace{-2pt}-\hspace{-2pt}\Theta_\omega\hspace{-1pt}(\hspace{-1pt}\tau_1\hspace{-1pt},\hspace{-1pt}\omega\hspace{-1pt})\frac{p_{\tau_2}}{p_{\tau_1}}\frac{G(\hspace{-1pt}\omega\hspace{-1pt})\hspace{-2pt}-\hspace{-2pt}G(\hspace{-1pt}\sigma^*\hspace{-1pt}(\hspace{-1pt}\tau_2\hspace{-1pt};\hspace{-1pt}{\tau_1}\hspace{-1pt},\hspace{-1pt}\omega\hspace{-1pt})\hspace{-1pt})\hspace{-1pt}}{g(\omega)}c_\theta\hspace{-1pt}(\hspace{-1pt}\hat{q};\hspace{-1pt}\Theta(\hspace{-1pt}\tau_1\hspace{-1pt},\hspace{-1pt}\omega\hspace{-1pt})\hspace{-1pt})$ is increasing in $\hat{q}$, for all $\omega$, the optimal allocation $q(\tau,\omega)$ is given by the unique solution to the following equations,
\begin{align}
	v(\hspace{-1pt}q(\hspace{-1pt}\tau_2\hspace{-1pt},\hspace{-1pt}\omega\hspace{-1pt})\hspace{-1pt})\hspace{-2pt}=& c(q(\tau_2,\omega);\Theta(\tau_2,\omega))\label{S-NM-P-qL}\\
	v(\hspace{-1pt}q(\hspace{-1pt}\tau_1\hspace{-1pt},\hspace{-1pt}\omega\hspace{-1pt})\hspace{-1pt})\hspace{-2pt}=& c(q(\tau_1,\omega);\Theta(\tau_1,\omega))\hspace{-2pt}-\hspace{-2pt}\Big[\blue{(\hspace{-1pt}1\hspace{-2pt}-\hspace{-2pt}\alpha\hspace{-1pt})}\frac{p_{\tau_2}}{p_{\tau_1}}\hspace{-1pt}\Theta_\omega\hspace{-1pt}(\hspace{-1pt}\tau_1\hspace{-1pt},\hspace{-1pt}\omega\hspace{-1pt})\nonumber\\&\hspace{22pt}\frac{G(\hspace{-1pt}\omega\hspace{-1pt})\hspace{-2pt}-\hspace{-2pt}G(\hspace{-1pt}\sigma^*\hspace{-1pt}(\hspace{-1pt}\tau_2\hspace{-1pt};\hspace{-1pt}{\tau_1}\hspace{-1pt},\hspace{-1pt}\omega\hspace{-1pt})\hspace{-1pt})\hspace{-1pt}}{g(\omega)}c_\theta\hspace{-1pt}(\hspace{-1pt}q(\hspace{-1pt}\tau_1\hspace{-1pt},\hspace{-1pt}\omega\hspace{-1pt})\hspace{-1pt};\hspace{-1pt}\Theta(\hspace{-1pt}\tau_1\hspace{-1pt},\hspace{-1pt}\omega\hspace{-1pt})\hspace{-1pt})\Big];\label{S-NM-P-qH}
\end{align}
the optimal payment function $t(\tau_i,\omega)$, is given by,
\begin{align}
t(\tau_2,\omega)\hspace{-2pt}:=&C(q(\tau_2,\omega);\Theta(\tau_2,\omega))\nonumber\\&\hspace{-2pt}-\int{\hspace*{-6pt}\int_{\sigma^*(\tau_2;{\tau_1},\omega)}^{\omega}\hspace{-25pt}C_\theta(q(\tau_1,\omega);\hspace{-2pt}\Theta(\tau_1,\omega))\Theta_\omega(\tau_1,\omega)dG(\hat{\omega}) d\omega }\nonumber\\
&\hspace{-2pt}-\hspace{-2pt}\int_{\underline{\omega}}^{\omega}\hspace{-2pt}C_\theta(q(\tau_2,\hat{\omega});\Theta(\tau_2,\hat{\omega})) \Theta_\omega(\tau_2,\hat{\omega})d\hat{\omega}\nonumber\\
&\hspace{-2pt}+\hspace{-2pt}\int_{\underline{\omega}}^{\overline{\omega}}\hspace{-3pt}[1\hspace{-2pt}-\hspace{-2pt}G(\hat{\omega})]C_\theta(q(\tau_2\hspace{-1pt},\hat{\omega});\hspace{-2pt}\Theta(\tau_2\hspace{-1pt},\hat{\omega})\hspace{-1pt}) \Theta_\omega(\tau_2\hspace{-1pt},\hat{\omega})d\hat{\omega},\label{S-NM-P-tL}\\
t(\tau_1,\omega)\hspace{-2pt}:=&C(q(\tau_1,\omega);\hspace{-2pt}\Theta(\tau_1\hspace{-1pt},\omega))\nonumber\\
&\hspace{-2pt}-\hspace{-2pt}\int_{\underline{\omega}}^{\omega}\hspace{-2pt}C_\theta(q(\tau_1,\hat{\omega});\Theta(\tau_1,\hat{\omega})) \Theta_\omega(\tau_1,\hat{\omega})d\hat{\omega}\nonumber\\
&\hspace{-2pt}+\hspace{-2pt}\int_{\underline{\omega}}^{\overline{\omega}}\hspace{-3pt}[1\hspace{-2pt}-\hspace{-2pt}G(\hat{\omega})]C_\theta(q(\tau_1\hspace{-1pt},\hat{\omega});\hspace{-2pt}\Theta(\tau_1\hspace{-1pt},\hat{\omega})\hspace{-1pt}) \Theta_\omega(\tau_1\hspace{-1pt},\hat{\omega})d\hat{\omega}.\label{S-NM-P-tH}
\end{align}

\vspace{5pt}

\begin{proof}[\textbf{Proof.}] 
	\blue{The designer's objective is $\mathcal{W}=\mathcal{S}-(1-\alpha)\mathcal{R}$. Therefore, }given an allocation function $q(\tau,\omega)$, the \blue{designer} wants to minimize the information rent $\mathcal{R}_\tau$ such that it satisfies the conditions of part (c) of Theorem \ref{thm-rent}. For $M=2$, we have $\mathcal{R}_{\tau_1}\hspace{-2pt}=\hspace{-2pt}0$ and $\mathcal{R}_{\tau_2}\hspace{-2pt}=\hspace{-2pt}\mathcal{R}^T(\tau_1,\tau_2;q)\hspace{-2pt}+\hspace{-2pt}\mathcal{R}^W(\tau_1,\tau_2;q)$. Using the results of Lemma \ref{lemma-IC2} and \ref{lemma-S-NM}, we can rewrite the \blue{dynamic} mechanism design problem as, 
	\begin{align}
	&\max_{q(\cdot,\cdot),t(\cdot,\cdot)} p_{\tau_2}\mathcal{S}_{\tau_2}-\blue{(1-\alpha)}p_{\tau_2}\mathcal{R}_{\tau_2}+p_{\tau_1}\mathcal{S}_{\tau_1}\nonumber\\
	&\hspace{-25pt}\text{subject to}\nonumber\\
	&\mathcal{R}_{\tau_2}\hspace{-2pt}=\hspace{-2pt}\mathcal{R}^T(\tau_1,\tau_2;q)\hspace{-2pt}+\hspace{-2pt}\mathcal{R}^W(\tau_1,\tau_2;q),\nonumber\\
	&\frac{\partial\mathcal{R}_{\tau_2,\omega}}{\partial \omega}=C_\theta(q(\tau_2,\omega);\Theta(\tau_2,\omega))\Theta_\omega(\tau_2,\omega), \label{appC-eq3}\\
	&\frac{\partial\mathcal{R}_{\tau_1,\omega}}{\partial \omega}=C_\theta(q(\tau_1,\omega);\Theta(\tau_1,\omega))\Theta_\omega(\tau_1,\omega),\label{appC-eq4}\\
	&q(\tau_2,\omega) \text{ and } q(\tau_1,\omega) \text{ are increasing in } \omega.\label{appC-eq5}
	\end{align}
	Consider the following relaxation of the above problem, where the last three constraints are omitted.
	\begin{align}
	&\max_{q(\cdot,\cdot),t(\cdot,\cdot)} p_{\tau_2}\mathcal{S}_{\tau_2}-\blue{(1-\alpha)}p_{\tau_2}\mathcal{R}_{\tau_2}+p_{\tau_1}\mathcal{S}_{\tau_1}\label{appC-eq1}\\
	&\hspace{-25pt}\text{subject to}\nonumber\\
	&\mathcal{R}_{\tau_2}\hspace{-2pt}=\hspace{-2pt}\mathcal{R}^T(\tau_1,\tau_2;q)\hspace{-2pt}+\hspace{-2pt}\mathcal{R}^W(\tau_1,\tau_2;q)\label{appC-eq2}.
	\end{align}  
	Below, we determine the solution to the above relaxed problem and show that its optimal solution also solves the original \blue{dynamic} mechanism design problems. 
	Using (\ref{thm-rent-c-eq3}) and (\ref{appC-eq2}), we obtain,
	\begin{align*}
	\mathcal{R}_{\tau_2}\hspace{-3pt}&=\hspace{-2pt}-\int{\hspace*{-6pt}\int_{\hat{\omega}}^{\sigma^*(\tau_1;{\tau_2},\hat{\omega})}\hspace{-25pt}C_\theta(q(\tau_1,\omega);\Theta(\tau_1,\omega))\Theta_\omega(\tau_1,\omega)d\omega dG(\hat{\omega})}\\
	&=\hspace{-2pt}-\int{\hspace*{-6pt}\int_{\sigma^*(\tau_2;{\tau_1},\omega)}^{\omega}\hspace{-25pt}C_\theta(q(\tau_1,\omega);\Theta(\tau_1,\omega))\Theta_\omega(\tau_1,\omega)dG(\hat{\omega}) d\omega }\\
	&=\hspace{-2pt}-\hspace{-4pt}\int{\hspace{-3pt}\left[G(\hspace{-1pt}\omega\hspace{-1pt})\hspace{-2pt}-\hspace{-2pt}G(\hspace{-1pt}\sigma^*\hspace{-1pt}(\hspace{-1pt}\tau_2\hspace{-1pt};\hspace{-1pt}{\tau_1}\hspace{-1pt},\hspace{-1pt}\omega\hspace{-1pt})\hspace{-1pt})\hspace{-1pt}\right]\hspace{-1pt}C_\theta\hspace{-1pt}(\hspace{-1pt}q(\hspace{-1pt}\tau_1\hspace{-1pt},\hspace{-1pt}\omega\hspace{-1pt})\hspace{-1pt};\hspace{-1pt}\Theta(\hspace{-1pt}\tau_1\hspace{-1pt},\hspace{-1pt}\omega\hspace{-1pt})\hspace{-1pt})\Theta_\omega\hspace{-1pt}(\hspace{-1pt}\tau_1\hspace{-1pt},\hspace{-1pt}\omega\hspace{-1pt})d\omega}\\
	&=\hspace{-2pt}-\hspace{-4pt}\int{\hspace{-3pt}\left[G(\hspace{-1pt}\omega\hspace{-1pt})\hspace{-2pt}-\hspace{-2pt}G(\hspace{-1pt}\sigma^*\hspace{-1pt}(\hspace{-1pt}\tau_2\hspace{-1pt};\hspace{-1pt}{\tau_1}\hspace{-1pt},\hspace{-1pt}\omega\hspace{-1pt})\hspace{-1pt})\hspace{-1pt}\right]\hspace{-1pt}C_\theta\hspace{-1pt}(\hspace{-1pt}q(\hspace{-1pt}\tau_1\hspace{-1pt},\hspace{-1pt}\omega\hspace{-1pt})\hspace{-1pt};\hspace{-1pt}\Theta(\hspace{-1pt}\tau_1\hspace{-1pt},\hspace{-1pt}\omega\hspace{-1pt})\hspace{-1pt})\Theta_\omega\hspace{-1pt}(\hspace{-1pt}\tau_1\hspace{-1pt},\hspace{-1pt}\omega\hspace{-1pt})\frac{dG(\omega)}{g(\omega)}}\hspace{-1pt},
	\end{align*}
	where the second equality results from changing the order of integration and $\sigma^*\hspace{-1pt}(\hspace{-1pt}\tau_1\hspace{-1pt};\hspace{-1pt}\tau_2\hspace{-1pt},\hspace{-1pt}\sigma^*\hspace{-1pt}(\hspace{-1pt}\tau_2\hspace{-1pt};\hspace{-1pt}\tau_1\hspace{-1pt},\hspace{-1pt}\omega\hspace{-1pt})\hspace{-1pt})\hspace{-2pt}=\hspace{-2pt}\omega$. Therefore, we can write the objective function (\ref{appC-eq1}) as,
	\begin{align*}
	&p_{\tau_2}\mathcal{S}_{\tau_2}\hspace{-2pt}-\hspace{-2pt}\blue{(\hspace{-1pt}1\hspace{-2pt}-\hspace{-2pt}\alpha\hspace{-1pt})}p_{\tau_2}\mathcal{R}_{\tau_2}\hspace{-2pt}+\hspace{-2pt}p_{\tau_1}\mathcal{S}_{\tau_1}\hspace{-2pt}=\\
	&p_{\tau_2}\hspace{-1pt}\hspace{-2pt}\int\hspace{-4pt} \left[\mathcal{V}(\hspace{-1pt}q(\hspace{-1pt}\tau_2\hspace{-1pt},\hspace{-1pt}\omega\hspace{-1pt})\hspace{-1pt})\hspace{-2pt}-\hspace{-2pt}C(\hspace{-1pt}q(\hspace{-1pt}\tau_2\hspace{-1pt},\hspace{-1pt}\omega\hspace{-1pt})\hspace{-1pt};\hspace{-1pt}\Theta(\hspace{-1pt}\tau_2\hspace{-1pt},\hspace{-1pt}\omega\hspace{-1pt})\hspace{-1pt}\right]\hspace{-1pt}dG(\hspace{-1pt}\omega\hspace{-1pt})+\\
	&\blue{(\hspace{-1pt}1\hspace{-2pt}-\hspace{-2pt}\alpha\hspace{-1pt})}\hspace{-1pt}p_{\tau_2}\hspace{-4pt}\int{\hspace{-4pt}\left[\hspace{-1pt}G(\hspace{-1pt}\omega\hspace{-1pt})\hspace{-2pt}-\hspace{-2pt}G(\hspace{-1pt}\sigma^*\hspace{-2pt}(\hspace{-1pt}\tau_2\hspace{-1pt};\hspace{-1pt}{\tau_1}\hspace{-1pt},\hspace{-1pt}\omega\hspace{-1pt})\hspace{-1pt})\hspace{-1pt}\right]\hspace{-1pt}C_\theta\hspace{-1pt}(\hspace{-1pt}q(\hspace{-1pt}\tau_1\hspace{-1pt},\hspace{-1pt}\omega\hspace{-1pt})\hspace{-1pt};\hspace{-1pt}\Theta(\hspace{-1pt}\tau_1\hspace{-1pt},\hspace{-1pt}\omega\hspace{-1pt})\hspace{-1pt})\Theta_\omega\hspace{-1pt}(\hspace{-1pt}\tau_1\hspace{-1pt},\hspace{-1pt}\omega\hspace{-1pt})\frac{dG(\omega)}{g(\omega)}}\hspace{-1pt}\\
	&+\hspace{-1pt}p_{\tau_1}\hspace{-4pt}\int\hspace{-4pt} \left[\mathcal{V}(\hspace{-1pt}q(\hspace{-1pt}\tau_1\hspace{-1pt},\hspace{-1pt}\omega\hspace{-1pt})\hspace{-1pt})\hspace{-2pt}-\hspace{-2pt}C(\hspace{-1pt}q(\hspace{-1pt}\tau_1\hspace{-1pt},\hspace{-1pt}\omega\hspace{-1pt})\hspace{-1pt};\hspace{-1pt}\Theta(\hspace{-1pt}\tau_1\hspace{-1pt},\hspace{-1pt}\omega\hspace{-1pt})\hspace{-1pt}\right]\hspace{-1pt}dG(\hspace{-1pt}\omega\hspace{-1pt})\\
	&\hspace{1pt}=p_{\tau_2}\hspace{-1pt}\hspace{-2pt}\int\hspace{-4pt} \left[\mathcal{V}(\hspace{-1pt}q(\hspace{-1pt}\tau_2\hspace{-1pt},\hspace{-1pt}\omega\hspace{-1pt})\hspace{-1pt})\hspace{-2pt}-\hspace{-2pt}C(\hspace{-1pt}q(\hspace{-1pt}\tau_2\hspace{-1pt},\hspace{-1pt}\omega\hspace{-1pt})\hspace{-1pt};\hspace{-1pt}\Theta(\hspace{-1pt}\tau_2\hspace{-1pt},\hspace{-1pt}\omega\hspace{-1pt})\hspace{-1pt}\right]\hspace{-1pt}dG(\hspace{-1pt}\omega\hspace{-1pt})\\
	&\hspace{3pt}+\hspace{-1pt}p_{\tau_1}\hspace{-4pt}\int\hspace{-3pt} \bigg[\hspace{-1pt}\mathcal{V}(\hspace{-1pt}q(\hspace{-1pt}\tau_1\hspace{-1pt},\hspace{-1pt}\omega\hspace{-1pt})\hspace{-1pt})\hspace{-2pt}-\hspace{-2pt}C(\hspace{-1pt}q(\hspace{-1pt}\tau_1\hspace{-1pt},\hspace{-1pt}\omega\hspace{-1pt})\hspace{-1pt};\hspace{-1pt}\Theta(\hspace{-1pt}\tau_1\hspace{-1pt},\hspace{-1pt}\omega\hspace{-1pt})\hspace{-1pt})\hspace{-1pt}\\
	&\hspace{17pt}+\hspace{-2pt}\blue{(\hspace{-1pt}1\hspace{-2pt}-\hspace{-2pt}\alpha\hspace{-1pt})}\frac{p_{\tau_2}}{p_{\tau_1}}\frac{G(\hspace{-1pt}\omega\hspace{-1pt})\hspace{-2pt}-\hspace{-2pt}G(\hspace{-1pt}\sigma^*\hspace{-1pt}(\hspace{-1pt}\tau_2\hspace{-1pt};\hspace{-1pt}{\tau_1}\hspace{-1pt},\hspace{-1pt}\omega\hspace{-1pt})\hspace{-1pt})\hspace{-1pt}}{g(\omega)}C_\theta\hspace{-1pt}(\hspace{-1pt}q(\hspace{-1pt}\tau_1\hspace{-1pt},\hspace{-1pt}\omega\hspace{-1pt})\hspace{-1pt};\hspace{-1pt}\Theta(\hspace{-1pt}\tau_1\hspace{-1pt},\hspace{-1pt}\omega\hspace{-1pt})\hspace{-1pt})\Theta_\omega\hspace{-1pt}(\hspace{-1pt}\tau_1\hspace{-1pt},\hspace{-1pt}\omega\hspace{-1pt})\hspace{-2pt}\bigg]\hspace{-1pt}\\
	&\hspace{228pt}dG(\hspace{-1pt}\omega\hspace{-1pt})\hspace{-1pt}.
	\end{align*}
	By maximizing the integrands point-wise with respect to $q(\tau_i,\omega)$ and using the first order condition, $i=1,2$, we obtain equations (\ref{S-NM-P-qL}) and (\ref{S-NM-P-qH}). We note that the allocation function $q(\tau_1,\omega)$, given by  (\ref{S-NM-P-qH}), is increasing  in $\omega$ since by assumption $c(q;\Theta(\tau_1,\omega))\hspace{-1pt}-\hspace{-1pt}\Theta_\omega\hspace{-1pt}(\hspace{-1pt}\tau_1\hspace{-1pt},\hspace{-1pt}\omega\hspace{-1pt})\frac{p_{\tau_2}}{p_{\tau_1}}\frac{G(\hspace{-1pt}\omega\hspace{-1pt})\hspace{-1pt}-\hspace{-1pt}G(\hspace{-1pt}\sigma^*\hspace{-1pt}(\hspace{-1pt}\tau_2\hspace{-1pt};\hspace{-1pt}{\tau_1}\hspace{-1pt},\hspace{-1pt}\omega\hspace{-1pt})\hspace{-1pt})\hspace{-1pt}}{g(\omega)}C_\theta\hspace{-1pt}(\hspace{-1pt}q;\hspace{-1pt}\Theta(\hspace{-1pt}\tau_1\hspace{-1pt},\hspace{-1pt}\omega\hspace{-1pt})\hspace{-1pt})$ is increasing in $q$. Moreover, the allocation function $q(\tau_2,\omega)$, given by  (\ref{S-NM-P-qL}), is increasing  in $\omega$ by Assumption \ref{assump-FSD}. Therefore, the omitted constraint (\ref{appC-eq5}) in the original optimization problem is satisfied automatically.
	
	Below, we construct payment function $t(\tau_i,\omega)$, $i=1,2,$ such that the omitted constraints (\ref{appC-eq3}) and (\ref{appC-eq4}) are satisfied, $\hspace{-2pt}\mathbb{E}_\omega\hspace{-1pt}\{t(\hspace{-1pt}\tau_2\hspace{-1pt},\hspace{-1pt}\omega\hspace{-1pt})\hspace{-2pt}-\hspace{-2pt}C(\hspace{-1pt}q(\hspace{-1pt}\tau_2\hspace{-1pt},\hspace{-1pt}\omega\hspace{-1pt})\hspace{-1pt};\hspace{-2pt}\Theta(\hspace{-1pt}\tau_2\hspace{-1pt},\hspace{-1pt}\omega\hspace{-1pt})\hspace{-1pt})\hspace{-1pt}\}=\mathcal{R}^T\hspace{-1pt}(\hspace{-1pt}\tau_2\hspace{-1pt},\hspace{-1pt}\tau_1\hspace{-1pt};\hspace{-1pt}q\hspace{-1pt})\hspace{-2pt}+\hspace{-2pt}\mathcal{R}^W\hspace{-1pt}(\hspace{-1pt}\tau_2\hspace{-1pt},\hspace{-1pt}\tau_1\hspace{-1pt};\hspace{-1pt}q\hspace{-1pt})$, and $\mathbb{E}_\omega\hspace{-1pt}\{\hspace{-1pt}t(\hspace{-1pt}\tau_1\hspace{-1pt},\hspace{-1pt}\omega\hspace{-1pt})\hspace{-2pt}-\hspace{-2pt}C(\hspace{-1pt}q\hspace{-1pt}(\hspace{-1pt}\tau_1\hspace{-1pt},\hspace{-1pt}\omega\hspace{-1pt})\hspace{-1pt};\hspace{-2pt}\Theta(\hspace{-1pt}\tau_1\hspace{-1pt},\hspace{-1pt}\omega\hspace{-1pt})\hspace{-1pt})\hspace{-1pt}\}\hspace{-2pt}=\hspace{-2pt}0$. Define,
	\begin{align}
	t(\tau_2,\omega)\hspace{-2pt}:=&C(q(\tau_2,\omega);\Theta(\tau_2,\omega))+\mathcal{R}^T\hspace{-1pt}(\hspace{-1pt}\tau_2\hspace{-1pt},\hspace{-1pt}\tau_1\hspace{-1pt};\hspace{-1pt}q\hspace{-1pt})\hspace{-2pt}+\hspace{-2pt}\mathcal{R}^W\hspace{-1pt}(\hspace{-1pt}\tau_2\hspace{-1pt},\hspace{-1pt}\tau_1\hspace{-1pt};\hspace{-1pt}q\hspace{-1pt})\nonumber\\
	&\hspace{-2pt}-\hspace{-2pt}\int_{\underline{\omega}}^{\omega}\hspace{-2pt}C_\theta(q(\tau_2,\hat{\omega});\Theta(\tau_2,\hat{\omega})) \Theta_\omega(\tau_2,\hat{\omega})d\hat{\omega}\nonumber\\
	&\hspace{-2pt}+\hspace{-2pt}\int_{\underline{\omega}}^{\overline{\omega}}\hspace{-3pt}[1\hspace{-2pt}-\hspace{-2pt}G(\hat{\omega})]C_\theta(q(\tau_2\hspace{-1pt},\hat{\omega});\hspace{-1pt}\Theta(\tau_2\hspace{-1pt},\hat{\omega})\hspace{-1pt}) \Theta_\omega(\tau_2\hspace{-1pt},\hat{\omega})d\hat{\omega},\label{lemmaC-eq1}\\
	t(\tau_1,\omega)\hspace{-2pt}:=&C(q(\tau_1,\omega);\Theta(\tau_1,\omega))\nonumber\\
	&\hspace{-2pt}-\hspace{-2pt}\int_{\underline{\omega}}^{\omega}\hspace{-2pt}C_\theta(q(\tau_1,\hat{\omega});\Theta(\tau_1,\hat{\omega})) \Theta_\omega(\tau_1,\hat{\omega})d\hat{\omega}\nonumber\\
	&\hspace{-2pt}+\hspace{-2pt}\int_{\underline{\omega}}^{\overline{\omega}}\hspace{-3pt}[1\hspace{-2pt}-\hspace{-2pt}G(\hat{\omega})]C_\theta(q(\tau_1\hspace{-1pt},\hat{\omega});\hspace{-1pt}\Theta(\tau_1\hspace{-1pt},\hat{\omega})\hspace{-1pt}) \Theta_\omega(\tau_1\hspace{-1pt},\hat{\omega})d\hat{\omega}.
	\end{align}
	By the above definition, we have $\frac{\partial \mathcal{R}_{\tau_i,\omega}}{\partial \omega}=C_\theta(q(\tau_i,\omega);\Theta(\tau_i,\omega))\Theta_\omega(\tau_i,\omega)$ for $i=1,2$. Thus, the omitted constraints (\ref{appC-eq3}) and (\ref{appC-eq4}) are satisfied. Moreover,
	\begin{align*}
	\mathbb{E}_\omega\hspace{-1pt}\{\hspace{-1pt}t(\hspace{-1pt}\tau_1\hspace{-1pt},\hspace{-1pt}\omega\hspace{-1pt})\hspace{-2pt}-\hspace{-2pt}C(\hspace{-1pt}q(\hspace{-1pt}\tau_1\hspace{-1pt},\hspace{-1pt}\omega\hspace{-1pt})\hspace{-1pt};\hspace{-2pt}\Theta(\hspace{-1pt}\tau_1\hspace{-1pt},\hspace{-1pt}\omega\hspace{-1pt})\hspace{-1pt})\hspace{-1pt}\}\hspace{-2pt}=&\\
	&\hspace{-45pt}
	-\hspace{-3pt}\int\hspace{-7pt}\int_{\underline{\omega}}^{\omega}\hspace{-8pt}C_\theta\hspace{-1pt}(\hspace{-1pt}q(\hspace{-1pt}\tau_2\hspace{-1pt},\hspace{-1pt}\hat{\omega}\hspace{-1pt})\hspace{-1pt};\hspace{-2pt}\Theta(\hspace{-1pt}\tau_2\hspace{-1pt},\hspace{-1pt}\hat{\omega}\hspace{-1pt})\hspace{-1pt}) \hspace{-1pt}\Theta_\omega(\hspace{-1pt}\tau_2\hspace{-1pt},\hspace{-1pt}\hat{\omega}\hspace{-1pt})\hspace{-1pt}d\hat{\omega}dG(\hspace{-1pt}\omega\hspace{-1pt})\\
	&\hspace{-45pt}
	-\hspace{-3pt}\int_{\underline{\omega}}^{\overline{\omega}}\hspace{-8pt}[\hspace{-1pt}1\hspace{-2pt}-\hspace{-2pt}G(\hspace{-1pt}\omega\hspace{-1pt})\hspace{-1pt}]C_\theta\hspace{-1pt}(\hspace{-1pt}q(\hspace{-1pt}\tau_2\hspace{-1pt},\hspace{-1pt}\hat{\omega}\hspace{-1pt})\hspace{-1pt};\hspace{-2pt}\Theta(\hspace{-1pt}\tau_2\hspace{-1pt},\hspace{-1pt}\hat{\omega}\hspace{-1pt})\hspace{-1pt}) \hspace{-1pt}\Theta_\omega(\hspace{-1pt}\tau_2\hspace{-1pt},\hspace{-1pt}\hat{\omega}\hspace{-1pt})\hspace{-1pt}d\hat{\omega}\\
	=&\\
	&\hspace{-45pt}
	-\hspace{-3pt}\int\hspace{-7pt}\int_{\hat{\omega}}^{\overline{\omega}}\hspace{-8pt}C_\theta\hspace{-1pt}(\hspace{-1pt}q(\hspace{-1pt}\tau_2\hspace{-1pt},\hspace{-1pt}\hat{\omega}\hspace{-1pt})\hspace{-1pt};\hspace{-2pt}\Theta(\hspace{-1pt}\tau_2\hspace{-1pt},\hspace{-1pt}\hat{\omega}\hspace{-1pt})\hspace{-1pt}) \hspace{-1pt}\Theta_\omega(\hspace{-1pt}\tau_2\hspace{-1pt},\hspace{-1pt}\hat{\omega}\hspace{-1pt})\hspace{-1pt}dG(\hspace{-1pt}\omega\hspace{-1pt})d\hat{\omega}\\
	&\hspace{-45pt}
	-\hspace{-3pt}\int_{\underline{\omega}}^{\overline{\omega}}\hspace{-8pt}[\hspace{-1pt}1\hspace{-2pt}-\hspace{-2pt}G(\hspace{-1pt}\omega\hspace{-1pt})\hspace{-1pt}]C_\theta\hspace{-1pt}(\hspace{-1pt}q(\hspace{-1pt}\tau_2\hspace{-1pt},\hspace{-1pt}\hat{\omega}\hspace{-1pt})\hspace{-1pt};\hspace{-2pt}\Theta(\hspace{-1pt}\tau_2\hspace{-1pt},\hspace{-1pt}\hat{\omega}\hspace{-1pt})\hspace{-1pt}) \hspace{-1pt}\Theta_\omega(\hspace{-1pt}\tau_2\hspace{-1pt},\hspace{-1pt}\hat{\omega}\hspace{-1pt})\hspace{-1pt}d\hat{\omega}\\
	=&0,
	\end{align*}
	 where the second equality results from changing the order of integration. Similarly, 
	 \begin{align*}
	 \mathbb{E}_\omega\hspace{-1pt}\{t(\hspace{-1pt}\tau_2\hspace{-1pt},\hspace{-1pt}\omega\hspace{-1pt})\hspace{-2pt}-\hspace{-2pt}C(\hspace{-1pt}q(\hspace{-1pt}\tau_2\hspace{-1pt},\hspace{-1pt}\omega\hspace{-1pt})\hspace{-1pt};\hspace{-2pt}\Theta(\hspace{-1pt}\tau_2\hspace{-1pt},\hspace{-1pt}\omega\hspace{-1pt})\hspace{-1pt})\hspace{-1pt}\}=\mathcal{R}^T\hspace{-1pt}(\hspace{-1pt}\tau_2\hspace{-1pt},\hspace{-1pt}\tau_1\hspace{-1pt};\hspace{-1pt}q\hspace{-1pt})\hspace{-2pt}+\hspace{-2pt}\mathcal{R}^W\hspace{-1pt}(\hspace{-1pt}\tau_2\hspace{-1pt},\hspace{-1pt}\tau_1\hspace{-1pt};\hspace{-1pt}q\hspace{-1pt}).
	 \end{align*}
	 Substituting $\mathcal{R}^T\hspace{-1pt}(\hspace{-1pt}\tau_2\hspace{-1pt},\hspace{-1pt}\tau_1\hspace{-1pt};\hspace{-1pt}q\hspace{-1pt})\hspace{-2pt}+\hspace{-2pt}\mathcal{R}^W\hspace{-1pt}(\hspace{-1pt}\tau_2\hspace{-1pt},\hspace{-1pt}\tau_1\hspace{-1pt};\hspace{-1pt}q\hspace{-1pt})$ in (\ref{lemmaC-eq1}) using (\ref{thm-rent-c-eq3}), we obtain (\ref{S-NM-P-tL}) and (\ref{S-NM-P-tH}).
\end{proof}

\vspace{10pt}

\subsection{The \blue{dynamic} mechanism \blue{with no penalty}} 

\vspace{5pt}

Similar to the \blue{dynamic} mechanism, for the \blue{dynamic} mechanisms \blue{with no penalty} there exists no closed form solution for \blue{ arbitrary $M$ and parameters of the model}, since the set of binding constraints from the inequality constraints, given by (\ref{ICbound-S-NM-P-np}), cannot be determined a priori, and it depends on allocation function $q(\tau,\omega)$. Moreover, for the \blue{dynamic} mechanisms \blue{with no penalty}, we face additional difficulties by imposing ex-post individual rationality, which results in the additional set of constraints (\ref{IRbound-S-NM-P-np}) on the information rent. As a result, unlike the \blue{dynamic} mechanism, we cannot determine a priori the set of binding constraints from the ones given by (\ref{ICbound-S-NM-P-np}) and (\ref{IRbound-S-NM-P-np}) even for $M=2$. 

\vspace*{5pt}

For $M=2$, the \blue{dynamic} mechanisms \blue{with no penalty is given by the solution to the following optimization problem}.
\begin{align}
&\max_{q(\cdot,\cdot),t(\cdot,\cdot)} p_{\tau_2}\hspace{-4pt}\int\hspace{-4pt} \left[\hspace{-1pt}\mathcal{V}(\hspace{-1pt}q(\hspace{-1pt}\tau_2\hspace{-1pt},\hspace{-1pt}\omega\hspace{-1pt})\hspace{-1pt})\hspace{-2pt}-\hspace{-2pt}C(\hspace{-1pt}q(\hspace{-1pt}\tau_2\hspace{-1pt},\hspace{-1pt}\omega\hspace{-1pt})\hspace{-1pt};\hspace{-2pt}\Theta(\hspace{-1pt}\tau_2\hspace{-1pt},\hspace{-1pt}\omega\hspace{-1pt})\hspace{-1pt})\hspace{-1pt}\right]\hspace{-2pt}dG(\omega)\hspace{-2pt}-\hspace{-2pt}\blue{(\hspace{-1pt}1\hspace{-2pt}-\hspace{-2pt}\alpha\hspace{-1pt})}p_{\tau_2}\mathcal{R}_{\tau_2}\nonumber\\
&\hspace{25pt}+p_{\tau_1}\hspace{-4pt}\int\hspace{-4pt} \left[\hspace{-1pt}\mathcal{V}(\hspace{-1pt}q(\hspace{-1pt}\tau_1\hspace{-1pt},\hspace{-1pt}\omega\hspace{-1pt})\hspace{-1pt})\hspace{-2pt}-\hspace{-2pt}C(\hspace{-1pt}q(\hspace{-1pt}\tau_1\hspace{-1pt},\hspace{-1pt}\omega\hspace{-1pt})\hspace{-1pt};\hspace{-2pt}\Theta(\hspace{-1pt}\tau_1\hspace{-1pt},\hspace{-1pt}\omega\hspace{-1pt})\hspace{-1pt})\hspace{-1pt}\right]\hspace{-2pt}dG(\omega)\hspace{-2pt}-\hspace{-2pt}\blue{(\hspace{-1pt}1\hspace{-2pt}-\hspace{-2pt}\alpha\hspace{-1pt})}p_{\tau_1}\mathcal{R}_{\tau_1}\hspace{-5pt}\label{appC-eq6}\\
&\hspace{-5pt}\text{subject to}\nonumber\\
&\mathcal{R}_{\tau_2}-\mathcal{R}_{\tau_1}\geq \mathcal{R}^T(\tau_2,\tau_1;q)+\mathcal{R}^W(\tau_2,\tau_1;q)\geq0,\label{appC-eq7}\\
&\mathcal{R}_{\tau_1}= -\hspace{-3pt}\int_{\underline{\omega}}^{\overline{\omega}}\hspace{-8pt}[\hspace{-1pt}1\hspace{-2pt}-\hspace{-2pt}G(\hspace{-1pt}\omega\hspace{-1pt})\hspace{-1pt}]C_\theta\hspace{-1pt}(\hspace{-1pt}q(\hspace{-1pt}\tau_1\hspace{-1pt},\hspace{-1pt}\hat{\omega}\hspace{-1pt})\hspace{-1pt};\hspace{-2pt}\Theta(\hspace{-1pt}\tau_1\hspace{-1pt},\hspace{-1pt}\hat{\omega}\hspace{-1pt})\hspace{-1pt}) \hspace{-1pt}\Theta_\omega(\hspace{-1pt}\tau_1\hspace{-1pt},\hspace{-1pt}\hat{\omega}\hspace{-1pt})\hspace{-1pt}d\hat{\omega}\geq 0,\label{appC-eq8}\\
&\mathcal{R}_{\tau_2}\geq -\hspace{-3pt}\int_{\underline{\omega}}^{\overline{\omega}}\hspace{-8pt}[\hspace{-1pt}1\hspace{-2pt}-\hspace{-2pt}G(\hspace{-1pt}\omega\hspace{-1pt})\hspace{-1pt}]C_\theta\hspace{-1pt}(\hspace{-1pt}q(\hspace{-1pt}\tau_2\hspace{-1pt},\hspace{-1pt}\hat{\omega}\hspace{-1pt})\hspace{-1pt};\hspace{-2pt}\Theta(\hspace{-1pt}\tau_2\hspace{-1pt},\hspace{-1pt}\hat{\omega}\hspace{-1pt})\hspace{-1pt}) \hspace{-1pt}\Theta_\omega(\hspace{-1pt}\tau_2\hspace{-1pt},\hspace{-1pt}\hat{\omega}\hspace{-1pt})\hspace{-1pt}d\hat{\omega}\geq 0\label{appC-eq9}.
\end{align}
The last two constraints result from  using the results of part (d) of Theorem \ref{thm-rent} (equation (\ref{IRbound-S-NM-P-np})) along with the result of Lemma \ref{lemma-IC2}, and setting $\mathcal{R}_{\tau_i,\underline{\omega}}\geq 0$, for $i=1,2$. We note that constraint (\ref{appC-eq8}) is written as an equality (binding) constraint (unlike (\ref{appC-eq9})) since for any value of $\mathcal{R}_{\tau_1}$ that does not bind (\ref{appC-eq8}), we can reduce the value of $\mathcal{R}_{\tau_1}$ by $\epsilon>0$ without violating any other constraints and improve the objective function  (\ref{appC-eq6}).

In the following, we determine the solution to the above optimization problem for the specific example considered in Section \ref{sec-example}. We have, $C(q;\theta)=\theta q$, 
$\Theta(\tau_1,\omega)=(1-\omega)$, $\Theta(\tau_1,\omega)=(1-\omega)^2$, $\mathcal{V}(q)=q-\frac{1}{2}q^2$, and $G(\omega)=\omega$ for $\omega\in[0,1]$.

From (\ref{appC-eq8}), we obtain,
 \begin{align}
 \mathcal{R}_{\tau_1}&=\int_0^1q(\tau_1,\omega)(1-\omega)d\omega.\label{appC-eq11}
 \end{align}
 
 It can be shown that constraints (\ref{appC-eq9}) is binding at the optimal solution.\footnote{To show this, one can solve the optimization problem (\ref{appC-eq6}) relaxing the constraint (\ref{appC-eq9}), and show that the optimal solution violates the constraint (\ref{appC-eq9}).} When (\ref{appC-eq9}) is binding, we obtain,
\begin{align}
\mathcal{R}_{\tau_2}&=\int_0^1q(\tau_2,\omega)2(1-\omega)^2d\omega.\label{appC-eq12}
\end{align}

Moreover, from (\ref{appC-eq7}),  we obtain
\begin{align}
\mathcal{R}_{\tau_2}-\mathcal{R}_{\tau_1}&\geq\int_0^1\hspace{-5pt}\int_{\omega}^{1-(1-\omega)^2}q(\tau_1,\hat{\omega})d\hat{\omega}d\omega\nonumber\\
&=\int_0^1\hspace{-5pt}\int^{\hat{\omega}}_{1-\sqrt{1-\hat{\omega}}}q(\tau_1,\hat{\omega})d\omega d\hat{\omega}\nonumber\\
&=\int_0^1\left[\sqrt{1-\hat{\omega}}-(1-\hat{\omega})\right]q(\tau_1,\hat{\omega})d\hat{\omega},
\label{appC-eq13}
\end{align}
where  the first equality results from (\ref{thm-rent-c-eq3}), and the second equality results from changing  the order of integration. 

Using (\ref{appC-eq11}) and (\ref{appC-eq12}), we can rewrite (\ref{appC-eq13}) as,
\begin{align}
\int_0^1\left[q(\tau_2,\omega)2(1-\omega)^2-q(\tau_1,\omega)\sqrt{1-\omega}\hspace{1pt}\right]d\omega\geq 0. \label{appC-eq13b}
\end{align} 

Substituting (\ref{appC-eq11}) and (\ref{appC-eq12}) in (\ref{appC-eq6}), we can rewrite the optimization problem as,
\begin{align}
&\hspace{-10pt}\max_{q(\cdot,\cdot),t(\cdot,\cdot)} p_{\tau_2}\hspace{-4pt}\int_0^1\hspace{-4pt}\Bigg[\mathcal{V}(\hspace{-1pt}q(\hspace{-1pt}\tau_2\hspace{-1pt},\hspace{-1pt}\omega\hspace{-1pt})\hspace{-1pt})\hspace{-2pt}-\hspace{-2pt}(\hspace{-1pt}1\hspace{-2pt}-\hspace{-2pt}\omega\hspace{-1pt})^2q(\hspace{-1pt}\tau_2\hspace{-1pt},\hspace{-1pt}\omega\hspace{-1pt})\hspace{-2pt}\nonumber\\
&\hspace{55pt}-\hspace{-2pt}\blue{(\hspace{-1pt}1\hspace{-2pt}-\hspace{-2pt}\alpha\hspace{-1pt})}2(\hspace{-1pt}1\hspace{-2pt}-\hspace{-2pt}\omega\hspace{-1pt})^2q(\hspace{-1pt}\tau_2\hspace{-1pt},\hspace{-1pt}\omega\hspace{-1pt})\hspace{-1pt}\Bigg]\hspace{-1pt}dG(\omega)\nonumber\\
&\hspace{-10pt}\hspace{35pt}+p_{\tau_1}\hspace{-4pt}\int_0^1\hspace{-4pt}\Bigg[\hspace{-1pt}\mathcal{V}(\hspace{-1pt}q(\hspace{-1pt}\tau_1\hspace{-1pt},\hspace{-1pt}\omega\hspace{-1pt})\hspace{-1pt})\hspace{-2pt}-\hspace{-2pt}(\hspace{-1pt}1\hspace{-2pt}-\hspace{-2pt}\omega\hspace{-1pt})q(\hspace{-1pt}\tau_1\hspace{-1pt},\hspace{-1pt}\omega\hspace{-1pt})\hspace{-2pt}\nonumber\\
&\hspace{60pt}-\hspace{-2pt}\blue{(\hspace{-1pt}1\hspace{-2pt}-\hspace{-2pt}\alpha\hspace{-1pt})}(\hspace{-1pt}1\hspace{-2pt}-\hspace{-2pt}\omega\hspace{-1pt})q(\hspace{-1pt}\tau_1\hspace{-1pt},\hspace{-1pt}\omega\hspace{-1pt})\hspace{-1pt}\Bigg]\hspace{-2pt}dG(\omega)\hspace{-10pt}\\
&\hspace{-5pt}\text{subject to}\nonumber\\
&\int_0^1\left[q(\tau_2,\omega)2(1-\omega)^2-q(\tau_1,\omega)\sqrt{1-\omega}\hspace{1pt}\right]d\omega\geq 0,\nonumber
\end{align} 
where we replaced constraint (\ref{appC-eq7}) by (\ref{appC-eq13b}).

The Lagrangian for the above optimization problem is given by,
\begin{align}
&p_{\tau_2}\hspace{-4pt}\int_0^1\hspace{-4pt}\left[\mathcal{V}(\hspace{-1pt}q(\hspace{-1pt}\tau_2\hspace{-1pt},\hspace{-1pt}\omega\hspace{-1pt})\hspace{-1pt})\hspace{-2pt}-\hspace{-2pt}\left[\blue{(3\hspace{-2pt}-\hspace{-2pt}2\alpha)}(\hspace{-1pt}1\hspace{-2pt}-\hspace{-2pt}\omega\hspace{-1pt})^2\hspace{-2pt}-\hspace{-2pt}2\lambda (\hspace{-1pt}1\hspace{-2pt}-\hspace{-2pt}\omega\hspace{-1pt})^2\right]q(\hspace{-1pt}\tau_2\hspace{-1pt},\hspace{-1pt}\omega\hspace{-1pt})\hspace{-1pt}\right]\hspace{-2pt}dG(\omega)\nonumber\\
&+p_{\tau_1}\hspace{-4pt}\int_0^1\hspace{-4pt}\left[\hspace{-1pt}\mathcal{V}(\hspace{-1pt}q(\hspace{-1pt}\tau_1\hspace{-1pt},\hspace{-1pt}\omega\hspace{-1pt})\hspace{-1pt})\hspace{-2pt}-\hspace{-2pt}\left[\blue{(2\hspace{-2pt}-\hspace{-2pt}\alpha)}\hspace{-1pt}(\hspace{-1pt}1\hspace{-2pt}-\hspace{-2pt}\omega\hspace{-1pt})\hspace{-2pt}+\hspace{-2pt}\lambda \sqrt{1-\omega}\hspace{1pt}\right]q(\hspace{-1pt}\tau_1\hspace{-1pt},\hspace{-1pt}\omega\hspace{-1pt})\hspace{-1pt}\right]\hspace{-2pt}dG(\omega)\nonumber.
\end{align}

Maximizing the integrands point-wise with respect to $q(\tau_i,\omega)$, $i=1,2$, using the first order conditions and setting $\mathcal{V}(q)=q-\frac{1}{2}q^2$, we obtain,
\begin{align}
&q(\tau_2,\omega)=\max\{1\hspace{-1pt}-\hspace{-1pt}(3\hspace{-1pt}-\hspace{-1pt}\blue{2\alpha}\hspace{-1pt}-\hspace{-1pt}2\lambda)(\hspace{-1pt}1\hspace{-2pt}-\hspace{-2pt}\omega\hspace{-1pt})^2\hspace{-2pt},0\},\label{LemmaC-eq2}\\
&q(\tau_1,\omega)=\max\{1-\blue{(2\hspace{-1pt}-\hspace{-1pt}\alpha)}(\hspace{-1pt}1\hspace{-2pt}-\hspace{-2pt}\omega\hspace{-1pt})\hspace{-2pt}-\hspace{-2pt}\lambda \sqrt{1-\omega},0\}.
\end{align}

The value of $\lambda$ must be such that, 
\begin{align}
\lambda \int_0^1\left[q(\tau_2,\omega)2(1-\omega)^2-q(\tau_1,\omega)\sqrt{1-\omega}\hspace{1pt}\right]d\omega= 0.\label{Lemma-eq3}
\end{align}

By numerical evaluation, $\lambda=0$ for $\alpha\geq 0.07$ and $\lambda>0$ for $\alpha\leq 0.07$.

{\begin{center} \includegraphics[width=0.3\textwidth]{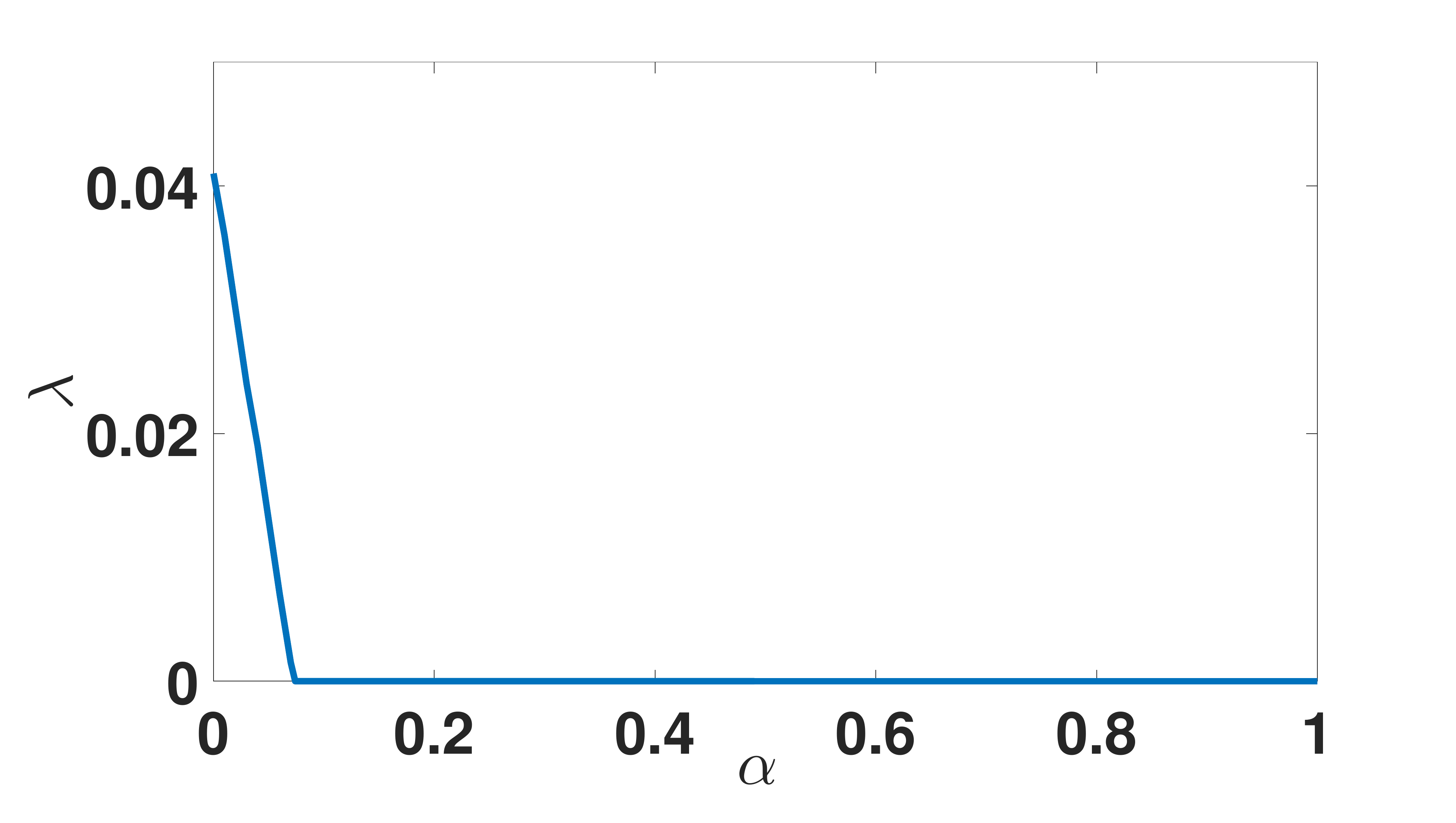} \end{center}}

Therefore, for $\alpha=0.5$, we have,
\begin{align}
&q(\tau_2,\omega)=\max\{1\hspace{-1pt}-\hspace{-1pt}\blue{2}(\hspace{-1pt}1\hspace{-2pt}-\hspace{-2pt}\omega\hspace{-1pt})^2\hspace{-2pt},0\},\label{LemmaC-eq2}\\
&q(\tau_1,\omega)=\max\{1-\blue{1.5}(\hspace{-1pt}1\hspace{-2pt}-\hspace{-2pt}\omega\hspace{-1pt}),0\}.
\end{align}

Using Lemma \ref{lemma-IC2}, the payment functions $t(\tau_i,\omega)$ for $i=1,2$, is  given by,
\begin{align}
&\hspace{-5pt}t(\hspace{-1pt}\tau_2\hspace{-1pt},\hspace{-1pt}\omega\hspace{-1pt})\hspace{-2pt}=\hspace{-2pt}C(\hspace{-1pt}q(\hspace{-1pt}\tau_2\hspace{-1pt},\hspace{-1pt}\omega\hspace{-1pt})\hspace{-1pt};\hspace{-2pt}\Theta(\hspace{-1pt}\tau_2\hspace{-1pt},\hspace{-1pt}\omega\hspace{-1pt})\hspace{-1pt})\hspace{-2pt}+\hspace{-4pt}\int_{\underline{\omega}}^{\omega}\hspace{-6pt}C_\theta\hspace{-1pt}(\hspace{-1pt}q(\hspace{-1pt}\tau_2\hspace{-1pt},\hspace{-1pt}\hat{\omega}\hspace{-1pt})\hspace{-1pt};\hspace{-2pt}\Theta(\hspace{-1pt}\tau_2\hspace{-1pt},\hspace{-1pt}\omega\hspace{-1pt})\hspace{-1pt})\Theta_\omega\hspace{-1pt}(\hspace{-1pt}\tau_2\hspace{-1pt},\hspace{-1pt}\hat{\omega}\hspace{-1pt})d\hat{\omega}\hspace{-1pt},\hspace{-5pt}\\
&\hspace{-5pt}t(\hspace{-1pt}\tau_1\hspace{-1pt},\hspace{-1pt}\omega\hspace{-1pt})\hspace{-2pt}=\hspace{-2pt}C(\hspace{-1pt}q(\hspace{-1pt}\tau_1\hspace{-1pt},\hspace{-1pt}\omega\hspace{-1pt})\hspace{-1pt};\hspace{-2pt}\Theta(\hspace{-1pt}\tau_1\hspace{-1pt},\hspace{-1pt}\omega\hspace{-1pt})\hspace{-1pt})\hspace{-2pt}+\hspace{-4pt}\int_{\underline{\omega}}^{\omega}\hspace{-6pt}C_\theta\hspace{-1pt}(\hspace{-1pt}q(\hspace{-1pt}\tau_1\hspace{-1pt},\hspace{-1pt}\hat{\omega}\hspace{-1pt})\hspace{-1pt};\hspace{-2pt}\Theta(\hspace{-1pt}\tau_1\hspace{-1pt},\hspace{-1pt}\omega\hspace{-1pt})\hspace{-1pt})\Theta_\omega\hspace{-1pt}(\hspace{-1pt}\tau_1\hspace{-1pt},\hspace{-1pt}\hat{\omega}\hspace{-1pt})d\hat{\omega}\hspace{-1pt}.\hspace{-5pt}
\end{align}

\end{document}